\documentclass[apj]{emulateapj}
\usepackage{natbib,graphicx,float,enumerate,amssymb,amsfonts,amsmath,amstext,amsgen,amsopn,amsxtra,esvect,array,tabu,mathtools}
\usepackage[section]{placeins}
\usepackage[normalem]{ulem}
\usepackage[usenames,dvipsnames]{xcolor}
\usepackage{cancel}
\usepackage{rotating}
\usepackage{chngcntr}
\usepackage{enumitem}
\newcommand{\tcross}{{t_{\rm cross}}}
\newcommand{\tgap}{t_{\rm gap}}
\newcommand{\Mstar}{M_\star}
\newcommand{\MJup}{M_{\rm Jup}}

\begin{document}

\title{On the gap-opening criterion of migrating planets in protoplanetary disks}
\author{M. Malik\altaffilmark{1,2}, F. Meru\altaffilmark{1,3}, L. Mayer\altaffilmark{4} and M. Meyer\altaffilmark{1}}
\altaffiltext{1}{ETH Z\"{u}rich, Institute for Astronomy, Wolfgang-Pauli-Strasse 27, CH-8093, Z\"{u}rich, Switzerland}
\altaffiltext{2}{University of Bern, Center for Space and Habitability, Hochschulstrasse 5, CH-3012, Bern, Switzerland}
\altaffiltext{3}{University of Cambridge, Institute of Astronomy, Madingley Road, Cambridge, CB3 0HA, United Kingdom}
\altaffiltext{4}{University of Z\"{u}rich, Institute for Computational Science, Winterthurerstrasse 190, CH-8057 Z\"{u}rich, Switzerland}

\shorttitle{On the gap-opening criterion of migrating planets}
\shortauthors{M. Malik, F. Meru \& L. Mayer}

%%%%%%%%%%%%%%%%%%%%%%%%%%%%%%%%%%%%%%%%%%%%%%%%%%%%%%%%
%%%%%%%%%%%%%%%%%%%% Abstract %%%%%%%%%%%%%%%%%%%%%%%%%%
%%%%%%%%%%%%%%%%%%%%%%%%%%%%%%%%%%%%%%%%%%%%%%%%%%%%%%%%

\begin{abstract}
We perform two-dimensional hydrodynamical simulations to quantitatively explore the torque balance criterion for gap-opening (as formulated by \citealt{cr06}) in a variety of disks when considering a migrating planet. We find that even when the criterion is satisfied, there are instances when planets still do not open gaps. We stress that gap-opening is not only dependent on whether a planet has the ability to open a gap, but whether it can do so \emph{quickly} enough. This can be expressed as an additional condition on the gap-opening timescale, $\tgap$, versus the crossing time, $\tcross$, i.e. the time it takes the planet to cross the region which it is carving out. While this point has been briefly made in the previous literature, our results quantify it for a range of protoplanetary disk properties and planetary masses, demonstrating how crucial it is for gap-opening. This additional condition has important implications for the survival of planets formed by core accretion in low mass disks as well as giant planets or brown dwarfs formed by gravitational instability in massive disks. It is particularly important for planets with \emph{intermediate} masses susceptible to Type III-like migration. For some observed transition disks or disks with gaps, we expect that estimates on the potential planet masses based on the torque balance gap-opening criterion alone may not be sufficient. With consideration of this additional timescale criterion theoretical studies may find a reduced planet survivability or that planets may migrate further inwards before opening a gap.
\end{abstract}


%%%%%%%%%%%%%%%%%%%%%%%%%%%%%%%%%%%%%%%%%%%%%%%%%%%%%%%%
%%%%%%%%%%%%%%%%%%%% Introduction %%%%%%%%%%%%%%%%%%%%%%
%%%%%%%%%%%%%%%%%%%%%%%%%%%%%%%%%%%%%%%%%%%%%%%%%%%%%%%%
\section{Introduction}
\label{introduction}

To predict the evolution of a planet's orbital separation in a gas disk it is crucial to know the conditions leading to different types of migration. Moreover, for a planet to survive the lifetime of the disk ($\approx 1-10$ Myr; \citealt{ha01}; \citealt{ma09}) without falling into the central star the radial motion of the planet needs to be slow enough. This can be achieved, for example, through the formation of a gap around the planet, which through a deficit of disk material in the planet's co-orbital region decreases the angular momentum exchange between the planet and the disk.

\subsection{Gaps in observations}
Annular regions with locally lower surface mass densities have been inferred based on the analysis of spectral energy distributions (SEDs) for decades (e.g. \citealt{ma92}).  However, these interpretations of the data are ambiguous given uncertainties in the opacity and temperature distributions.  In recent years, inner holes (e.g. CoKu Tau 4) and gaps (e.g. GM Auriga) have been confirmed by detailed high signal-to-noise SEDs from the Spitzer Space Telescope (e.g. \citealt{ca05}). Inner holes can be understood either as a natural consequence of disk evolution due to grain growth (and accompanying changes in opacity) or photoevaporation \citep{al13} or as an indication for the presence of planets. As the midplane temperature falls gradually in circumstellar disks, gaps are hard to identify from SEDs unless they are very large.  Gaps are more often identified by direct imaging, either in scattered light or thermal emission.  One caveat to these interpretations is that the appearance of a gap in scattered light could also be explained by shadows in the outer disk due to structures in the inner disk. In most cases of resolved gaps, the innermost disk is unresolved, but inferred from SEDs and confirmed by on-going accretion onto the star.  The inner edge of the outer disk is often $> 20$ AU and can be resolved (e.g. \citealt{ga13}).  Thus there is a growing list of objects with resolved outer disks and strong evidence for an inner disk with gaps $\sim 10$ AU or more.  There are far fewer objects with direct observational evidence for gaps where the outer edge of the inner disk and the inner edge of the outer disk are both resolved in multiple bands (e.g. HD 100546 \citealt{wa14} \& \citealt{pi14} and HD 169142 \citealt{qu13}).

\subsection{Theory of gap formation}

 In the past several criteria have been proposed that quantify the planet and disk conditions leading to gap formation. A pressure stability condition was proposed by \cite{li93} and is given by

\begin{equation}
\label{shock}
 H \lesssim R_H \approx R_{\rm p} (q/3)^{1/3},
\end{equation}
stating that the pressure scale height $H$ in the disk at the planet's location, $R_{\rm p}$, should be smaller than the planet's Hill radius, $R_H$, for a gap to form. As $R_H$ depends on the planet to primary star mass ratio, $q$, a sufficiently large planet mass is needed for gravitationally induced density wakes to result in shocks repelling the disk material from the planet. This condition is also known as the ''strong shock limit'' or the thermal criterion for gap opening. Furthermore \cite{li93} stated that for a gap to form the disk viscosity should be small enough so that the disk's viscous diffusion does not completely negate the mechanism of gap clearing. This condition on the viscosity can be written as

\begin{equation}
M_{\rm p} \gtrsim \frac{40 \nu_{\rm p} M_\star}{\Omega_p R_p^2},
\end{equation}
where $M_\star$ is the primary mass, $\Omega_{\rm p}$, $\nu_{\rm p}$ and $M_{\rm p}$ are the angular Keplerian velocity and the kinematic viscosity at the planet's location and the planet's mass, respectively.  The most general criterion for gap-opening is provided by \cite{cr06}. 
\begin{equation}
\label{cridacrit}
\frac{3}{4}\frac{H}{R_H}+\frac{50 \nu_{\rm p}}{q \Omega_{\rm p} R_{\rm p}^2} \lesssim 1.
\end{equation}
It is a semi-analytical criterion based on the balance between pressure, gravitational and viscous torques for a planet on a fixed circular orbit. Hence we refer to it as the \emph{torque balance criterion}. \cite{cr06} define a gap when the mass density drops to 10\% of the unperturbed density at the planet's location. 

In addition to the effects of the pressure and viscosity which act against gap formation, there is also the time aspect to consider. \cite{ho84} stated that the time required for gap clearing, $t_{gap}$, must be smaller than the planet's migration timescale across its own horseshoe region, $t_{cross}$, i.e. $t_{cross} \gtrsim t_{gap}$. \cite{wa89} refer to this criterion as the ''inertial mass limit'' given by

\begin{equation}
q_{\rm limit} \approx \frac{7}{8}k^2\left(\frac{H}{R_{\rm p}}\right)\left(\frac{\pi\Sigma_{\rm p} H^2}{M_\star}\right)
\label{masslimit}
\end{equation}
as this inequality imposes a lower limit for the planetary mass required for gap-opening. $\Sigma_{\rm p} \propto R^{-k}$ represents the unperturbed surface mass density that would be found at the planet's location. It should be noted that these timescale considerations were calculated for low planetary masses in low-viscosity disks. Hence the problem could be solved analytically using linear approximations for the tidal torque between the disk and planet. \cite{li86} studied the same timescale problem numerically and found the less strict limit $t_{cross} \gtrsim t_\Delta = (H/R) t_{gap}$ by arguing that it is not necessary to open a full gap during the crossing time but it suffices to significantly perturb the surface density to considerably alter the planet's migration regime.

Comparing the pressure stability condition on the planet's mass (eq. \ref{shock}) and \cite{ho84}'s inertial mass limit (eq. \ref{masslimit}) one finds that for a disk with profile $k=3/2$ the pressure stability condition is stricter if

\begin{equation}
M_\star \gtrsim 0.7 \pi R_{\rm p}^2 \Sigma_{\rm p} .
\end{equation}
For typical circumstellar disks this inequality is satisfied and thus planets satisfying the pressure criterion and consequently the stricter torque balance criterion of \cite{cr06} will automatically satisfy \cite{ho84}'s inertial mass limit. However, both criteria use linear analysis for the planet's interaction with the disk which may break down when a planet grows in mass.

We are curious that the inertial limit is dependent on the disk mass and surface mass density profile (eq. \ref{masslimit}), but both parameters are missing in the torque balance criterion (eq. \ref{cridacrit}). In addition \cite{cr06} neglected planet migration. However, since these parameters directly affect the angular momentum exchange between a planet and a disk, and hence planet migration, they may affect the planet's ability to open a gap.

In the last few years the pressure stability criterion (eq. \ref{shock}) or the torque balance criterion (eq. \ref{cridacrit}) have been widely used to predict or explain numerical studies of planet formation/synthesis using various disk models with some investigations of high-mass companions in massive disks (e.g. \citealt{al04}; \citealt{id04}; \citealt{zh12}; ; \citealt{fo13}; \citealt{vo13}; \citealt{ga14}). 

%%%%%%%%%%%%%%%%%%%%
\subsection{Outline of this paper}%%%
%%%%%%%%%%%%%%%%%%%%

As both the torque balance criterion and inertial mass limit are based on linear analyses of planet-disk interactions we are hence intrigued by the following questions: 

\vspace{0.15cm}

\begin{itemize}[nosep]
\item{Is the timescale condition still implicitly covered by the torque criterion in migration scenarios where non-linear torques prevail?} 
\item{Can the torque balance criterion of \cite{cr06} be safely applied to disk models that do not necessarily offer the conditions usually assumed for a ''pure'' Type I migration?}
\item{Is gap-opening affected by the disk mass?}
\end{itemize}

\vspace{0.15cm}

We note that when considering gap-opening in disks (e.g. when interpreting observations) the criterion based on the balance of torques is always used but that based on the timescale is often not considered.  In this work we point out that one needs to consider not just whether a planet can open a gap (based on the balance of torques) but also whether it can do it quickly enough in a variety of disks expected to form companions by either core accretion or gravitational instability.

As a broad distribution of disk and companion masses exists we choose two regimes for our studies: massive self-gravitating (SG) disks with migrating companions of the (sub-)brown dwarf regime and disks based on the Minimun Mass Solar Nebula (MMSN) model where we introduce Jupiter-mass planets. We let these companions migrate freely and compare our findings to the prediction of the torque balance criterion given by eq. \ref{cridacrit}.

In Section~\ref{sec:notation} we outline our notation.  We describe our numerical method and simulations performed in Sections \ref{Code} and \ref{simulations}, respectively.  We then present our results in Section~\ref{results}.  Finally, in Section~\ref{discussion} we discuss our results in the context of both theories and observations, and present our conclusions in Section~\ref{conclusion}.

%%%%%%%%%%%%%%%%%%%%%%%%%%%%%%%%%
\subsection{Notation}%%%%%%%%%%%%
%%%%%%%%%%%%%%%%%%%%%%%%%%%%%%%%%
\label{sec:notation}

We refer to our migrating objects as ''companions'' regardless of whether they would commonly fit into the brown dwarf or giant planet mass regime.

The companion to primary mass ratio is defined as $q \equiv M_{\rm c}/M_\star$ and the companion's Hill radius is described by $R_H \approx R_{\rm c} (q/3)^{1/3}$. The subscript ''$c$'' indicates that the parameter is evaluated for the companion or at the companion's position. $H=c_s/\Omega $ is the pressure scale height in the disk with $\Omega = \sqrt{G M_\star/ R^3}$ the Keplerian angular velocity for a circular orbit and $h=H/R$ is the disk aspect ratio. Finally, we write the physical disk viscosity $\nu$ according to the $\alpha$-disk model $\nu = \alpha c_s H$, with the dimensionless stress parameter $\alpha$ \citep{SS_viscosity}.

%%%%%%%%%%%%%%%%%%%%%%%%%%%%%%%%%%%%%%%%%%%%%%%%%%%%%%%%
%%%%%%%%%%%%%%%%%%%% Method %%%%%%%%%%%%%%%%%%%%%%%%%%%%
%%%%§§§§§§§§%%%%%%%%%%%%%%%%%%%%%%%%%%%%%%%%%%%%%%%%%%%%%%%%%%%%
\section{Numerical method}
\label{Code}

The simulations are carried out using the polar 2D grid-based hydrodynamics code, {\sc fargo-adsg}\footnote{See fargo.in2p3.fr}, originally developed by \cite{ma00} and later extended to implement a self-gravity solver based on fast Fourier transforms \citep{baru08}. The disk's grid is described by polar coordinates ($R,\Theta$) with the star at the origin. An indirect term is included, which accounts for the frame acceleration due to the gravitational interaction of the bodies. The grid has open boundaries. The companions are represented by point masses and self-gravity is included in all of our simulations.

%%%%%%%%%%%%%%%%%%%%%%%
\subsection{Equations of state}%%%%
%%%%%%%%%%%%%%%%%%%%%%%

Our disk models include two distinctive regimes. For gravitationally unstable disks in which planets are expected to form by gravitational instability (see section \ref{SGdisks}) we use an adiabatic equation of state together with the energy equation

\begin{equation}
\frac{\partial u}{\partial t}+\overrightarrow{\nabla}(u\overrightarrow{v})=-p\overrightarrow{\nabla}\cdot\overrightarrow{v}+Q_+-Q_-,
\end{equation}
where $u$ is the thermal energy density, $\overrightarrow{v}$ is the flow velocity and $p$ is the vertically integrated pressure \citep{ba08}. The first term on the right hand side describes the compressional heating due to physical processes and $Q_+$ expresses the shock heating due to artificial bulk viscosity. $Q_- = u / t_{\rm cool}$ is the cooling term with $t_{\rm cool}$ being the cooling timescale. The equation of state is written by

 \begin{equation}
\label{1}
p=(\gamma - 1)u = \Sigma R_{spec} T.
\end{equation}
where $R_{spec}= R_{univ}/(\mu m_p N_A)$ is the specific gas constant ($R_{univ}$ is the universal gas constant, $\mu = 2.4$ is the mean molecular weight of the gas particles and $m_p$ is the proton mass), $N_A$ is the Avogadro Number and $T$ is the temperature. The sound speed, $c_s$, is given by $c_s^2 = (\partial p / \partial \Sigma)$, which for the adiabatic model ($p = const\cdot\Sigma^\gamma$) leads to $c_{s,adiab} = \sqrt{\gamma p/\Sigma}$. The adiabatic index is $\gamma=5/3$.

For MMSN-like disks (see section \ref{mmsndisks}) we use an isothermal equation of state without solving an energy equation. In these isothermal runs the viscosity is added in the momentum equation only. Here, the sound speed is given by $c_{s,iso} = c_{s,adiab}/\sqrt{\gamma}$.

%%%%%%%%%%%%%%%%%%%%%%%%%%%%%%%%%%%%%%%
\subsection{Gravitational instability}%
\label{sec:SG}
%%%%%%%%%%%%%%%%%%%%%%%%%%%%%%%%%%%%%%%

The gravitationally unstable disks are set up so that the self-gravitating structure is present but the disks do not fragment (so that the interactions between just a single companion and its disk can be considered).  The fragmentation of disks requires two conditions to be satisfied. Firstly the \citep{to64} criterion should be fulfilled, i.e.

\begin{equation}
 Q=\frac{c_s\Omega}{\pi G \Sigma} \lesssim 1,
\end{equation}
where Q is the dimensionless Toomre parameter. The second condition for fragmentation requires the disk to cool on a fast timescale \citep{ga01}. Expressing the cooling timescale as a dimensionless cooling parameter $\beta =  t_{cool} \Omega$ the critical value below which the disk fragments was estimated to be $\beta_{crit} \approx 3$ and 7 for a ratio of specific heats of 2 and 5/3, respectively (\citealt{ga01}; \citealt{ri05}). Recently, \cite{me11a} showed that the numerical simulations leading to the latter result were limited by the resolution. Upon increasing the resolution they found that $\beta_{crit} > 20$ and may even be as large as 30 as a larger resolution involves more heating from the artificial viscosity employed in numerical simulations to model shocks correctly \citep{Meru_Bate_convergence}.  While this is still an active area of research (e.g. \citealt{ro12}, \citealt{pa12}, \citealt{ri14}, \citealt{pa11}, \citealt{lo11}) we set the cooling rate per unit area to be $Q_- = u / t_{\rm cool} = (u\Omega) / \beta$, where $\beta = 30$, to ensure no fragmentation occurs, and the Toomre parameter stabilizes in these equations to $Q \approx 2$ throughout the disk.

If we assume that the transport of angular momentum due to gravitoturbulence occurs locally\footnote{This does not have to be generally true as was pointed out by \cite{ba99} and \cite{co09}.} we can connect the stress parameter $\alpha$ with the cooling parameter $\beta$ by \citep{ga01}
\begin{equation}
\label{beta}
\alpha = \frac{4}{9}\frac{1}{\gamma(\gamma-1)\Omega t_{cool}} = \frac{4}{9}\frac{1}{\gamma(\gamma-1)\beta}.
\end{equation}
This indicates that a more rapid cooling increases the viscous stress $\alpha$ in the disk as the self-gravity is less efficiently countered by heating.

%%%%%%%%%%%%%%%%%%%%%%%%%%%%%%%%%%%%%%%%
\subsubsection{Mechanical equilibrium}%%
\label{equi}%%%%%%%%%%%%%%%%%%%%%%%%%%%%
%%%%%%%%%%%%%%%%%%%%%%%%%%%%%%

A SG disk is self-regulating in the sense that internal cooling and shock heating due to gravitoturbulence leads to a Toomre $Q$ parameter near unity. The disk material rearranges naturally until a mechanical equilibrium with $\partial \Sigma / \partial t = 0$ is reached. Using the viscous evolution equation for Keplerian disks in circular motion from \cite{ly74} this condition becomes \citep{pr81}: 

\begin{equation}
\label{eq}
\frac{\partial \Sigma}{\partial t} = \frac{3}{R}\frac{\partial}{\partial R}\left[R^{1/2}\frac{\partial}{\partial R}\left(\nu\Sigma R^{1/2}\right)\right] = 0 .
\end{equation}
In our set-up the SG disks rearrange to the state $\Sigma \propto R^{-3/2}$ and $T = const$, which solves the above equation when inserting $\nu= \alpha c_s H$ and using the fact that the self-gravity mechanism leads to a uniform Toomre parameter, $Q=const$.

\vspace{0.4cm}

%%%%%%%%%%%%%%%%%%%%%%%%%%%%%%
\subsection{Softening length}%
%%%%%%%%%%%%%%%%%%%%%%%%%%%%%%

2D polar disk models do not strictly simulate the physical impact of the vertical dimension of a realistic disk. For this reason a softening length $\epsilon_{\rm c}$ is introduced which mimics the dampening of the companion's potential in a disk with finite thickness given by

\begin{equation}
\Phi_{\rm c} = - \frac{GM_{\rm c}}{(d^2+\epsilon_{\rm c}^2)^{1/2}},
\end{equation}
where $\Phi_{\rm c}$ is the companion's potential and $d$ is the distance from the companion \citep{kl12}. Furthermore the utilization of a softening length of the order of a significant fraction of the disk scale height avoids mesh singularities in numerical grid-based simulations. The self-gravity of the disk is treated analogously with the softening length $\epsilon_{\rm SG}$. \cite{mu12} extensively studied the effects of various softening lengths. In their Figure 4 they show which value of the softening length should be considered to obtain the best approximation of a realistic physical force at a given distance from the companion. At very close ranges the suggested value varies significantly with the distance from the companion.  However, beyond a distance $\approx 0.7H$ from the companion the value of $\epsilon_{\rm c} \approx 0.7H$ remains below an error of less than $10\%$ from the optimum value. Following this reasoning we choose $\epsilon_{\rm c} = 0.7H$ and analogously, based on their Figures 13 \& 15 we choose $\epsilon_{\rm SG}=0.8H$.

%%%%%%%%%%%%%%%%%%%%%%%%%%%%%%%%%%%%%%%%%%%%%%%%%%%%%%%%
%%%%%%%%%%%%%%%%%%%% Simulations %%%%%%%%%%%%%%%%%%%%%%%
%%%%%%%%%%%%%%%%%%%%%%%%%%%%%%%%%%%%%%%%%%%%%%%%%%%%%%%%
\section{Simulations}
\label{simulations}

\begin{table*}
	\caption{Initial parameters for the SG disk models.}
	\label{init}
	\vspace{-0.4cm}
\begin{center}
\bgroup
\def\arraystretch{1.5}
  \begin{tabular}{|l|l|l|}
    \hline
disk model & reference SG disk & lighter SG disk \\ \hline
\multicolumn{3}{|c|}{{\bf parameters common to both disks}} \\ \hline
grid size & \multicolumn{2}{c|}{$20 - 250$ AU} \\
number of grid cells & \multicolumn{2}{c|}{1536 (azimuthal) x 516 (radial,[log])} \\
softening lengths &\multicolumn{2}{c|}{$\epsilon_{\rm c} = 0.7 H$, $\epsilon_{\rm SG} = 0.8 H$} \\
primary mass & \multicolumn{2}{c|}{$M_\star = 1 M_\odot$}  \\
$\alpha$-stress parameter &\multicolumn{2}{c|}{$\alpha \approx 0.013$} \\
aspect ratio & \multicolumn{2}{c|}{$h=0.1$}  \\
temperature & \multicolumn{2}{c|}{$T  = 15 $ K $ \left(\frac{R}{100 {\rm AU}}\right)^{-1}$}  \\ \hline
\multicolumn{3}{|c|}{{\bf disk specific parameters}} \\ \hline
disk mass & $M_{\rm disk} = 0.4 M_\odot$ & $M_{\rm disk} = 0.2 M_\odot$ \\
surface mass density & $\Sigma = 25$ g cm$^{-2} \left(\frac{R}{100 {\rm AU}}\right)^{-2}$ & $\Sigma = 12.5$ g cm$^{-2} \left(\frac{R}{100 {\rm AU}}\right)^{-2}$ \\
Toomre $Q$-parameter & $Q = 1.7$& $Q = 3.3$ \\
    \hline
  \end{tabular}
	\egroup
	\end{center}
\end{table*}

In this section we describe the disk configurations used in our simulations. First, we focus on massive gravitationally unstable disks. Then we switch to the low-mass end of the common circumstellar disk distribution by setting up disks based on the MMSN model to examine whether our results hold for a variety of disks and companion masses.

\begin{table*}
\vspace{-0.4cm}
\caption{Parameters for the SG disk models at the time of companion introduction, when the self-gravity has already developed.}
	\label{rearr}
\vspace{-0.4cm}
\begin{center}
\bgroup
\def\arraystretch{1.5}
  \begin{tabular}{|l|l|l|}
    \hline
disk model & reference SG disk & lighter SG disk \\ \hline
\multicolumn{3}{|c|}{{\bf parameters at companion introduction}} \\ \hline
aspect ratio & $h \approx 0.11\left(\frac{R}{100 {\rm AU}}\right)^{1/2}$ & $h \approx 0.06\left(\frac{R}{100 {\rm AU}}\right)^{1/2}$ \\
temperature & $T  \approx 17$ K & $T  \approx 6.2$ K \\
disk mass & $M_{\rm disk} = 0.29 M_\odot$ & $M_{\rm disk} = 0.15 M_\odot$ \\
surface mass density & $\Sigma \approx 18.3$ g cm$^{-2} \left(\frac{R}{100 {\rm AU}}\right)^{-3/2}$ & $\Sigma \approx 9.6$ g cm$^{-2} \left(\frac{R}{100 {\rm AU}}\right)^{-3/2}$ \\
Toomre $Q$-parameter & $Q\approx 1.6$& $Q\approx 2.0$ \\
    \hline
  \end{tabular}
	\egroup
	\end{center}
\end{table*}

\begin{table*}
\vspace{-0.4cm}
	\caption{Parameters for the MMSN-like disk models at the time of companion introduction.}
	\label{mmsntab}
	\vspace{-0.4cm}
\begin{center}
\def\arraystretch{1.5}
  \begin{tabular}{|l|l|l|l|}
    \hline
disk model &  lighter MMSN-like disk & intermediate MMSN-like disk & heavier MMSN-like disk \\ \hline 
\multicolumn{4}{|c|}{{\bf common parameters}} \\ \hline
grid size & \multicolumn{3}{c|}{$2.5 - 30$ AU} \\
number of grid cells & \multicolumn{3}{c|}{1536 (azimuthal) x 508 (radial,[log])}   \\
softening length &\multicolumn{3}{c|}{$\epsilon_{\rm c} = 0.7 H$, $\epsilon_{\rm SG} = 0.8 H$} \\ 
primary mass & \multicolumn{3}{c|}{$M_\star = 1 M_\odot$} \\
$\alpha$-stress parameter &\multicolumn{3}{c|}{$\alpha = 0.0013$}  \\
aspect ratio &\multicolumn{3}{c|}{$h=0.05\left(\frac{R}{5 {\rm AU}}\right)^{1/2}$} \\
temperature &\multicolumn{3}{c|}{$T = 129$ K} \\ \hline
\multicolumn{4}{|c|}{{\bf disk specific parameters}} \\ \hline
disk mass & $M_{\rm disk} = 0.02 M_\odot$ & $M_{\rm disk} = 0.05 M_\odot$ & $M_{\rm disk} = 0.08 M_\odot$ \\
surface mass density & $\Sigma = 338$ g cm$^{-2} \left(\frac{R}{5 {\rm AU}}\right)^{-3/2}$ & $\Sigma = 845$ g cm$^{-2} \left(\frac{R}{5 {\rm AU}}\right)^{-3/2}$ & $\Sigma = 1352$ g cm$^{-2} \left(\frac{R}{5 {\rm AU}}\right)^{-3/2}$\\
Toomre $Q$-parameter & $Q = 18$& $Q = 7.0$ & $Q = 4.4$ \\ \hline
  \end{tabular}
	\end{center}	
\end{table*}

%%%%%%%%%%%%%%%%%%%%%
\subsection{Massive SG disk models}%%
\label{SGdisks}%%%%%%%%%%%%%
%%%%%%%%%%%%%%%%%%%%%
We use two different disk set-ups for our massive SG disk simulations: 

\begin{enumerate}

\item{The initial disk condition by \cite{ba11}. We refer to this as the ''reference'' configuration.}

\item{The same as (1) but with half the initial disk mass, herein after called the "lighter SG disk".}

\end{enumerate}

The initial conditions for the two SG disk models are described in Table \ref{init}. Most of the parameters are defined directly in the code. Others are evaluated indirectly, e.g. the temperature $T$, which is found by inserting the expression for the sound speed into eq. \ref{1}, solving for $T$ and using the relation $c_s = hR\Omega$. Further we estimate the $\alpha$-parameter using eq. \ref{beta} with $\beta=30$. In general $\alpha$ is measured as the sum of the gravitational and Reynolds stresses in the disk. As we employ a similar disk set-up as \cite{ba11}, we adopt their findings (see their Fig. 2) suggesting that the real value of $\alpha$ might deviate from theory (equation \ref{beta}) by up to factor of two. The implication of such a deviation is discussed in section \ref{add}.

We let the self-gravity develop over 30 orbits (orbital timescale at 100 AU) in the reference SG disk and over 90 orbits in the lighter SG disk\footnote{Due to the lower disk mass the self-gravity mechanism develops slower and the mechanical equilibrium is achieved at later times in the lighter SG disk compared to the reference SG disk.} so that the disk reaches a quasi-steady state without any companions. During this time the disk parameter profiles rearrange as described in subsection \ref{equi} and the resulting parameters can be seen in Table \ref{rearr}. Note that due to gravitoturbulence some of the parameters deviate somewhat from their values shown in Table \ref{rearr} as they suffer from stochastic fluctuations. The value of $Q$ at a certain radial separation is estimated by

\begin{equation}
\label{qplot}
Q=\frac{\left<c_s\right>_{\rm {azimuth}}\left<\Omega\right>_{\rm {azimuth}}}{\pi G \left<\Sigma\right>_{\rm {azimuth}}},
\end{equation}
where $\left\langle...\right\rangle_{\rm azimuth}$ denotes the azimuthal average of the physical quantity.

Note that for our numerical studies we use simplifying models of circumstellar disks and so their parameters are not necessarily of the form found through observations, but are chosen so that the results can be understood under controlled conditions.

%%%%%%%%%%%%%%%%%%%%%%%%%%%
\subsection{MMSN-like disk models}%
\label{mmsndisks}%%%%%%%%%%
%%%%%%%%%%%%%%%%%%%%%%%%%%%

For the simulations at the lower end of the disk mass distribution we set-up disks similar to a MMSN disk \citep{ha81} and the disk model utilized by \cite{cr06}, i.e. with typical T Tauri disk properties in which planets would be expected to form by core accretion. The disks extend between 2.5 - 30 AU with a surface mass density profile $\Sigma \propto R^{-3/2}$ and a constant temperature across the whole disk.

We model three disks with different masses all satisfying the above conditions as follows: 

\begin{enumerate}

\item{a disk with $M_{\rm disk}=0.02 M_\odot$, which we call herein the ''lighter MMSN disk''.}

\item{a disk with $M_{\rm disk}=0.05 M_\odot$, wich we call herein the ''intermediate MMSN disk''.}

\item{a disk with $M_{\rm disk}=0.08 M_\odot$, which we call herein the ''heavier MMSN disk''.}

\end{enumerate}

The detailed initial parameters for the disks are found in Table \ref{mmsntab}.

\begin{figure*}
\begin{center}
\begin{minipage}[t]{0.32\textwidth}
\includegraphics[width=\textwidth]{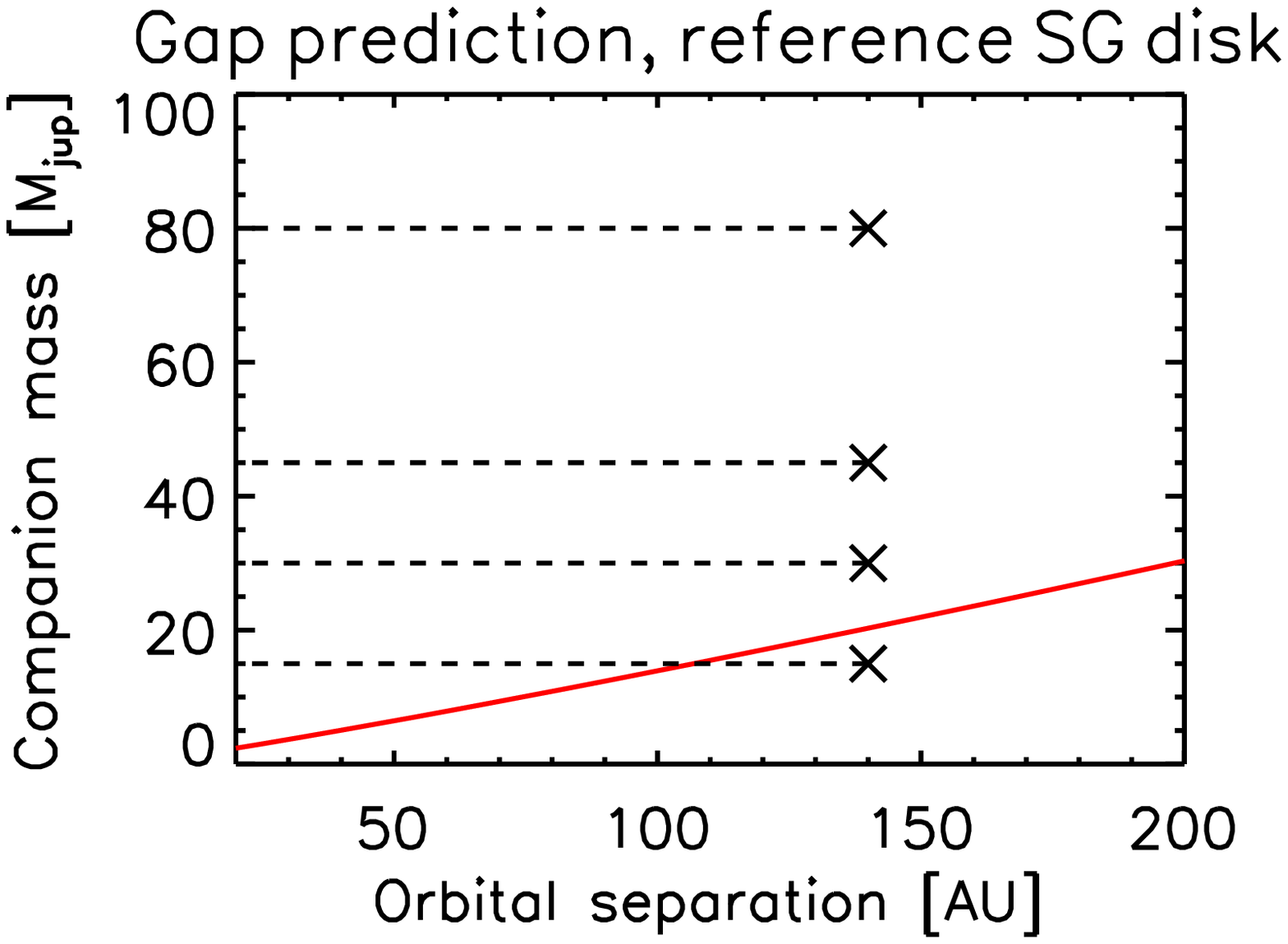}
\end{minipage}
\hfill
\begin{minipage}[t]{0.32\textwidth}
\includegraphics[width=\textwidth]{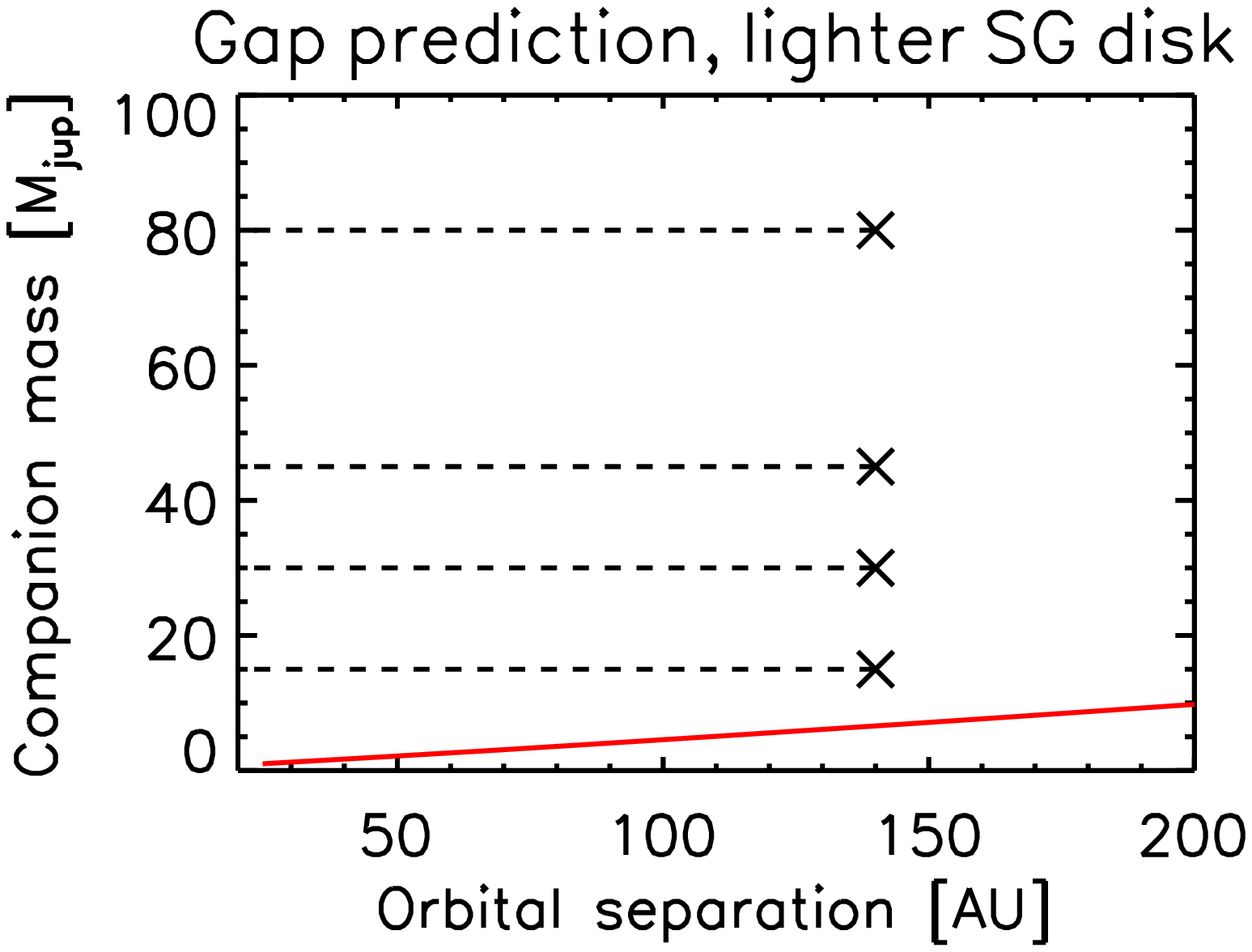}
\end{minipage}
\hfill
\begin{minipage}[t]{0.32\textwidth}
\includegraphics[width=\textwidth]{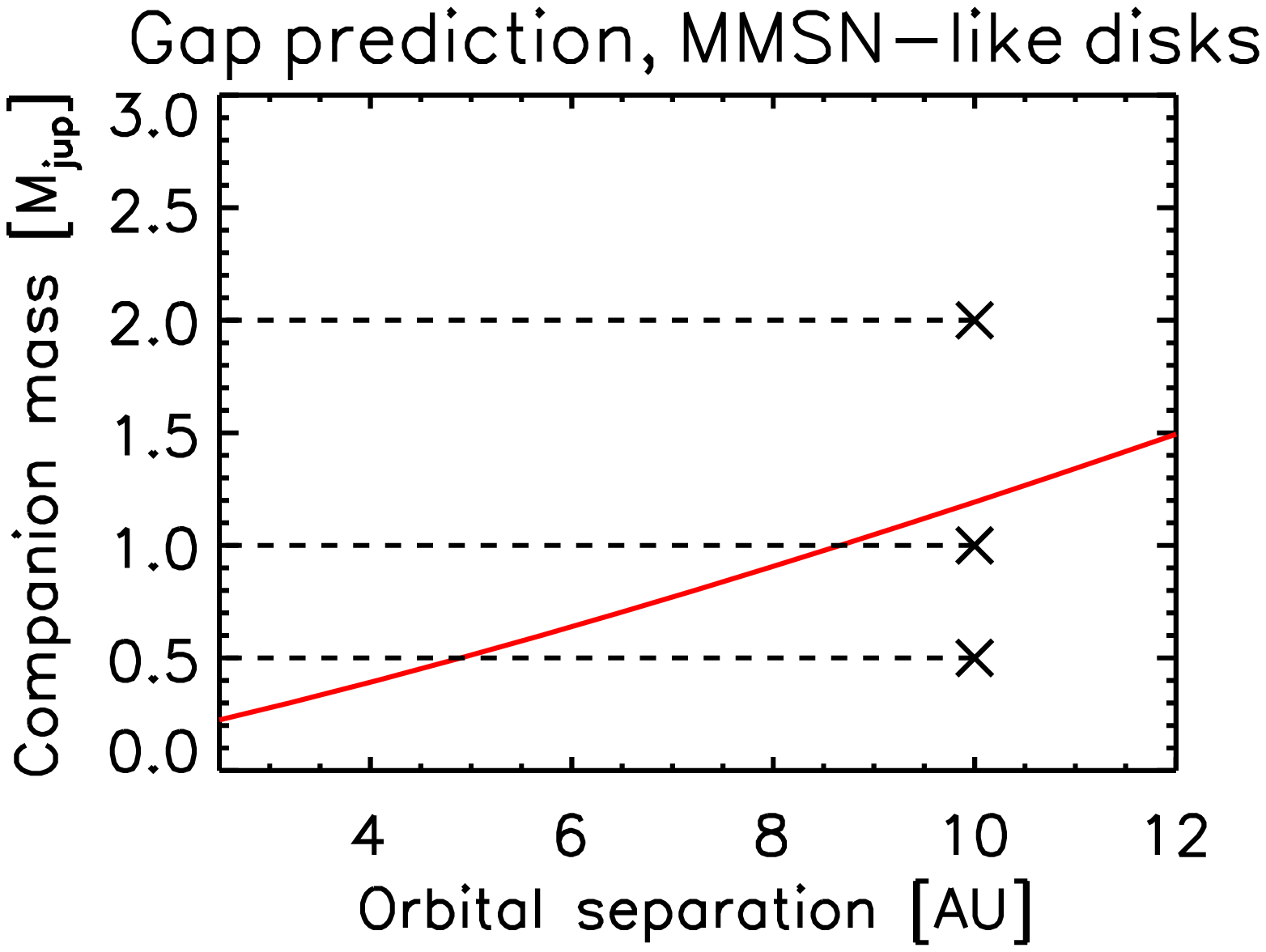}
\end{minipage}
\vspace{-0.3cm}
\caption{Companion's mass versus orbital radius indicating the torque balance criterion of \protect\cite{cr06} (solid red line). Companions are expected to open a gap when situated above the limit. The crosses indicate where in the disk the companions are introduced in our simulations while the dashed lines indicate the inward migration tracks. {\bf Left:} Prediction for the reference SG disk. {\bf Middle:} Prediction for the lighter SG disk. {\bf Right:} Prediction for the three MMSN-like disks. All simulations are expected to show gap-opening at some stage during an inwards migration.}
\label{criteria}
\end{center}
\end{figure*}

%%%%%%%%%%%%%%%%%%%%%%%%%%%%%%%%%%%%%%%%%%
\subsection{Predictions for gap-opening}%%
%%%%%%%%%%%%%%%%%%%%%%%%%%%%%%%%%%%%%%%%%%
\label{predictions}

Fig. \ref{criteria} shows the prediction of the semi-analytical criterion (eq. \ref{cridacrit}, \citealt{cr06}) in each of our disk set-ups in a companion mass versus orbital radius diagram. The solid red line represents the criterion's limit, i.e. eq. \ref{cridacrit} with an equality, calculated using the values in Tables \ref{rearr} and \ref{mmsntab} for the SG and MMSN-like disks, respectively. Companions situated above this line are expected to open a gap. Note that this criterion is \emph{not} dependent on the disk mass so the graph for the MMSN-like disks is applicable to all the disk masses considered since the thermal and viscous properties remain the same. In the SG disks the thermal properties vary with disk mass, hence the need for separate predictions for different disk masses.

We allow a single companion in each of our disk set-ups to migrate freely and compare its ability to open gaps with the prediction of the torque balance criterion of \cite{cr06} in equation (\ref{cridacrit}). To be consistent with their approach we regard a gap as a one order of magnitude decrease in the surface mass density in the companion's horseshoe region compared to the unperturbed density. The crosses in Fig. \ref{criteria} show the radial location where each companion is introduced and the dashed horizontal lines indicate the inward migration tracks. All companions are expected to start gap clearing eventually as they cross the criterion's line because of their inward migration.

To test the criterion using our reference SG disk model we choose 15, 30, 45 and 80 $\MJup$ for the companion masses. 
While these values are higher than the initial clump mass estimates of e.g. \cite{bo10} given our disk parameters, clumps may accrete gas and hence companions formed by gravitational instability could be much more massive during their migration than at the time they form (e.g. \citealt{zh12}).

Furthermore, we want to stress that the purpose of this study is a numerical experiment to test the gap-opening criterion. We therefore choose companion masses that are both above and below the critical mass limit given by the red line in Fig. \ref{criteria}, left panel. To highlight the significance of the disk mass on the simulation results we utilize the same companion masses in our lighter SG disk. In the MMSN-like disks we employ 0.5, 1 and 2 $\MJup$ mass companions with the same intention.

%%%%%%%%%%%%%%%%%%%%%%%%%%%%%%%%%%%%%%%%%%%%%%%%%%
\subsection{Introduction of the companions}%%%%%%%%%%%%%%%%%%%%%%%%%%%
\label{intro}%%%%%%%%%%%%%%%%%%%%%%%%%%%%%%%%%%%%%%%%%%
%%%%%%%%%%%%%%%%%%%%%%%%%%%%%%%%%%%%%%%%%%%%%%%%%

First, the companions are inserted into the SG disks at an orbital separation of 140 AU. A companion is introduced with a linear mass accumulation over one orbital period at 100 AU for the massive SG disk simulations (in Sect. \ref{SGdisks}) and one orbital period at 5 AU for the MMSN-like disk simulations (see Sect. \ref{mmsndisks}). Since the companions are not fixed to a particular radial separation they start to migrate immediately after introduction, as soon as they possess a sufficient amount of mass to exchange enough angular momentum with their surroundings to induce radial movement. Since the net torque acting on the companion is predominantly negative the direction of migration should be inwards in most cases. 

The companions are not allowed to accrete mass from the surrounding disk during the migration. However they may still trap disk material in their gravitational influence radius (i.e. the companion's circumplanetary disk), which might have an impact on the angular momentum transfer between the companion and the disk. Therefore while this additional mass is accounted for when computing the torques responsible for gap formation, the additional mass is not used to test the criterion directly. 

\begin{figure*}
\begin{center}
\begin{minipage}[t]{0.48\textwidth}
\includegraphics[width=\textwidth]{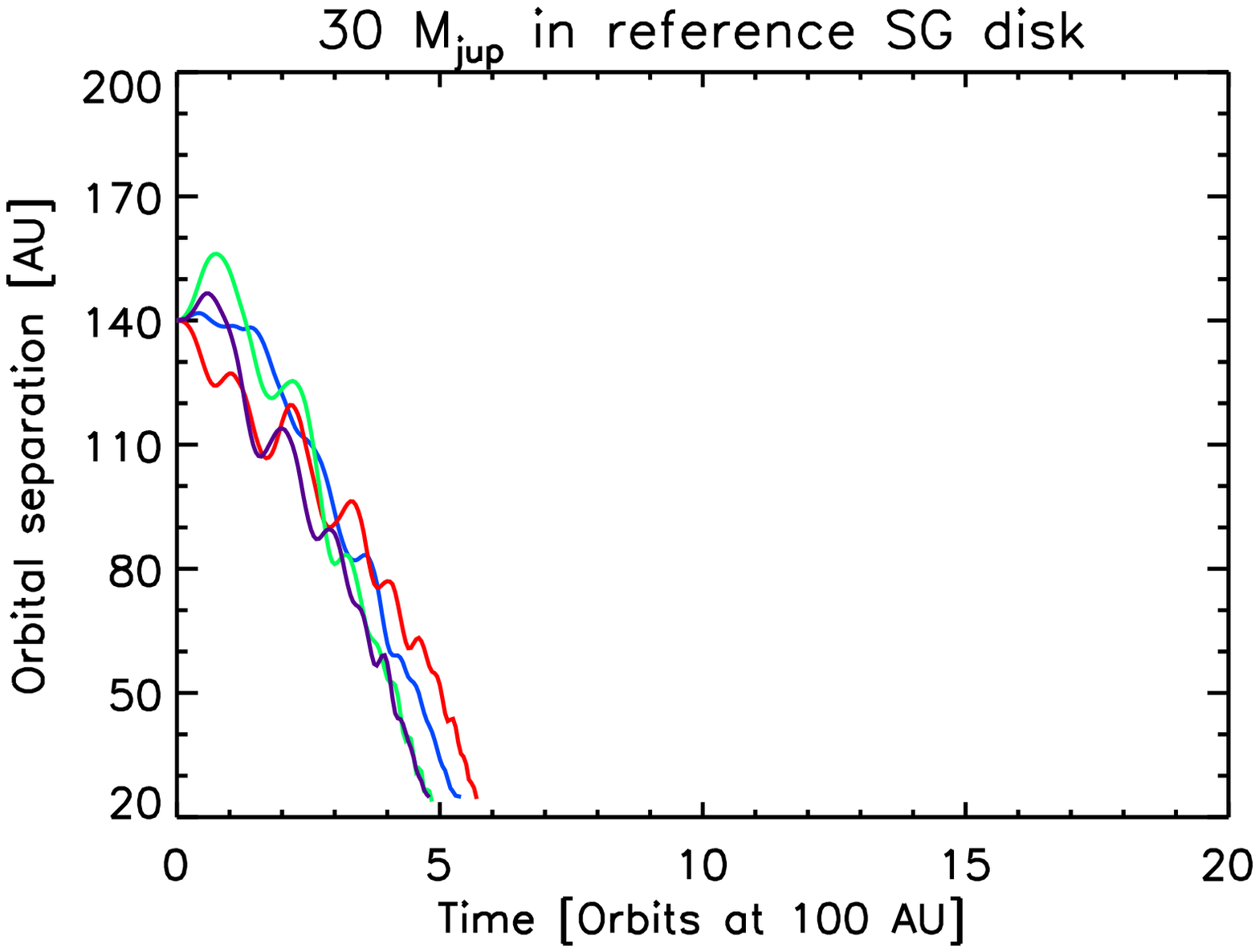}
\end{minipage}
\hfill
\begin{minipage}[t]{0.48\textwidth}
\includegraphics[width=\textwidth]{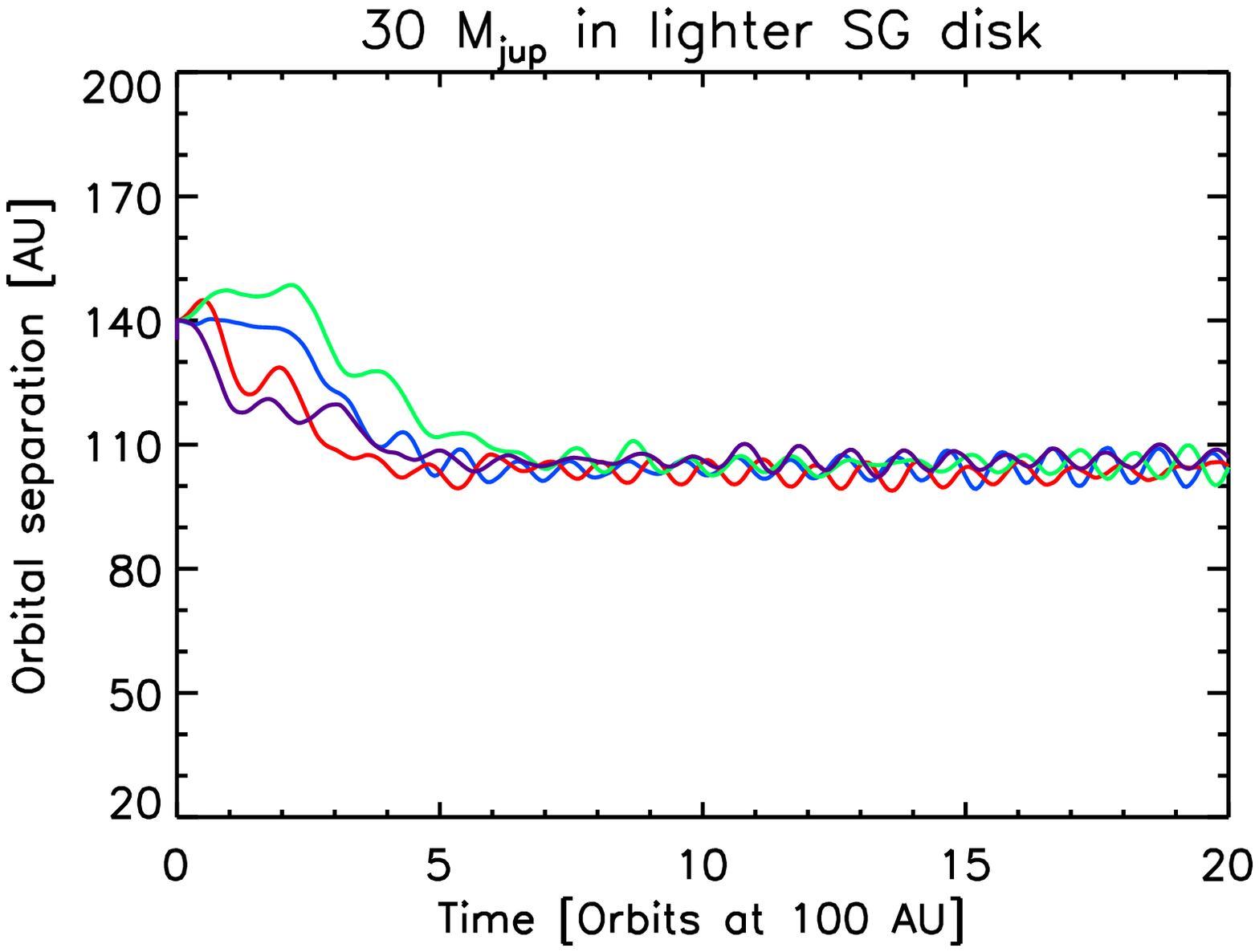}
\end{minipage}
\begin{minipage}[t]{0.48\textwidth}
\includegraphics[width=\textwidth]{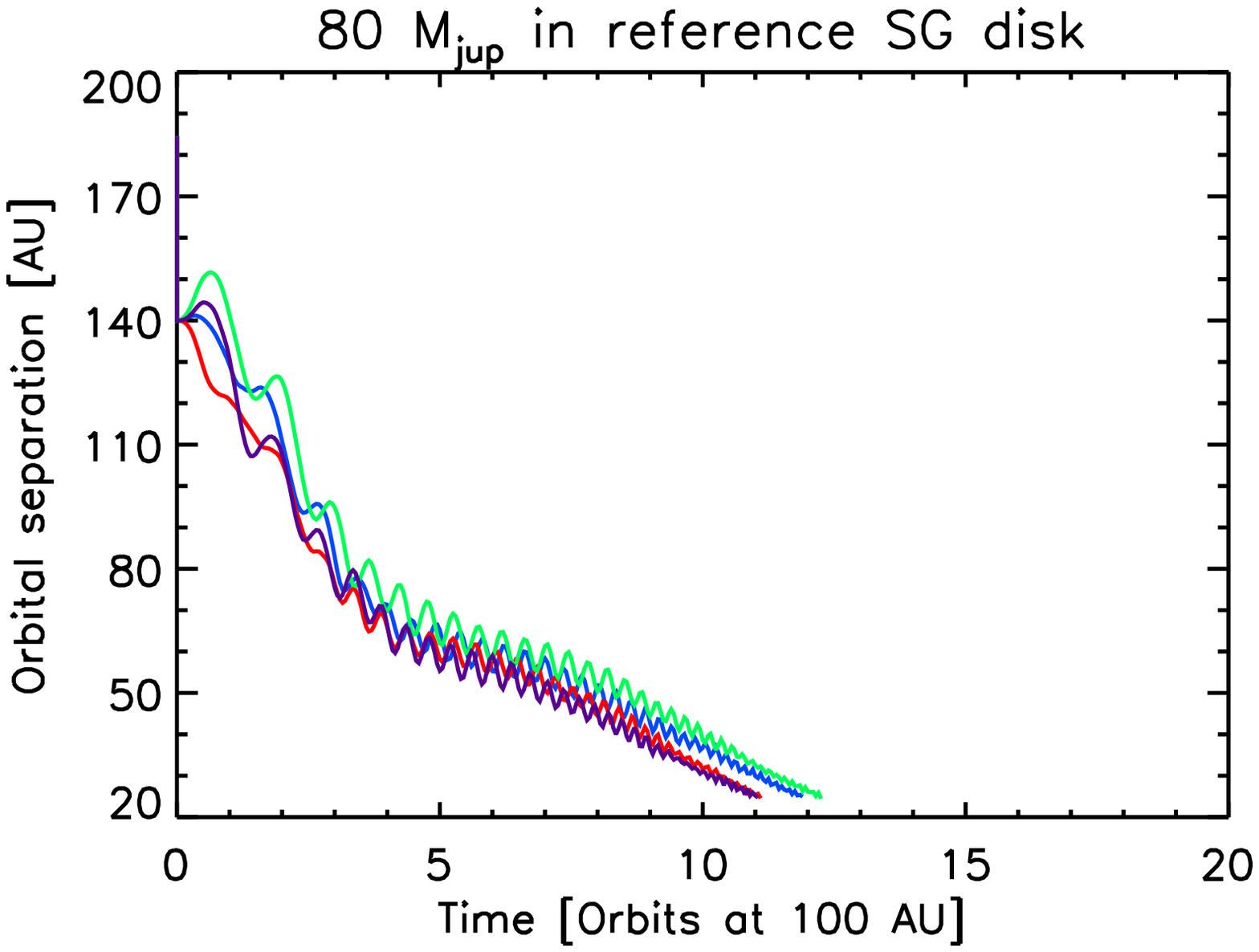}
\end{minipage}
\hfill
\begin{minipage}[t]{0.48\textwidth}
\includegraphics[width=\textwidth]{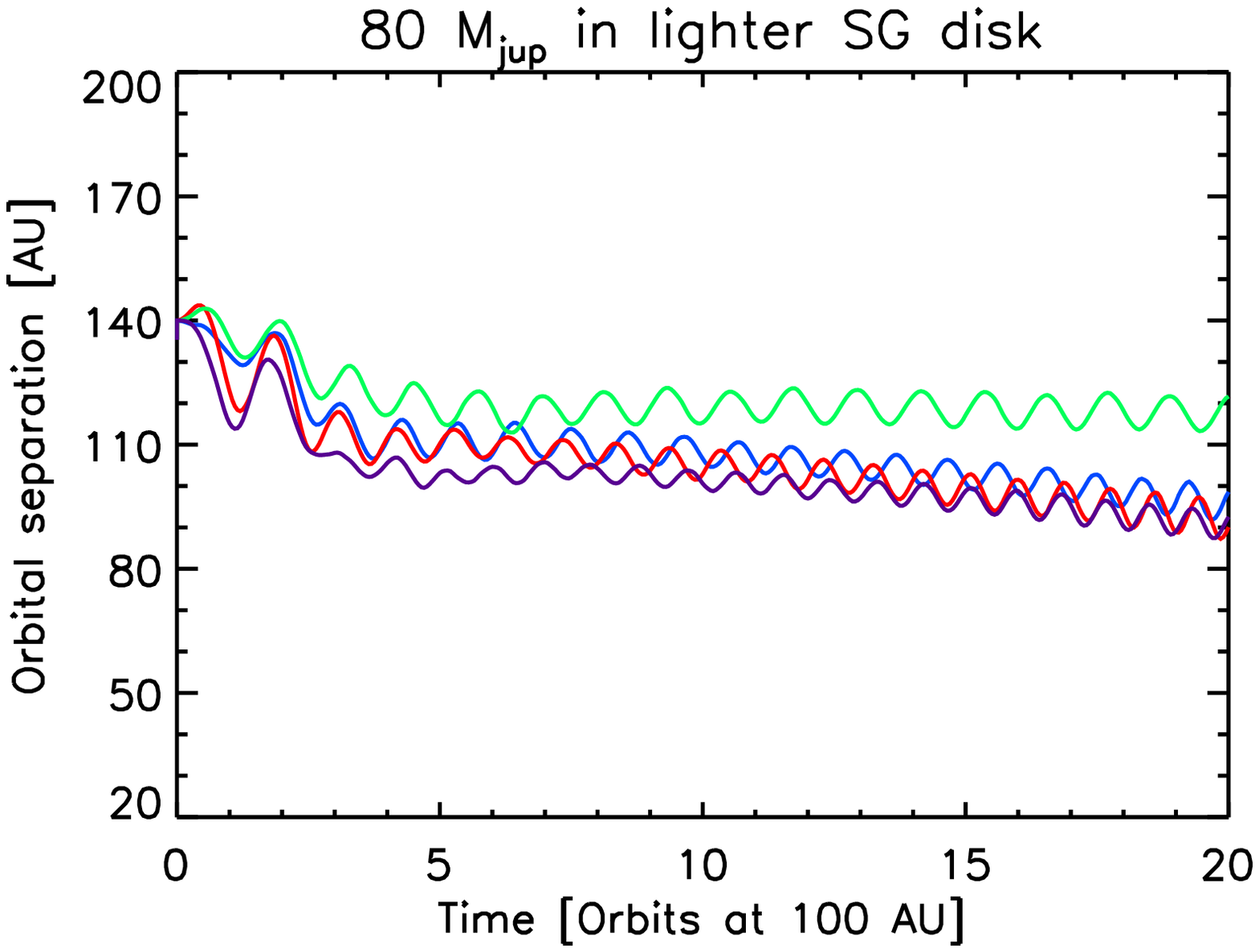}
\end{minipage}
\vspace{-0.3cm}
\caption{Evolution of the orbital separation of a migrating 30 $M_{\rm Jup}$ (top) and 80 $M_{\rm Jup}$ (bottom) companion in the reference (left) and lighter (right) SG disks. The simulations are performed four times starting at the same orbital radius but each run at a different azimuth leading to four separate planetary trails (displayed with different colors). In the reference SG disk the companions migrate rapidly through the disk preventing any potential gap-opening. In the lighter SG disk they decelerate enough or even remain at their introductory radial location so that a gap might evolve.}
\label{SGdisk}
\end{center}
\end{figure*}

For each prepared SG disk set-up we introduce a 15, 30, 45 and 80 $M_{\rm Jup}$ companion separately. The simulations are performed four times for each companion represented by different colors in Figure~\ref{SGdisk}. Each time the companions are introduced with the same orbital radius but at different azimuthal angles (0, $\pi/2$, $\pi$ and $3\pi/2$) to ensure the overall migration and gap-opening results are statistically significant. These simulations are run for 20 orbits.\footnote{This time period is a reasonable compromise between a short simulation time and acquiring sufficient data to distinctively characterize the companions' migrations.} 

In the MMSN-like disks we introduce companions with 0.5 $M_{\rm Jup}$, 1 $M_{\rm Jup}$ and 2 $M_{\rm Jup}$ at 10 AU and allow them to migrate freely. The companions' position is then tracked for 600 orbits (at 5 AU).

%%%%%%%%%%%%%%%%%%%%%%%%%%%%%%%%%%%%%%%%%%%%%%%%%%%%%%%%
%%%%%%%%%%%%%%%%%%%% Results %%%%%%%%%%%%%%%%%%%%%%%%%%%
%%%%%%%%%%%%%%%%%%%%%%%%%%%%%%%%%%%%%%%%%%%%%%%%%%%%%%%%
\section{Results}
\label{results}

The most direct indication for gap-opening comes as a drastic slow-down in the migration rate. Also an order of magnitude decrease in the surface density in the co-orbital region \citep{cr06} can confirm gap clearing. Through this approach we assess whether or not the companions open gaps, and if so, at which orbital separation.

%%%%%%%%%%%%%%%%%%%%%%%%%%%%%%%%%%
\subsection{SG disk simulations}%%
\label{SGsim}%%%%%%%%%%%%%%%%%%%%%
%%%%%%%%%%%%%%%%%%%%%%%%%%%%%%%%%%

In the reference SG disk the migrations are very rapid. In all cases the companions reach the inner grid boundary in a few orbits. The 80 $M_{\rm Jup}$  companion suffers a significant deceleration at an orbital separation of $\approx 70$ AU. However the subsequent radial motion is still much more rapid than would be expected from a gap-opening procedure. Figure \ref{SGdisk} shows the evolution of the orbital separation for the 30 $M_{\rm Jup}$ and 80 $M_{\rm Jup}$ companions in the two SG disk set-ups.  Appendix \ref{plots} contains the orbital evolution figures for the 15 $M_{\rm Jup}$ and the 45 $M_{\rm Jup}$ companions. 

\begin{figure}
\begin{center}
\begin{minipage}[t]{0.48\textwidth}
\includegraphics[width=\textwidth]{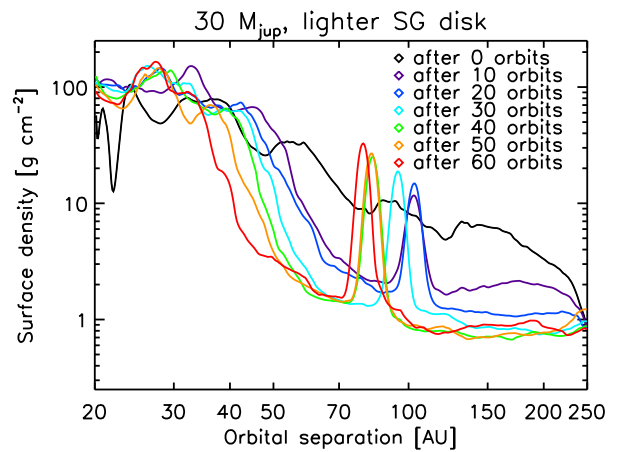}
\end{minipage}
\vspace{-0.3cm}
\caption{Evolution of the disk surface density perturbation due to a migrating 30 $M_{\rm Jup}$ companion in the lighter SG disk. A gap is believed to be established when the surface density drops by approximately one order of magnitude in the companion's horseshoe region, which is satisfied at approx. 30 - 40 orbits after introduction but not yet after 20 orbits. During the run the companion's orbital separation decreases only slowly due to a strong attenuation of torques in the cleared co-orbital region.}
\label{longlight}
\end{center}
\end{figure}

In contrast to the reference SG disk, the lighter SG disk offers conditions which cause the companions to migrate much slower right from the point of introduction. They often undergo a further decelaration until they reach a quasi-constant orbit. No companions reach the inner grid boundary during the first 20 orbits of their migration (except the 15 $M_{\rm Jup}$ companion once). Although they do not clear their horseshoe region to the extent of a full gap, the slow-down in the inward migration suggests that given enough time they would do so. This can be seen in Figure \ref{longlight}, which depicts the surface mass density distribution around the embedded 30 $\MJup$ companion in one of the lighter SG disk simulations. After 20 orbits the companion's co-orbital region still appears to be in the process of being cleared out. To investigate this further we run this particular simulation longer (until 60 orbits), and find that after $\sim$ 30 - 40 orbits the depth of the co-orbital region reaches a fully developed gap level. However, the companion's orbital separation still decreases significantly during the 60 orbits simulation suggesting that the companion might have not yet reached a type II migration regime, commonly associated with a fully developed gap. We discuss the possible reasons behind the slow migration rates in the lighter SG disk in section \ref{analysis}.

A snapshot of the 30 $M_{\rm Jup}$ companion in the reference and lighter SG disks after 3 orbits is provided by the upper panels of Fig. \ref{images}. The companion in the reference SG disk is located at a lower orbital radius due to its more rapid inward migration.

We conclude that in the lighter SG disk our empirical study appears to be generally consistent with the torque balance criterion of \cite{cr06} (see Fig. \ref{criteria}, middle panel). In the reference SG disk however, despite being expected to open gaps (Fig. \ref{criteria}, left panel), the companions migrate inwards rapidly. To investigate whether the time aspect might be a limiting constraint we conduct fixed orbit simulations and determine the gap-opening timescale for our SG disk set-ups empirically, as analytical estimates for the gap-opening timescale are based on linear analysis (e.g. \citealt{li86}, their eq. 9) while our simulations are in the non-linear regime. We then compare these with the migration timescale for the ''free'' companions.

\begin{figure*}
\begin{center}
\begin{minipage}[t]{0.48\textwidth}
\includegraphics[width=\textwidth]{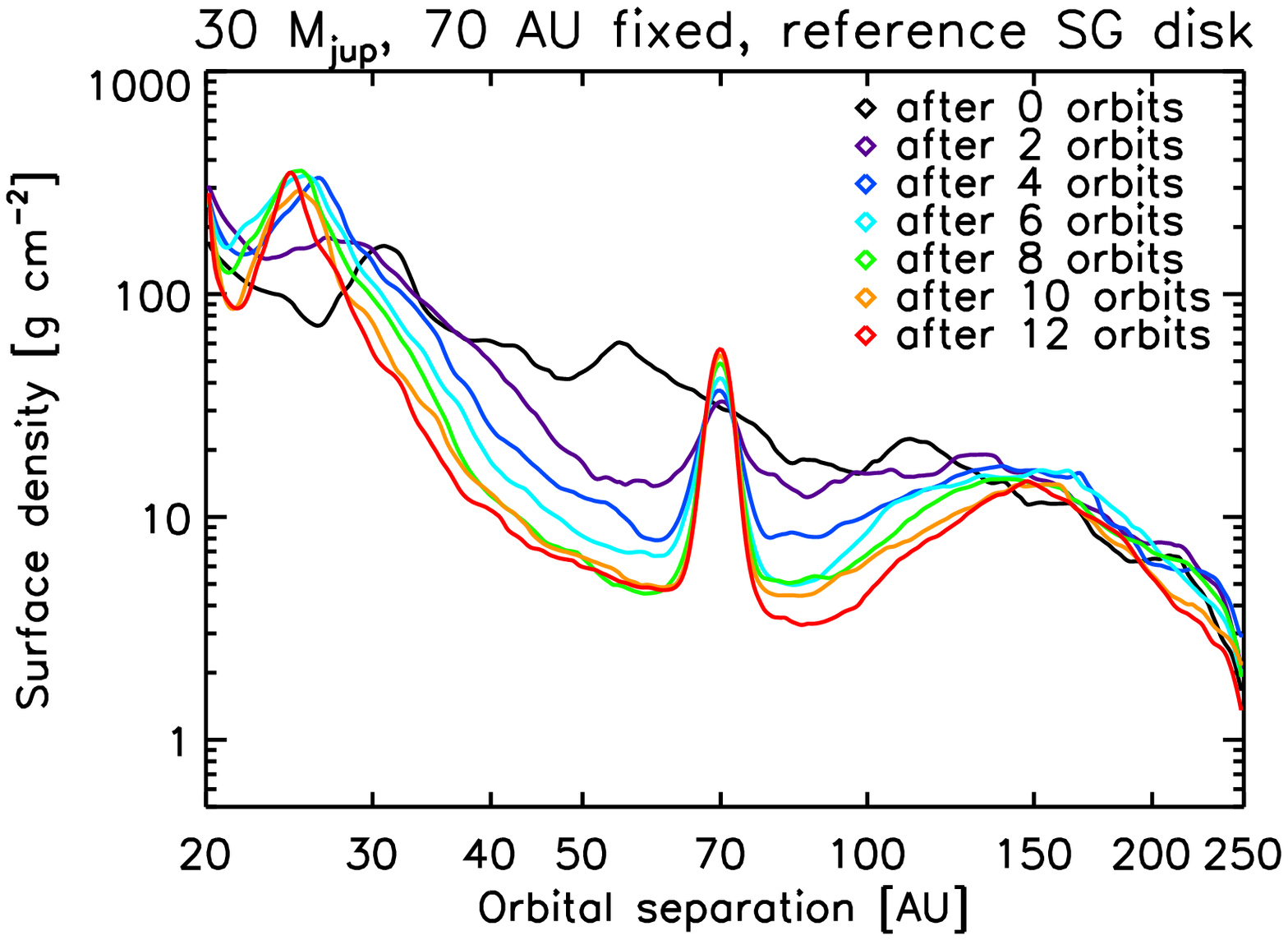}
\end{minipage}
\hfill
\begin{minipage}[t]{0.48\textwidth}
\includegraphics[width=\textwidth]{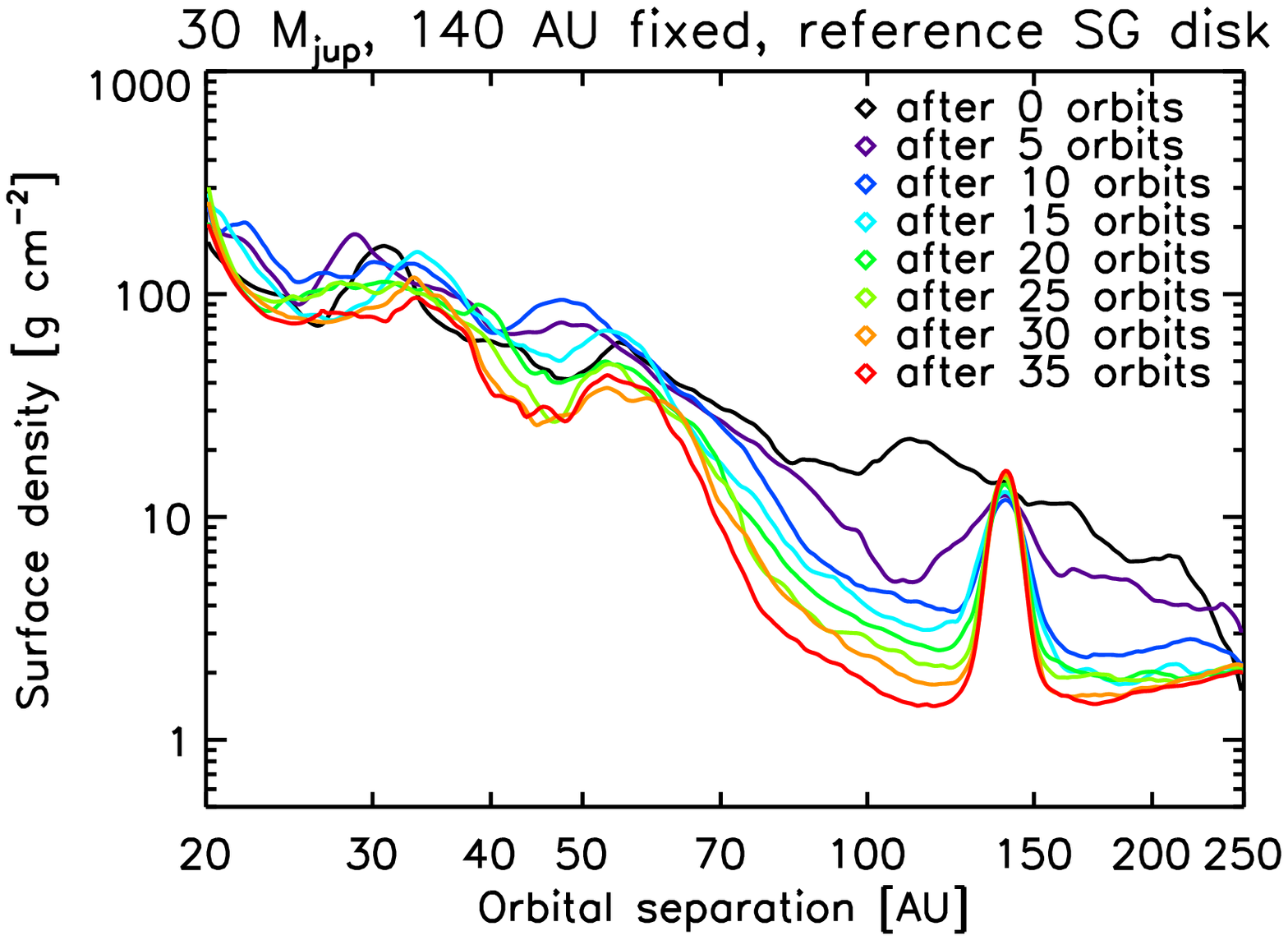}
\end{minipage}
\begin{minipage}[t]{0.48\textwidth}
\includegraphics[width=\textwidth]{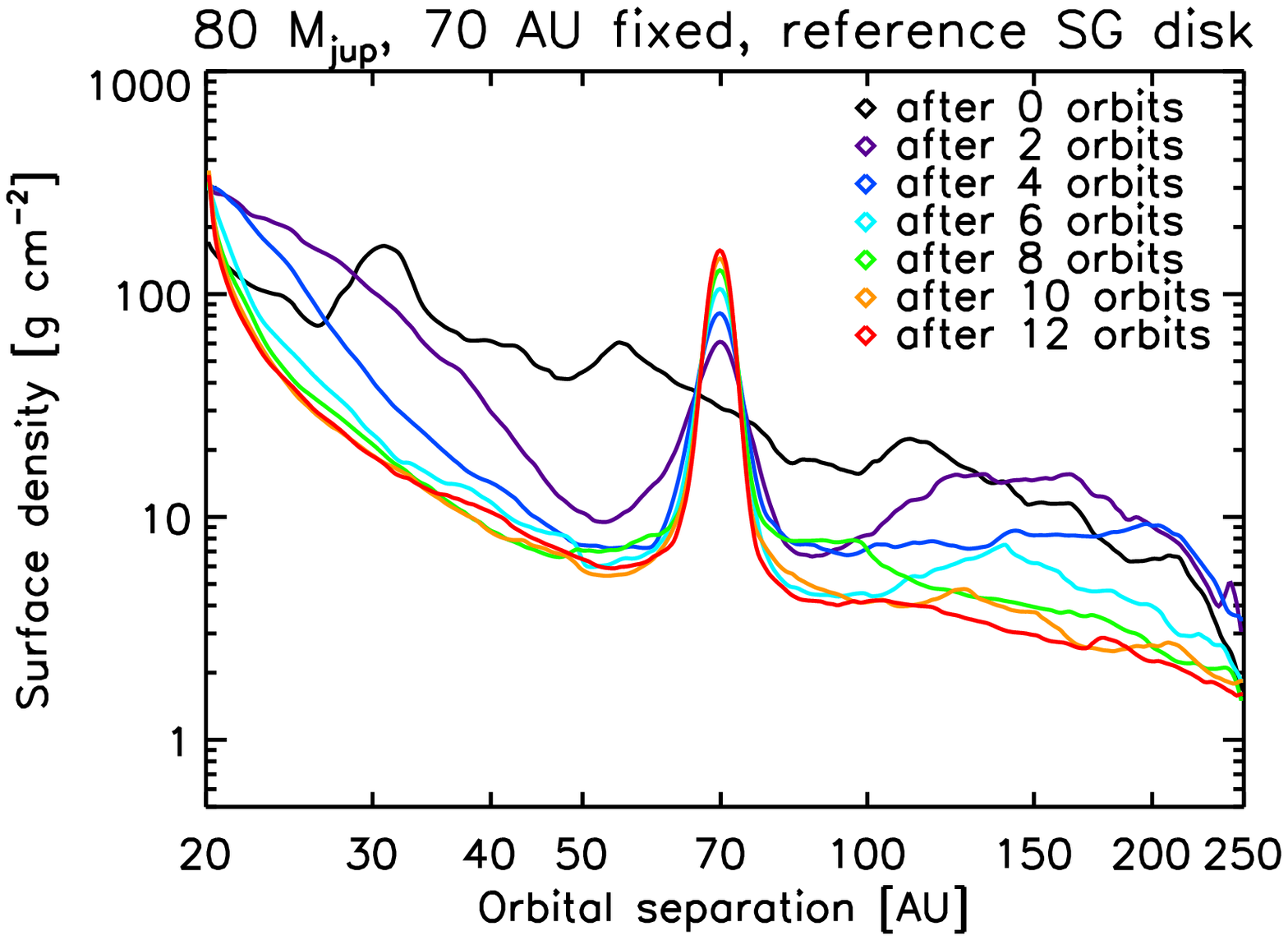}
\end{minipage}
\hfill
\begin{minipage}[t]{0.48\textwidth}
\includegraphics[width=\textwidth]{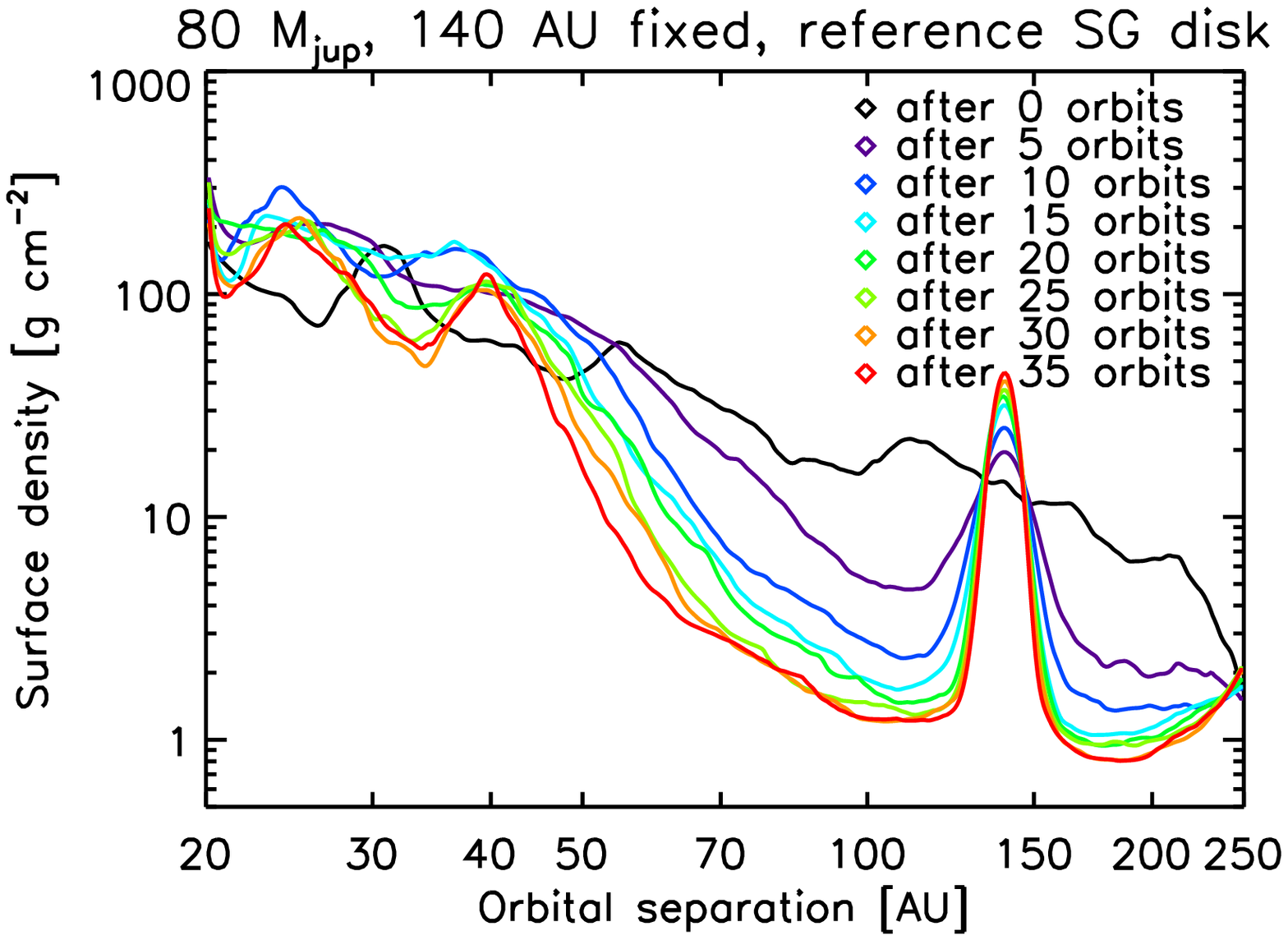}
\end{minipage}
\vspace{-0.3cm}
\caption{Evolution of the disk surface density perturbation due to a 30 $M_{\rm Jup}$ (top) and 80 $M_{\rm Jup}$ (bottom) companion held at a constant orbital radius at 70 AU (left) and 140 AU (right) in the reference SG disk. A gap is assumed to be fully established when the surface density drops by approximately one order of magnitude in the companion's horseshoe region. For the embedded 30 $M_{\rm Jup}$ (80 $M_{\rm Jup}$) companion a gap is formed after approximately 8 (6) orbits at 70 AU and after 35 (20) orbits at 140 AU. The time it takes for a gap to form is mostly larger than the respective migration timescale shown in Figure~\ref{SGdisk} (see also Table \ref{tab:timescales}) for both companions.}
\label{fixed}
\end{center}
\end{figure*}

\vspace{0.4cm}

%%%%%%%%%%%%%%%%%%%%%%%%%%%%%%%%%%%%%%%%%%%%
\subsubsection{Companions on a fixed orbit}%
\label{fixie}%%%%%%%%%%%%%%%%%%%%%%%%%%%%%%%
%%%%%%%%%%%%%%%%%%%%%%%%%%%%%%%%%%%%%%%%%%%%

Using the reference SG disk set-up we firstly employ as before 15 $M_{\rm Jup}$, 30 $M_{\rm Jup}$, 45 $M_{\rm Jup}$ and 80 $M_{\rm Jup}$ conpanions at fixed orbital separations of 70 AU and 140 AU. The results for the 30 $M_{\rm Jup}$ and 80 $M_{\rm Jup}$ companions are depicted in Fig. \ref{fixed} (and for the curious reader the results for the 15 and 45 $M_{\rm Jup}$ companions are presented in Appendix~\ref{plots}).

At both introduction locations all companions succeed eventually at opening gaps. This is consistent with the general criterion's prediction (see Fig. \ref{criteria}, left panel). The 15 $M_{\rm Jup}$ companion is special in the sense that it is not predicted to clear a gap at 140 AU but does so in our simulation. However, taking the whole perturbing mass $\sim 22.7$ $\MJup$, i.e. the companion mass and the mass of its circumplanetary disk (material within $\approx 2/3$ $R_{\rm H}$ of the companion; \citealt{cr09}), eliminates the discrepancy between simulation results and prediction as this value lies slightly above the mass limit shown in Fig. \ref{criteria}. We hence find that when radial migration is disabled and timescales are not considered, gaps open as predicted by the torque balance criterion when otherwise they do not. This provides a first indication that the timescale for gap-opening may be important.

\begin{table*}
	\caption{Comparison between the gap-opening timescale $\tgap$ versus the crossing timescale $\tcross$ empirically determined at 70 and 140 AU. The time is normalized to orbits at 100 AU for the SG disks and to orbits at 5 AU for the MMSN-like disks. Since in the vast majority of the simulations $\tgap \gg \tcross$ gap-opening might be prevented even though the companions are predicted to open gaps by the torque balance criterion of \protect\cite{cr06}.}
	\label{tab:timescales}
	\vspace{-0.4cm}
\begin{center}
\bgroup
\def\arraystretch{1.5}
  \begin{tabular}{|l|l|l|l|l|}
    \hline
disk model & planet mass [$\MJup$] & orbital separation [AU] & $\tgap$ [orbits] & $\tcross$ [orbits] \\
\hline
reference SG & 15 $\MJup$ & 70 & $\sim 12$ & $\sim 1$ \\ \cline{3-5}
& & 140 & $\sim 65$ & $\sim 2$ \\ \cline{2-5}
& 30 $\MJup$ & 70 & $\sim 8$ & $\sim 1.2$ \\ \cline{3-5}
& & 140 & $\sim 35$ & $\sim 3$ \\ \cline{2-5}
& 45 $\MJup$ & 70 & $\sim 8$ & $\sim 1.5$ \\ \cline{3-5}
& & 140 & $\sim 20$ & $\sim 2$ \\ \cline{2-5}
& 80 $\MJup$ & 70 & $\sim 6$ & $\sim 8$ \\ \cline{3-5}
& & 140 & $\sim 20$ & $\sim 5$ \\ \cline{1-5}
heavier MMSN-like & 0.5 $\MJup$ & 5 & $\sim 400$ & $\sim 3$ \\ \cline{3-5}
&  & 10 & $\gtrsim 600$ & $\sim 30$ \\ \cline{2-5}
& 1 $\MJup$ & 5 & $\sim 200$ & $\sim 2$ \\ \cline{3-5}
&  & 10 & $\gtrsim 600$ & $\sim 20$ \\ \cline{2-5}
& 2 $\MJup$ & 5 & $\sim 60$ & $\sim 3$ \\ \cline{3-5}
&  & 10 & $\sim 400$ & $\sim 15$ \\ \cline{1-5}
   \hline
  \end{tabular}
	\egroup
	\end{center}
\end{table*}

\begin{table*}
\vspace{-0.4cm}
	\caption{Comparison between the empirically determined gap-opening timescale $\tgap$ for the 30 $\MJup$ companion in the reference and lighter SG disks (top) and for the 2 $\MJup$ companion in the heavier and lighter MMSN-like disks (bottom). The time is normalized to orbits at 100 AU for the SG disks and to orbits at 5 AU for the MMSN-like disks. The gap-opening timescale appears to be independent of disk mass.}
	\label{tab:timecomparison}
		\vspace{-0.4cm}
\begin{center}
\bgroup
\def\arraystretch{1.5}
  \begin{tabular}{|l|l|l|l|}
    \hline
planet mass [$\MJup$] & orbital separation [AU] & $\tgap$ [orbits] & $\tgap$ [orbits] \\
&  & reference SG disk & lighter SG disk \\ \hline
30 $\MJup$ & 70 & $\sim 8$ & $\sim 8$ \\ \cline{2-4}
 & 140 & $\sim 35$ & $\sim 35$ \\ \cline{1-4}
planet mass [$\MJup$] & orbital separation [AU] & $\tgap$ [orbits] & $\tgap$ [orbits]\\
 &  &  heavier MMSN-like disk &  lighter MMSN-like disk\\ \hline
2 $\MJup$ & 5 & $\sim 60$ & $\sim 60$ \\ \cline{2-4}
 & 10 & $\sim 400$ & $\sim 450$ \\ \cline{1-4}
   \hline
  \end{tabular}
	\egroup
	\end{center}
\end{table*}

Secondly, all of the companions require more time for gap-clearing than a ''free'' migration would take. We estimate the gap-opening timescale to be the average time taken to clear the inner and outer horseshoe region such that the surface density drops by approximately one order of magnitude in the gap region - consistent with the definition of a gap in \cite{cr06}. This is sufficient to provide a reasonable estimate of the gap-opening timescale. Figure \ref{fixed} shows that the 30 $M_{\rm Jup}$ (80 $M_{\rm Jup}$) companion clears a gap after $\approx$ 8 (6) orbits at 70 AU and after $\approx$ 35 (20) orbits at 140 AU.  The migration timescales from 140 AU to the inner grid at 20 AU for the 30 $M_{\rm Jup}$ and 80 $M_{\rm Jup}$ planets are $\approx$ 5 - 6 orbits and $\approx$ 11 - 13 orbits, respectively, and so the crossing timescale across a gap region that a companion may try to form is much less than this.  We define the crossing timescale to be

\begin{equation}
\label{tcrossing}
\tcross = \frac{R_{\rm HS}}{\left | dR/dt \right |}
\end{equation}
where $R_{\rm HS}$ is the half-width of the companion's horseshoe region or equivalently the distance over which the companion would need to migrate to start the gap-opening process again. Although it is not clear what a reasonable estimate for $R_{\rm HS}$ is in our mass regime, we extrapolate the results of \cite{ma06} and \cite{pa09}, which found through streamline analysis that $R_{\rm HS} \approx$ 2.5 $R_{\rm H}$ for higher mass planets\footnote{\cite{ma06} fitted data points up to a planetary mass of 0.2 $\MJup$. \cite{pa09} used the dimensionless parameter $q/h^3$ to represent their planet masses and examined a mass regime below $q/h^3 = 5$. As a comparison: our MMSN-like simulations correspond to values $q/h^3 = 1.4 - 5.7$, which is almost fully in \cite{pa09}'s examined regime. The SG disk simulations are in the range $q/h^3 = 6.8 - 224$ and hence involve a large extrapolation of \cite{pa09}'s results.}. The migration rate $dR/dt$ is defined as the gradient of the migration curves in Figures~\ref{SGdisk}, \ref{tinylam} and \ref{fig:SG_15_45} at the relevant location. Table \ref{tab:timescales} compares the gap-opening timescale (for planets on a fixed orbit) to the crossing timescale determined from the migration simulations at particular disk radii and shows that the simulations in which gaps do not open are those in which the crossing is too fast. Since in most cases $\tcross < \tgap$, this may explain the absence of gap-opening  in the free migration simulations.

As an aside we also point out that the gap-opening timescale decreases with increasing companion mass and decreasing orbital separation. This is consistent with the prediction of the torque balance criterion, which is satisfied more easily by massive companions at smaller orbital radii (see Fig. \ref{criteria}). Furthermore we observe no significant deviation in the gap-opening timescale when changing the disk mass even in the non-linear regime -- consistent with the findings of the linear analysis by \cite{li86} (see Table \ref{tab:timecomparison} and compare the top panels of Figures \ref{fixed} and \ref{denslighters}).

%%%%%%%%%%%%%%%%%%%%%%%%%%%%%%%%%%%%%%%%%%%%
\subsection{MMSN-like disk simulations}%%%%%%%%%%
%%%%%%%%%%%%%%%%%%%%%%%%%%%%%%%%%%%%%%%%%%%%

\begin{figure*}
\begin{center}
\begin{minipage}[b]{0.32\textwidth}
\includegraphics[width=\textwidth]{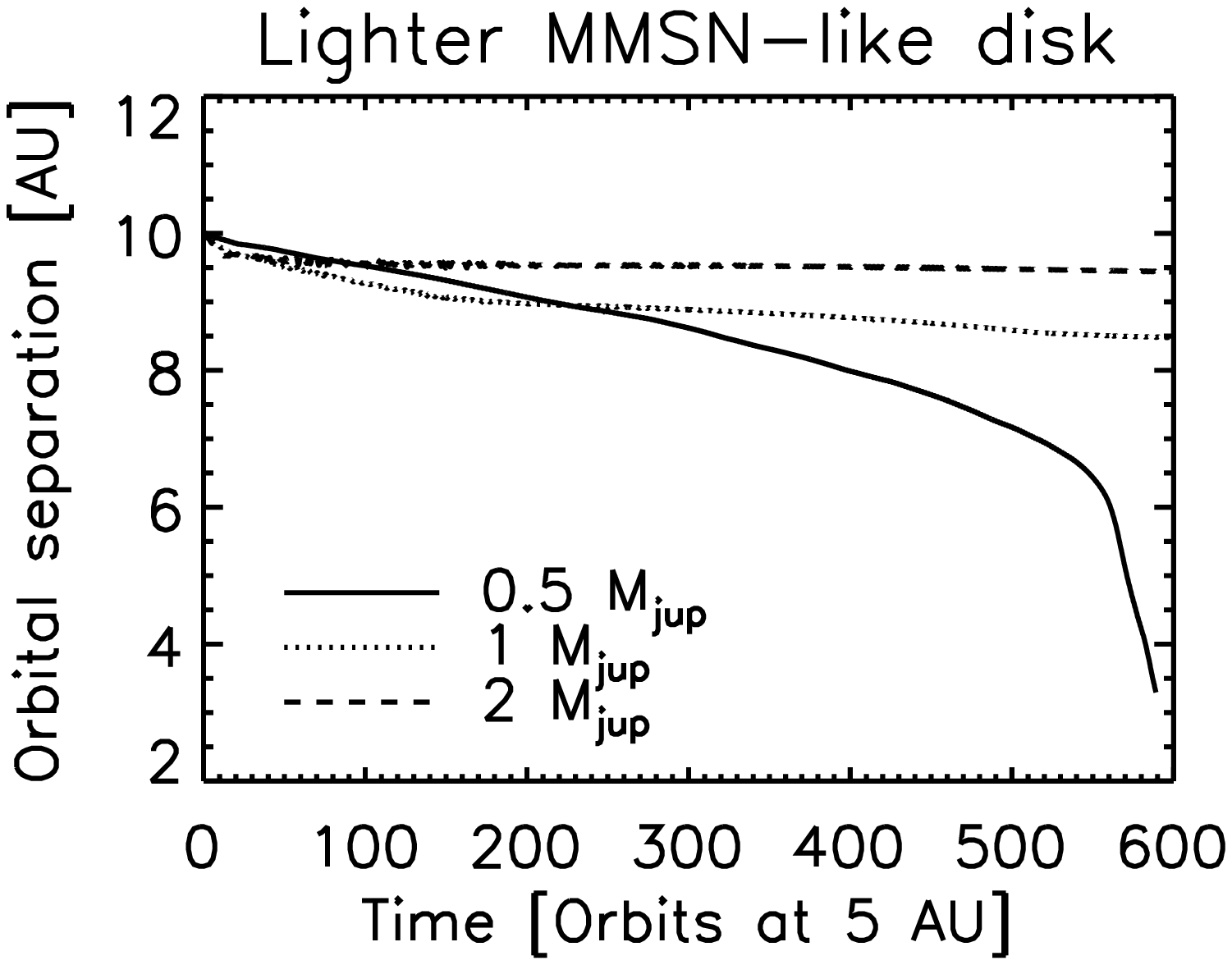}
\end{minipage}
\hfill
\begin{minipage}[b]{0.32\textwidth}
\includegraphics[width=\textwidth]{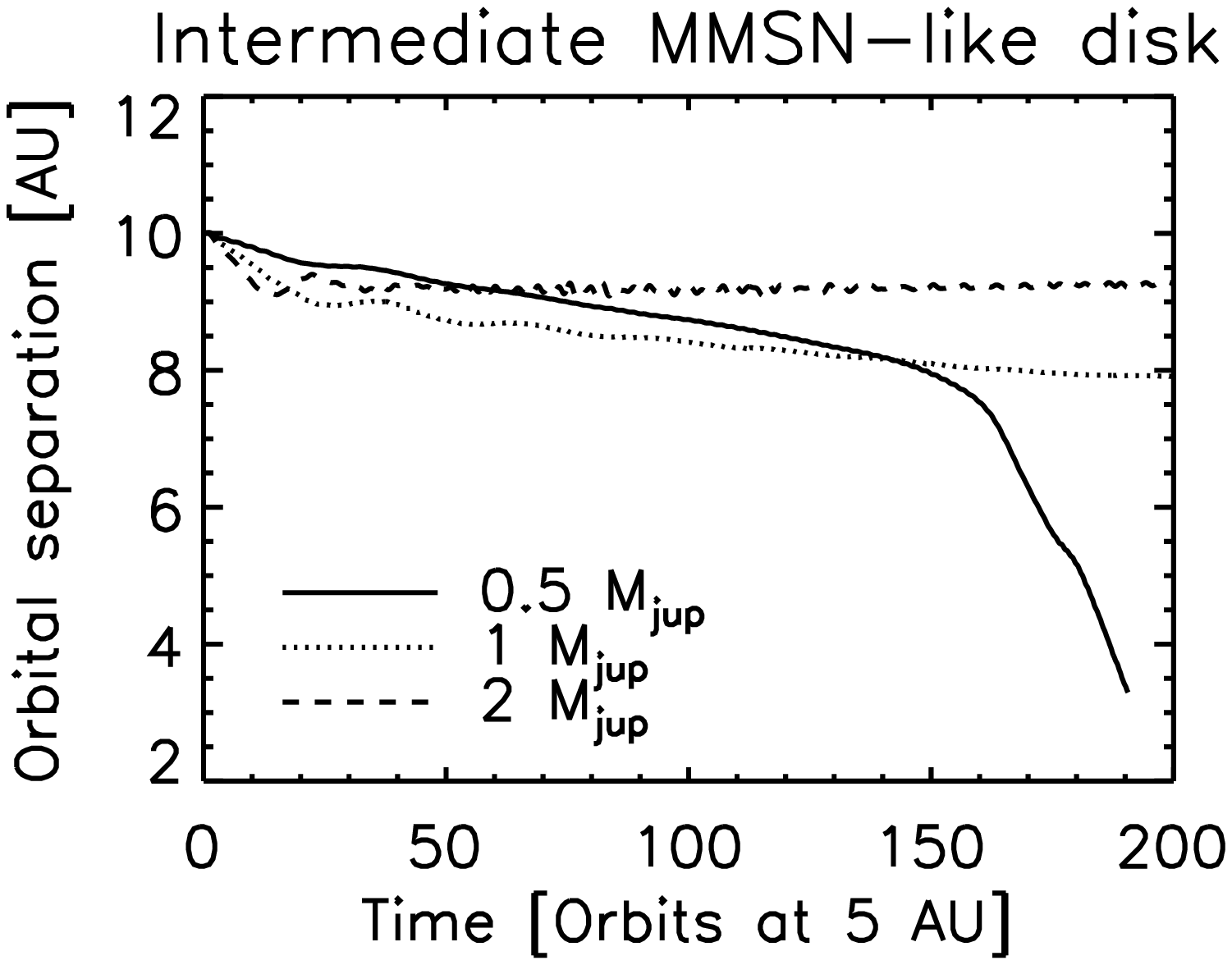}
\end{minipage}
\hfill
\begin{minipage}[b]{0.32\textwidth}
\includegraphics[width=\textwidth]{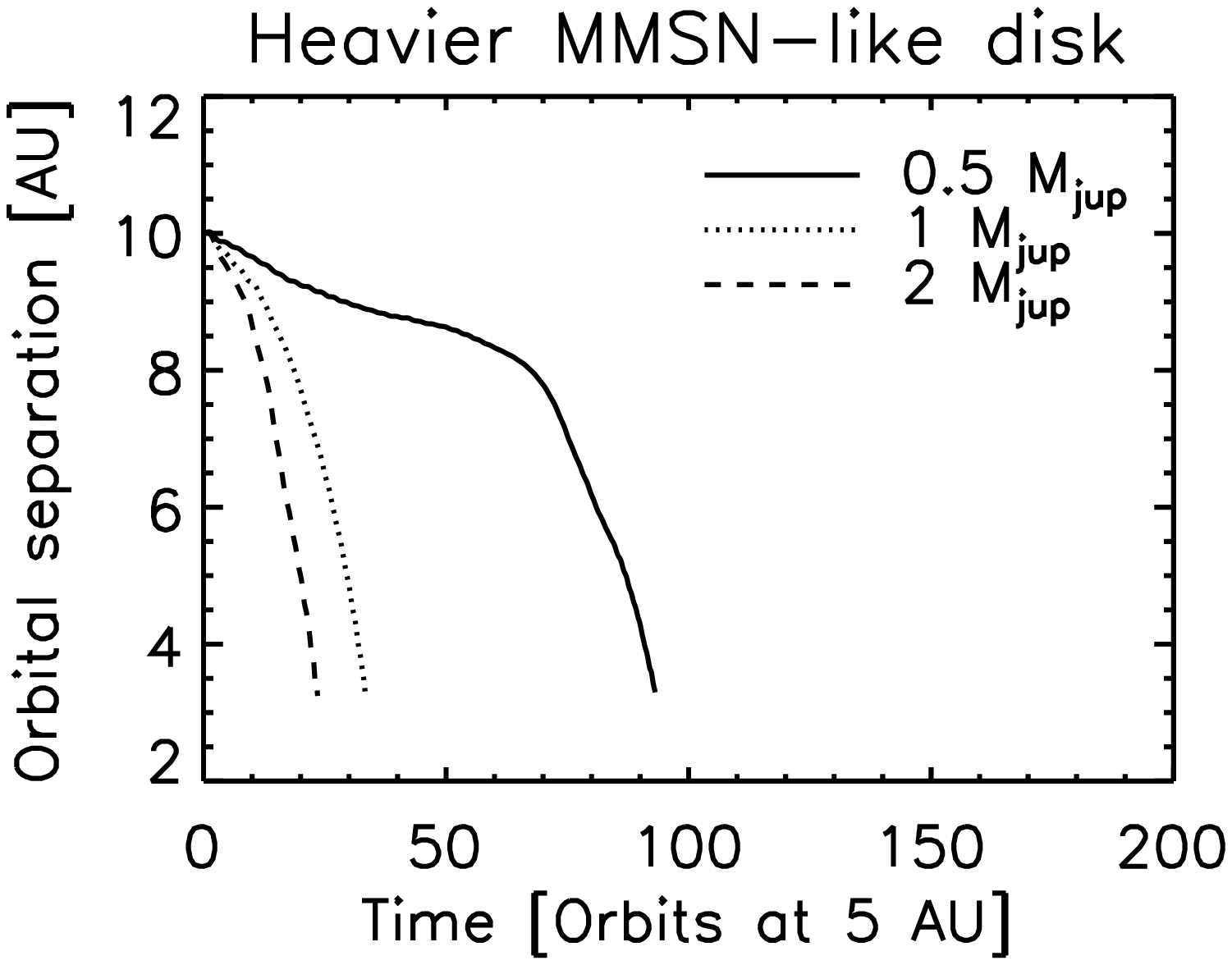}
\end{minipage}
\vspace{-0.3cm}
\caption{Evolution of the orbital separation of a migrating 0.5 $M_{\rm Jup}$ (solid line), 1 $M_{\rm Jup}$ (dotted line) and 2 $M_{\rm Jup}$ companion (dashed line) in the lighter (left), intermediate (middle) and heavier MMSN-like disks (right). The larger the disk mass the more rapid the inward migration of the companions. In the lighter and intermediate disks we observe the predicted gap-opening by the 2 $M_{\rm Jup}$ companion whereas the 1 $M_{\rm Jup}$ companion only partially manages to clear its co-orbital region during the simulated 600 orbits. The 0.5 $M_{\rm Jup}$ companion shows little signs of gap clearing due to its rapid inward migration in all set-ups. The companions' ability to open a gap depends significantly on the disk mass, although the torque balance criterion of \protect\cite{cr06} does not suggest this.}
\label{tinylam}
\end{center}
\end{figure*}

\begin{figure*}
\begin{center}
\begin{minipage}[t]{0.48\textwidth}
\includegraphics[width=\textwidth]{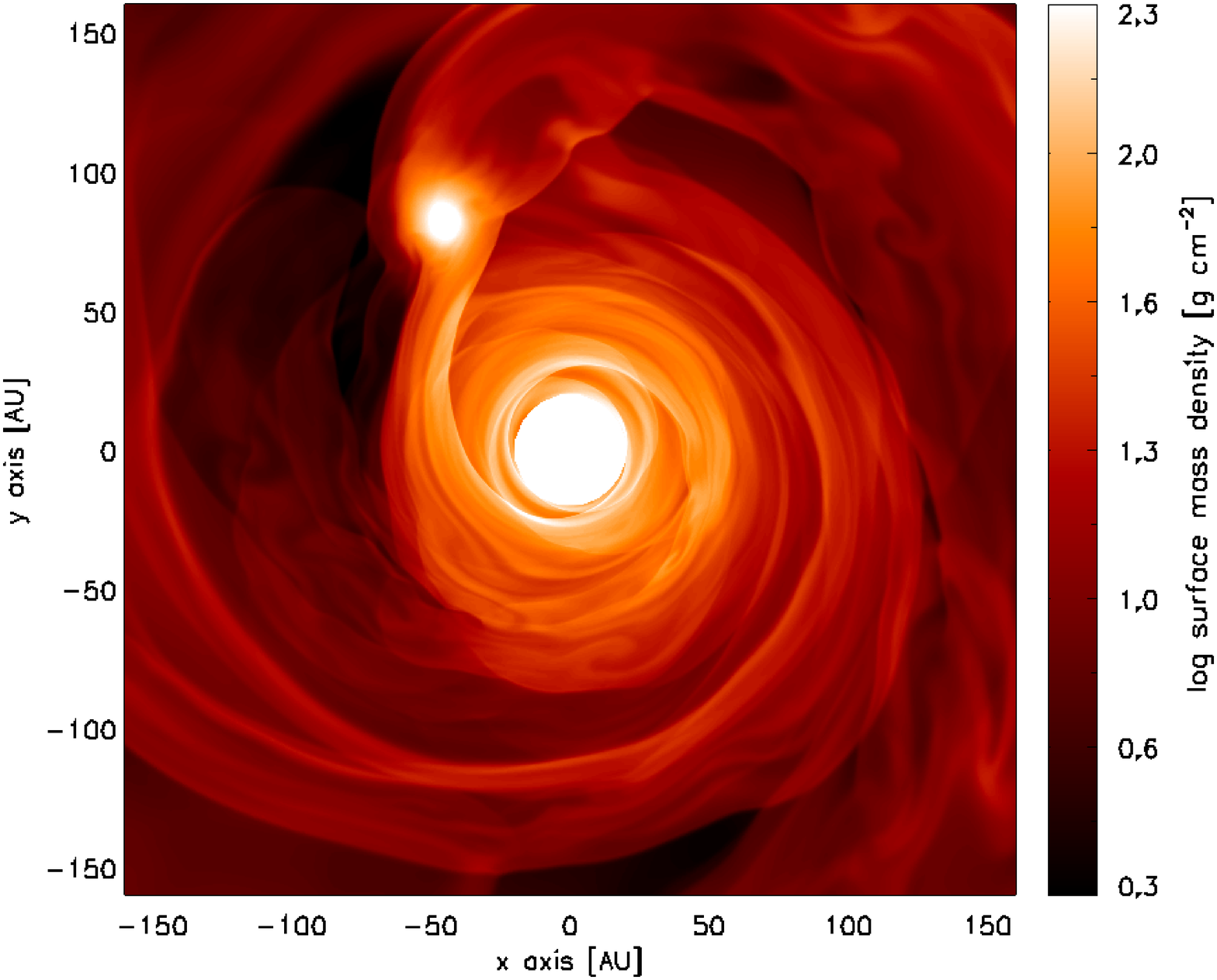}
\end{minipage}
\hfill
\begin{minipage}[t]{0.48\textwidth}
\includegraphics[width=\textwidth]{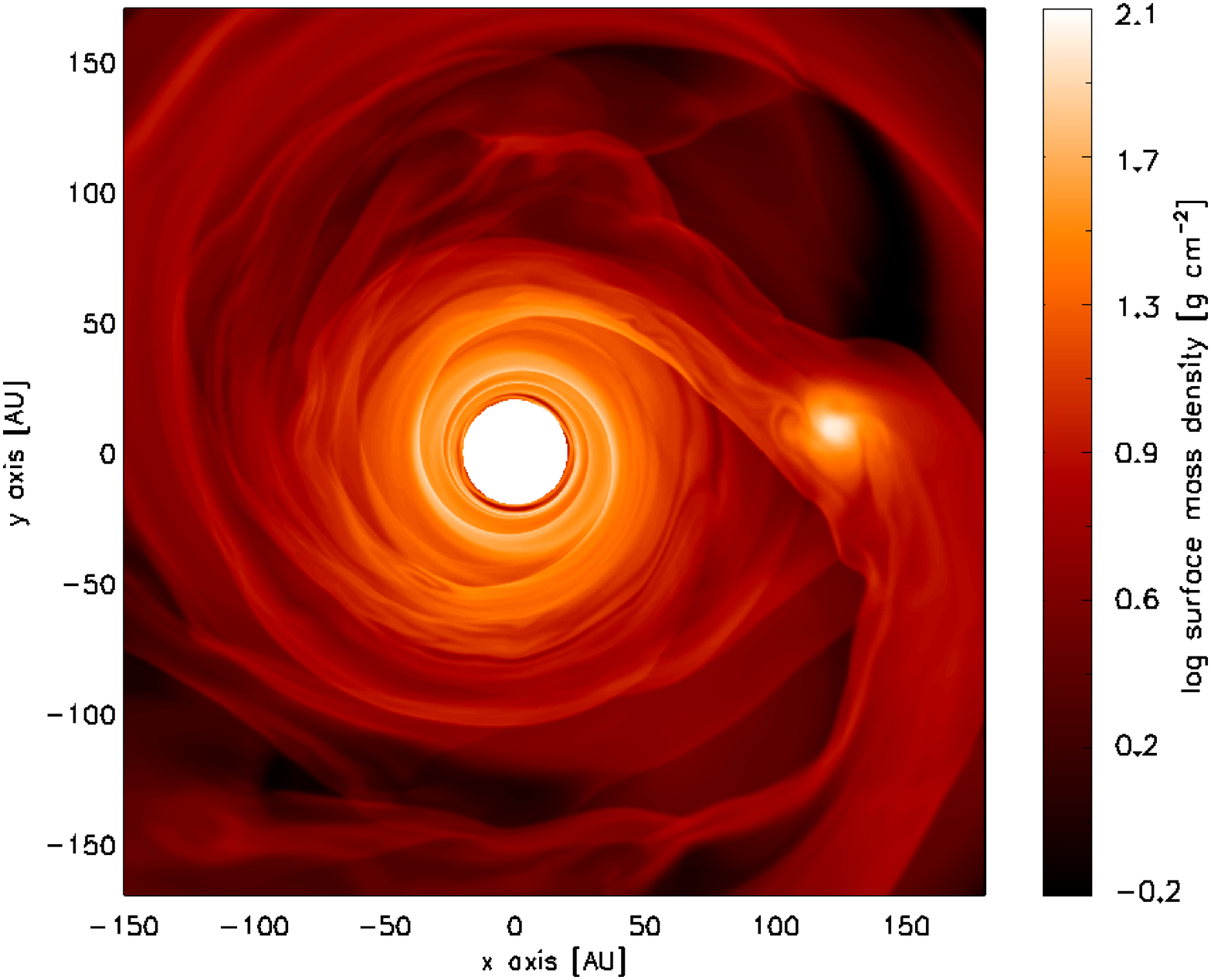}
\end{minipage}
\begin{minipage}[t]{0.48\textwidth}
\includegraphics[width=\textwidth]{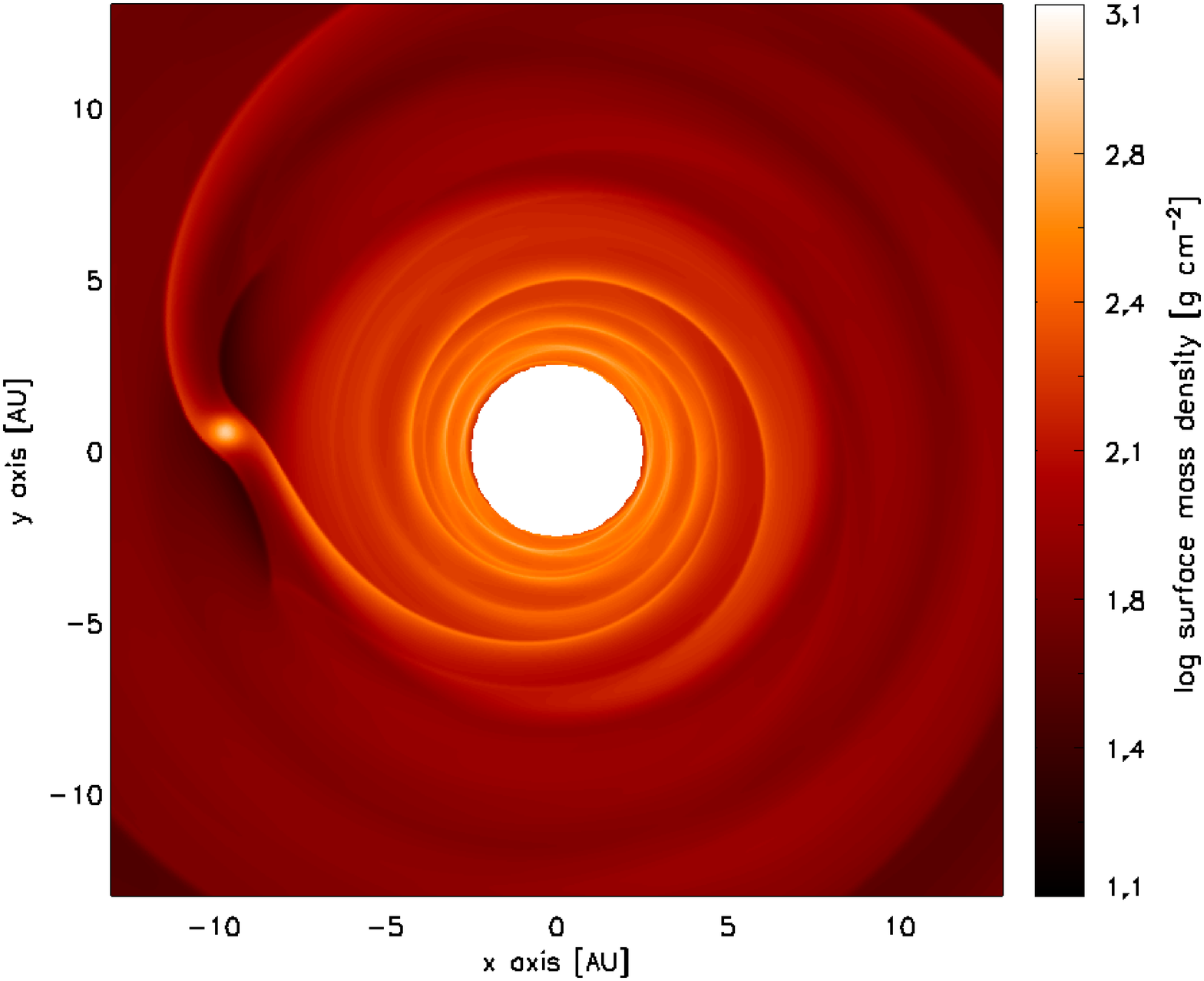}
\end{minipage}\
\hfill
\begin{minipage}[t]{0.48\textwidth}
\includegraphics[width=\textwidth]{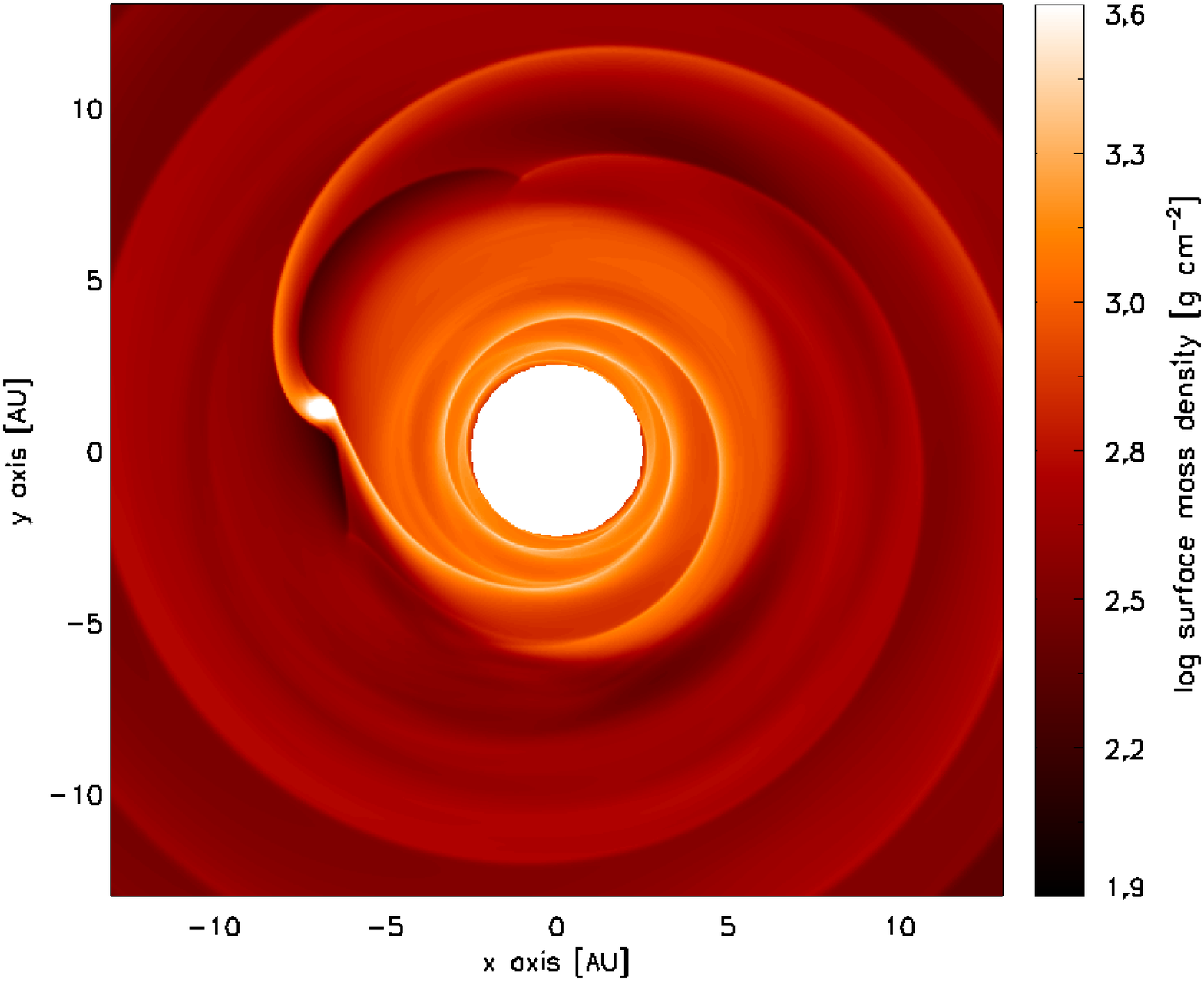}
\end{minipage}
\vspace{-0.1cm}
\caption{A snapshot of the inwards migration of a companion through the disk. {\bf Top left:} 30 $M_{\rm Jup}$ companion in the reference SG disk after 3 orbits (at 100 AU). {\bf Top right:} 30 $M_{\rm Jup}$ companion in the lighter SG disk after 3 orbits (at 100 AU). {\bf Bottom left:} 2 $M_{\rm Jup}$ companion in the lighter MMSN-like disk after 15 orbits (at 5 AU). {\bf Bottom right:} 2 $M_{\rm Jup}$ companion in the heavier MMSN disk after 15 orbits (at 5 AU). The lack of disk material in the co-orbital region is visible by the darker colors around the embedded companions. The companions in the reference SG and heavier MMSN-like disks are located further in than their counterparts in the lighter SG and MMSN-like disks since they migrate inwards more rapidly due to a stronger torque (see Figure \ref{torque}).}
\label{images}
\end{center}
\end{figure*}

Figure \ref{tinylam} shows the evolution of the companions' orbital separation in the lighter (left), intermediate (middle) and heavier MMSN-like disks (right). As shown in Fig. \ref{criteria} (right panel) the torque balance criterion predicts that the 2 $M_{\rm Jup}$ companion should immediately open a gap, the 1 $M_{\rm Jup}$ companion should do so somewhat further in and the 0.5 $M_{\rm Jup}$ companion should do so after having crossed large parts of the disk. In the lighter and intermediate disks the 1 $M_{\rm Jup}$ and 2 $M_{\rm Jup}$ companions confirm this prediction. They both start to slow their migration by clearing their co-orbital region early. However, the 0.5 $M_{\rm Jup}$ migrates to the inner grid boundary and indicates no gap clearing. In the heavier MMSN-like disk, the migration times of all the companions decrease substantially. The bottom two panels of Figure \ref{images} depict the 2 $M_{\rm Jup}$ companion at 15 orbits after introduction into the lighter (left) and the heavier disks (right). The 2 $M_{\rm Jup}$ companion is located at a smaller orbital radius in the heavier MMSN-like disk compared to the lighter MMSN-like disk due to its more rapid migration.

As with the high-mass SG simulations (see section \ref{SGsim}) we find that the disk mass is a non-negligible factor for migration and also affects the companion's ability to open a gap. This is surprising as the disk mass is not included in the torque balance criterion of \cite{cr06} and thus its prediction is the same for the different MMSN-like disks.

%%%%%%%%%%%%%%%%%%%%%%%%%%%%%%%%%%%%%%%%%%%%
\subsubsection{Companions on a fixed orbit}%
\label{fixieMMSN}%%%%%%%%%%%%%%%%%%%%%%%%%%%%%%%
%%%%%%%%%%%%%%%%%%%%%%%%%%%%%%%%%%%%%%%%%%%%

As with the SG disks in Section \ref{fixie} we perform simulations with the companions held on a fixed orbit at 10 AU and 5 AU and find that the gap-opening timescale is much larger than the migration timescale in the heavier MMSN-like disk (see Table \ref{tab:timescales}). This confirms that the timescale argument is the limiting factor for gap-opening. We also determine the gap-opening timescale for the 2 $\MJup$ companion in the lighter MMSN-like disk and find that the timescale is independent of the disk mass (see Table \ref{tab:timecomparison}) as with the SG disks simulations. The curious reader finds the according surface density evolutions shown in Figures \ref{densvhfix} and \ref{denslighters}.

%%%%%%%%%%%%%%%%%%%%%%%%%%%%%%%%
\subsection{Migration Analysis}%%%%%%%%%%%%%%%
\label{analysis}%%%%%%%%%%%%%%%%%%%%%%%%
%%%%%%%%%%%%%%%%%%%%%%%%%%%%%%%%

\begin{figure*}
\begin{center}
\begin{minipage}[b]{0.48\textwidth}
\includegraphics[width=\textwidth]{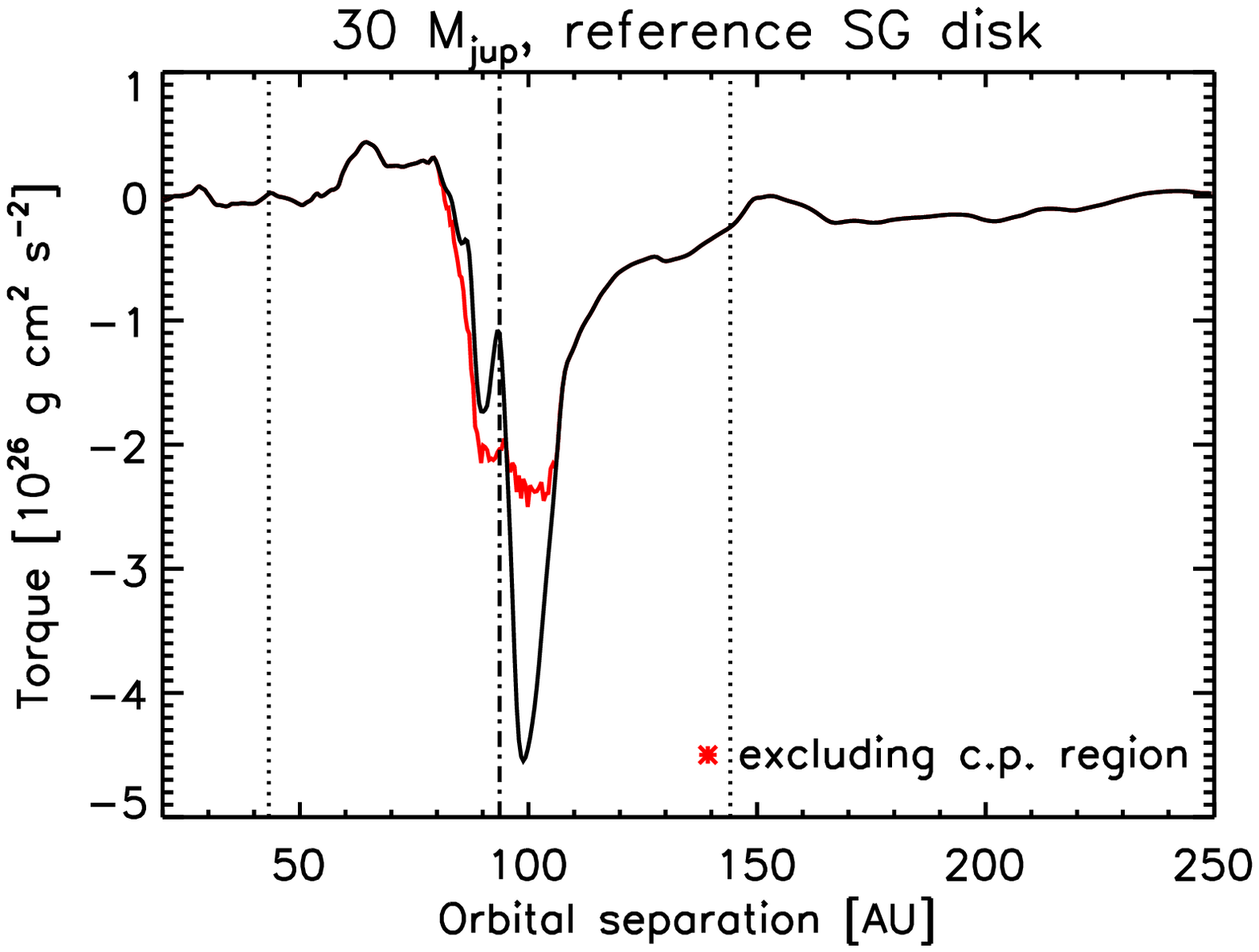}
\end{minipage}
\hfill
\begin{minipage}[b]{0.48\textwidth}
\includegraphics[width=\textwidth]{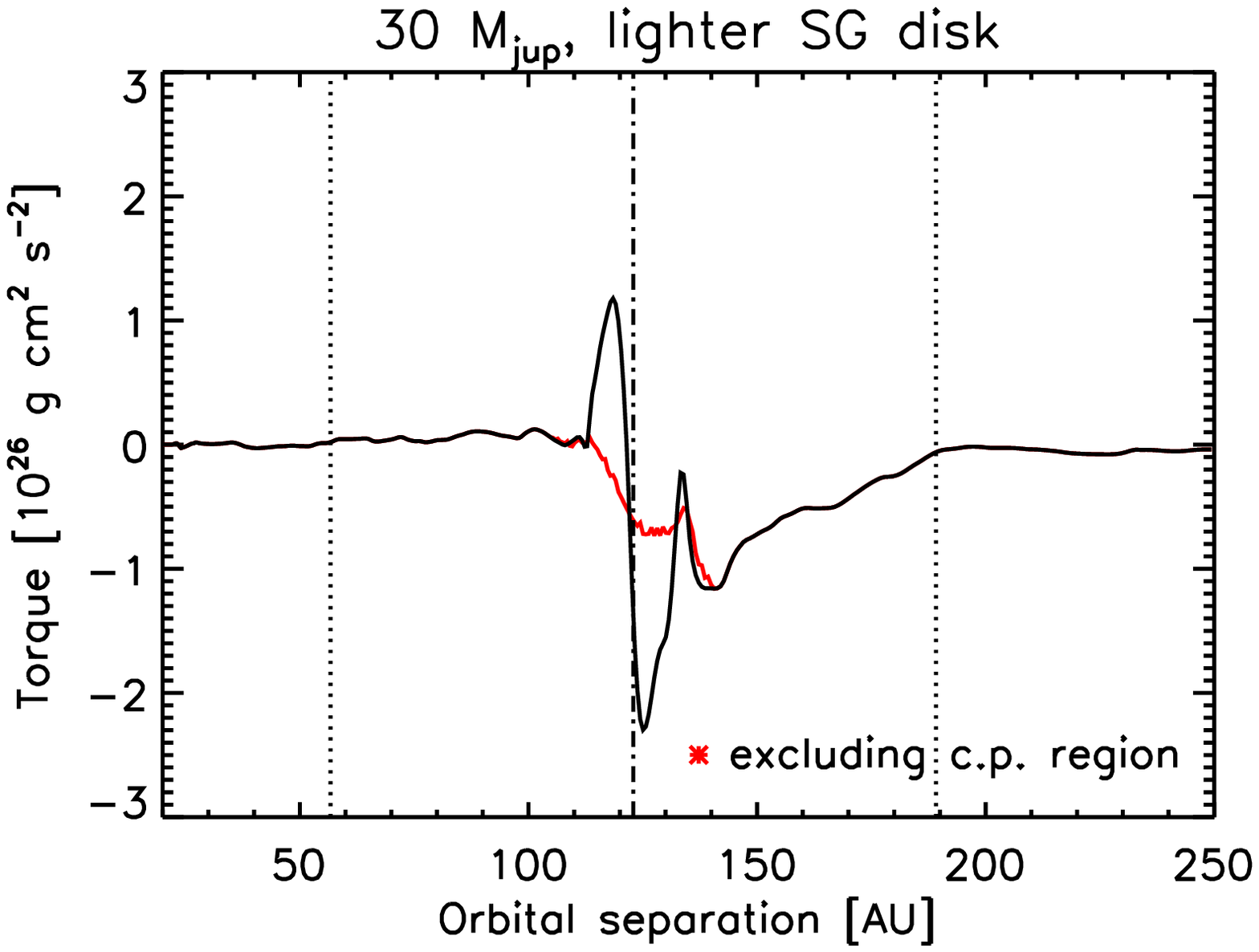}
\end{minipage}
\vspace{-0.6cm}
\begin{minipage}[b]{0.48\textwidth}
\includegraphics[width=\textwidth]{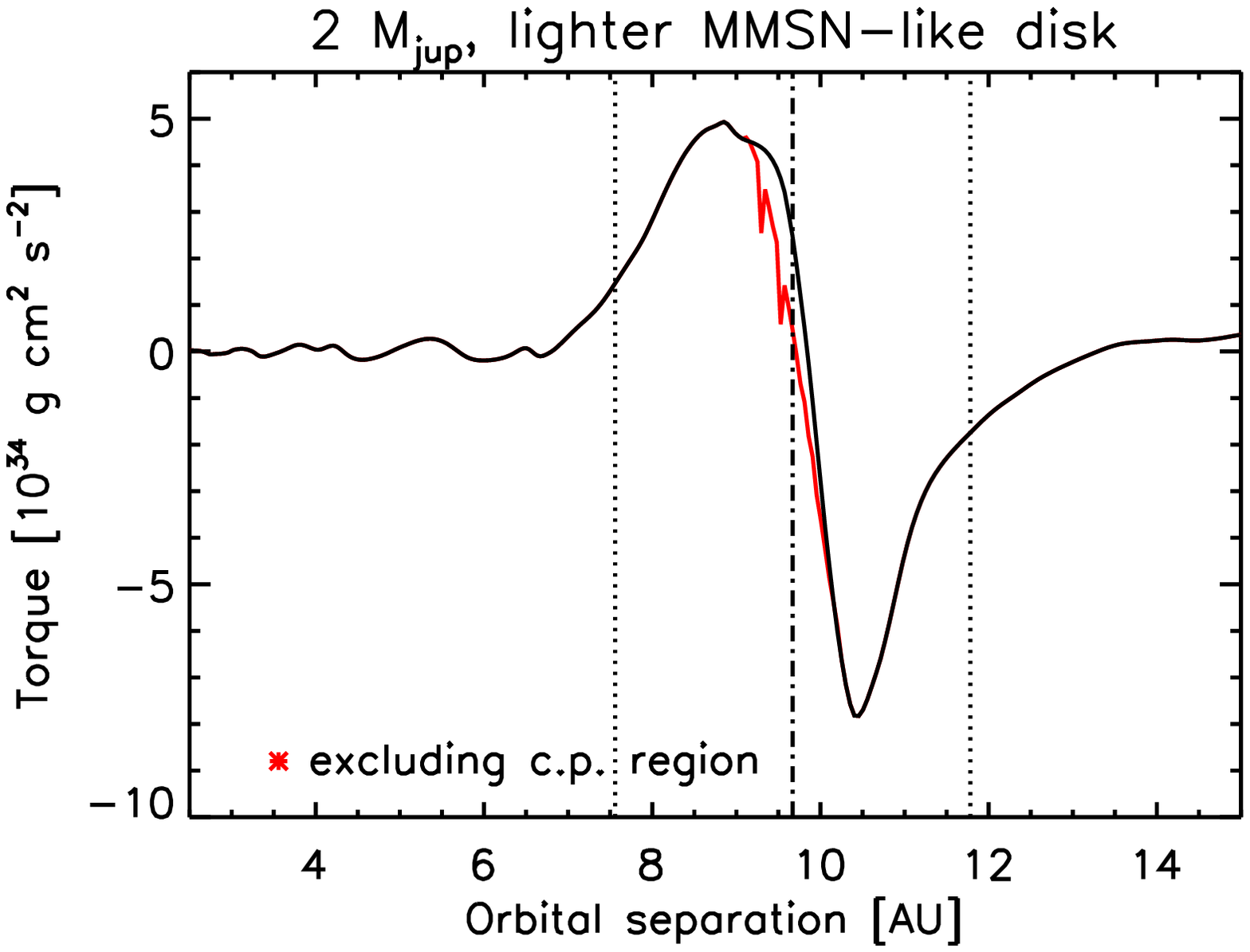}
\end{minipage}
\hfill
\begin{minipage}[b]{0.48\textwidth}
\includegraphics[width=\textwidth]{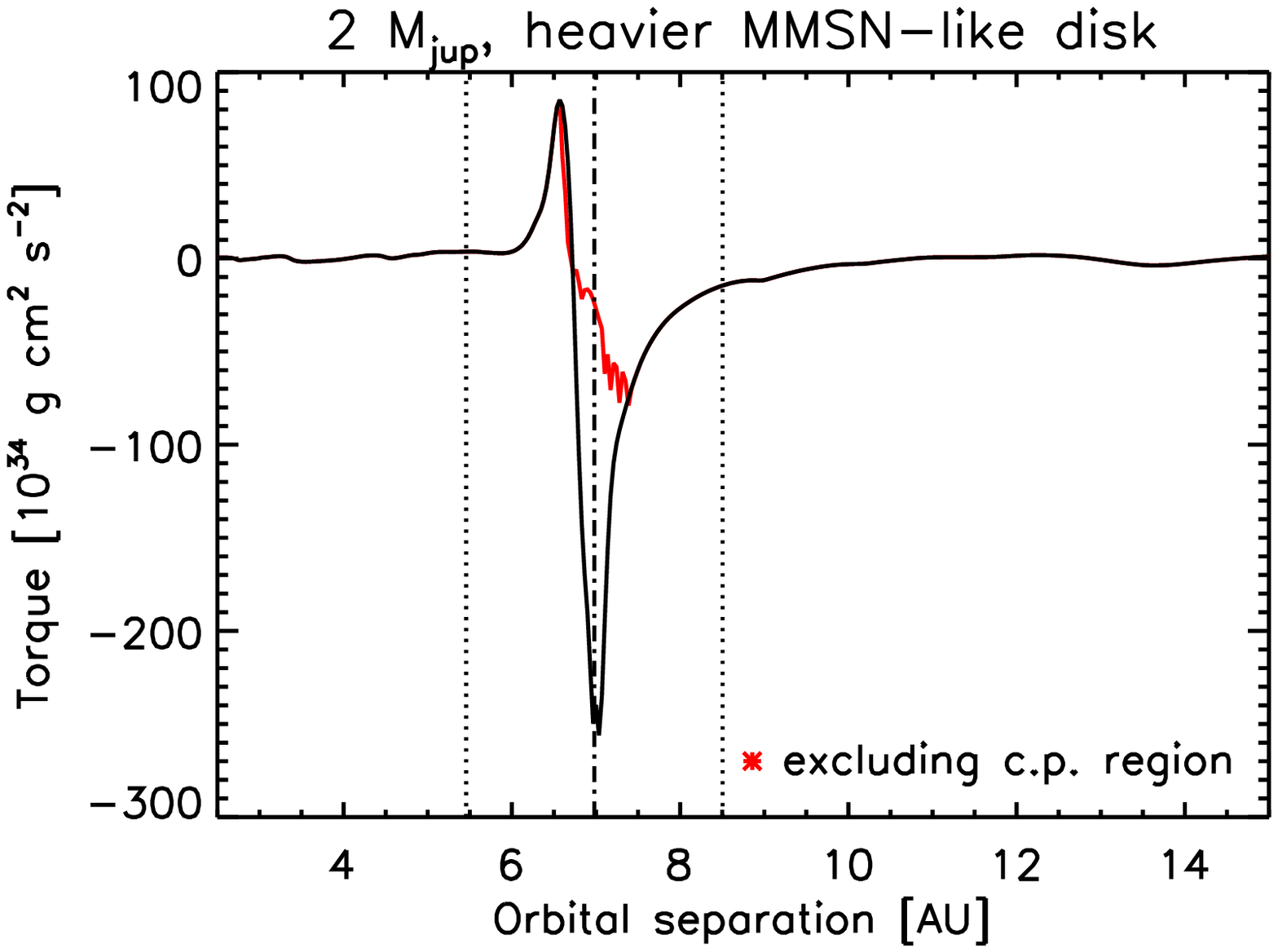}
\end{minipage}
\vspace{0.2cm}
\caption{Radial distribution of the total tidal torque acting on the companion due to the influence of the disk material. The effect of excluding the torque from the circumplanetary disk is shown in red. Vertical lines depict the companion's position (dot-dashed line) and the theoretical  outer and inner horseshoe region boundaries at $\pm$ 2.5 Hill radii (dotted lines). {\bf Top left:} 30 $M_{\rm Jup}$ companion in the reference SG disk after 3 orbits (at 100 AU). {\bf Top right:} 30 $M_{\rm Jup}$ companion in the lighter SG disk after 3 orbits (at 100 AU). {\bf Bottom left:} 2 $M_{\rm Jup}$ companion in the lighter MMSN-like disk after 15 orbits (at 5 AU). {\bf Bottom right:} 2 $M_{\rm Jup}$ companion in the heavier MMSN-like disk after 15 orbits (at 5 AU). In the reference SG and heavier MMSN-like disks the corotation torque is predominantly negative and the circumplanetary disk also provides a negative torque contribution. In the lighter SG disk the torque is generally weaker and the circumplanetary disk has a strong effect on the torque profile, providing both a positive and a negative torque. In the lighter MMSN-like disk the torque is weak, more equally distributed and the circumplanetary disk has only a small impact on the torque profile. This profile is consistent with a common Type I migration regime. These four graphs correspond to the same images displayed in Figure \ref{images}.}
\label{torque}
\end{center}
\end{figure*}

We have found that companions in disks massive enough migrate very rapidly inwards even if they are predicted to open gaps and slow down their migration, according to the torque balance criterion \citep{cr06}. The results in Table \ref{tab:timescales} show the reason behind this discrepancy: the companions simply lack the time to open gaps, as the gap-opening timescale is larger than the crossing timescale (eq. \ref{tcrossing}). We find that in both the SG and the MMSN-like disks the torque balance condition for gap-opening does not seem to sufficiently describe when gap-opening will occur. How can such rapid migrations in some of our disk models be explained?

%%%%%%%%%%%%%%%%%%%%%%%
\subsubsection{Torque}%%%%%%%%%%%
\label{secttorque}%%%%%%%%
%%%%%%%%%%%%%%%%%%%%%%%

\begin{table*}
	\caption{Companion mass including the mass of its circumplanetary disk, $M_{\rm c}$ + $M_{\rm cp}$, co-orbital mass deficit, $\delta m$, libration timescale, $\tau_{\rm lib}$, and migration timescale, $\tau_{\rm migr}$, for the simulations described in sect. \ref{secttorque}. We find that $\delta m$ is somewhat lower than $M_{\rm c}$ + $M_{\rm cp}$ in all the simulation runs but still of the same order of magnitude.  This indicates that the migrations might be within the general Type III regime but not in the runaway regime (see sect. \ref{massdef}). In contrast,  the migration time $\tau_{\rm migr}$ corresponds fairly well to the libration timescale $\tau_{\rm lib}$ in the reference SG and heavier MMSN-like disks suggesting indeed a Type III runaway migration.}
	\label{mhs}
	\vspace{-0.4cm}
\begin{center}
\bgroup
\def\arraystretch{1.5}
  \begin{tabular}{|l|l|l|l|l|}
    \hline
simulation run & $M_{\rm c}$ + $M_{\rm cp}$ & $\delta m$ &  $\tau_{\rm migr}$ [orbits] &   $\tau_{\rm lib}$ [orbits] \\ \hline 
30 $M_{\rm jup}$, reference SG disk, 3 orbits& $\sim 43.8 \MJup$ &  $\sim 17 M_{\rm jup}$ & $\sim 5$ & $\sim 4$ \\
30 $M_{\rm jup}$, lighter SG disk, 3 orbits  & $\sim 37.5 \MJup$ & $\sim 17 M_{\rm jup}$& $> 60$ & $\sim 4$ \\
2 $M_{\rm jup}$, heavier MMSN-like disk, 15 orbits & $\sim 3.2 \MJup$ & $\sim 2.1 M_{\rm jup}$& $\sim 24$ & $\sim 17$ \\
2 $M_{\rm jup}$, lighter MMSN-like disk, 15 orbits  & $\sim 2.07 \MJup$ & $\sim 0.7 M_{\rm jup}$&$\gg 600$ & $\sim 17$ \\
    \hline
  \end{tabular}
	\egroup
	\end{center}
\end{table*}

To provide an explanation for the very rapid migration timescales we investigate the tidal torque between the companion and disk material. We evaluate the torque from each grid cell on the companion given by \citep{de06}

\begin{equation}
\tau = R_{\rm c} G M_{\rm c} M_{\rm cell} \frac{d}{(d^{2}+\epsilon_{\rm c}^2)^{3/2}} {\rm sin}(R_{\rm c},d).
\end{equation}
Here $d$ represents the distance between the companion and the cell, ${\rm sin}(R_{\rm c},d)$ is the sine of the angle between the $R_{\rm c}$ and $d$ vectors and $\epsilon_{\rm c} = 0.7 H$ (as for the computation of the companion's gravitational potential). Figure \ref{torque} displays the azimuthally summed up contributions of the torque acting on the companion at a particular orbital radius in our migration simulations of the 30 $M_{\rm Jup}$ companion in the reference (top left) and lighter (top right) SG disks after 3 orbits, as well as the 2 $M_{\rm Jup}$ companion in the lighter (bottom left) and heavier (bottom right) MMSN-like disks after 15 orbits. To examine the impact of the circumplanetary disk on the torque we also show the torque profile with this region's exclusion in red. We find in general that the corotation torque, exerted by disk material within the horseshoe region, plays an important role on the magnitude and the orientation of the total torque. In the reference SG and heavier MMSN-like disks the torque is predominantly negative and the circumplanetary disk has also an additionally negative effect (compare red and black profiles in Fig. \ref{torque}). Such a strong negative torque would explain the rapid inwards migrations and very short migration times in these simulation runs. Torque profiles with a large peak in the immediate proximity of the companion may be indicative of a Type III migration regime \citep{ma03}. In the lighter SG disk the torque is generally weaker than in the reference SG disk. The circumplanetary disk has also a strong effect but in this case provides both positive and negative contributions, which may cancel each other out and thus provide a possible explanation for the slower migration of the 30 $\MJup$ companion in the lighter SG disk. In the lighter MMSN-like disk the torque is substantially weaker than in the heavier MMSN-like disk. Moreover, the torque is more equally distributed and hence the exclusion of the circumplanetary region has a significantly smaller effect on the torque profile, which is consistent with a common Type I migration regime.

Although not depicted here, examining the tidal torque profile of the 0.5 $M_{\rm Jup}$ companion at several stages of its inward migration in all three MMSN-like disks suggests a Type I to Type III regime change during the inward migration. This is visible in Fig. \ref{tinylam} by the companion's acceleration of the migration rate between an orbital separation of 6 - 9 AU (depending on the disk model).

%%%%%%%%%%%%%%%%%%%%%%%%%%%%%%%%%%%%%%%%%
\subsubsection{Mass deficit and timescales}%%
\label{massdef}%%%
%%%%%%%%%%%%%%%%%%%%%%%%%%%%%%%%%%%%%%%%%

There are two dynamical effects that play a dominant role in the corotation torque of a migrating companion: A rapid inward migration of a companion causes disk material, which otherwise would be trapped on a horseshoe orbit, to flow across the companion's corotation region. The associated negative angular momentum transfer from the crossing particles onto the companion scales with the drift rate. In contrast, material within the companion's circumplanetary region, dragged inwards with the companion, yields a positive torque on the companion. Hence this mechanism slows down the migration. From these considerations it is plausible, that the mass found within the horseshoe region as well as the mass trapped in the circumplanetary region impacts the migration rate. \cite{ma03} stated that for a very rapid migration (Type III migration) to establish ,the ''co-orbital mass deficit''\footnote{The reduction in disk mass in the companion's horseshoe region (with width 5 $R_H$) due to its presence}, $\delta m$, should be of the same order of magnitude as the companion's mass plus the mass within the circumplanetary disk, $M_{\rm c} + M_{\rm cp}$. Moreover, if $\delta m > M_{\rm c} + M_{\rm cp}$ the positive feedback of the crossing disk material causes the migration to become extremely fast, a regime known as "runaway" migration. If $\delta m < M_{\rm c} + M_{\rm cp}$ the negative feedback of the circumplanetary disk prevails and weakens the negative corotation torque enough to prevent any runaway. The migration timescale in the runaway case is thought to be a few times the libration timescale of disk particles on a horseshoe orbit around the companion \citep{li10}.

We perform a rough analysis of the mass deficit and the migration versus libration timescale for the simulations in question to determine whether our simulations fit the above stated conditions for a Type III migration regime in general or even specifically a runaway migration. First, we compute the mass deficit by

\begin{equation}
{\delta m} = 2\pi \int \limits_{R_{\rm c}-R_{\rm HS}}^{R_{\rm c}+R_{\rm HS}} \Delta \Sigma R \mathrm{d} R,
\label{eq:massdef}
\end{equation}
where $\Delta \Sigma$ is the difference between the unperturbed surface density (i.e. the "smoothed" density with the stochastic peaks and troughs flattened out at the time of the companion's introduction) and the actual surface density at a given $R$.\footnote{Our method of calculating the mass deficit is similar, but slightly different to that often used in the literature (e.g. \citealt{kl12}).  The latter assumes a uniform surface density over the horseshoe region equivalent to that at the inner separatrix.  We find that while this approach is accurate for light disks and small planets, it largely overestimates the total mass the horseshoe region would have without the influence of the companion in our simulation set-up.  For example, for the 30 $\MJup$ companion in our SG disks, this equates to the mass deficit being almost equivalent to the entire disk mass. Therefore a more accurate way of calculating the mass deficit in this region is needed rather than assuming a uniform surface density.} Second, we write the libration timescale as \citep{ba08}
\begin{equation}
\tau_{\rm lib} \approx \frac{8\pi R_{\rm p}}{3\Omega_{\rm p}R_{\rm HS}}.
\label{libtime}
\end{equation}
In addition, we estimate $M_{\rm cp}$, the mass of the circumplanetary disk.

We focus our attention on the simulations described in section \ref{secttorque} and shown in Fig. \ref{images} and Fig. \ref{torque}. For these we compare the companion's mass plus the mass of the circumplanetary disk, the mass deficit, the libration timescales, and the migration timescales in Table \ref{mhs}. The values for $M_{\rm c} + M_{\rm cp}$ and $\delta m$ are the average between the 2nd and the 4th orbit in the massive SG disk simulations and between the 11th and the 20th orbit in the MMSN-like disk simulations. The mass deficit is also visible in Figure \ref{images} (darker regions around the embedded companions).

The general condition for Type III, i.e. $\delta m \sim M_{\rm c} + M_{\rm cp}$, is satisfied roughly for all the depicted cases. Although not strongly, it is noticeable that the difference between $\delta m$ and $M_{\rm c} + M_{\rm cp}$ is smaller in the reference SG disk and heavier MMSN-like disk compared to the lighter SG disk and lighter MMSN-like disk, respectively. The condition for runaway migration, i.e. $\delta m > M_{\rm c} + M_{\rm cp}$, is not satisfied for any of our simulations. In contrast to the condition imposed on the mass deficit, the timescale comparison suggests that the reference SG and heavier MMSN-like disks might be in the runaway regime as their migration times are comparable to the respective libration timescale. Overall it remains unclear if the very rapid migration regime observed in many of our runs belongs to the categories referred to as "runaway" or simply "Type III migration" in the literature.

\subsection{Numerical tests}

\subsubsection{Resolution}
Since rapid migration depends on the local torques near the corotation region, inadequate resolution can affect the migration results. We repeat the simulation involving the 2 $\MJup$ mass companion in the heavier MMSN-like disk, as well as the 30 $\MJup$ companion in the reference SG disk using twice as many grid cells in both the radial and azimuthal directions and find that this does not affect the migration timescales.

\subsubsection{Softening length}

In their Fig. 7, \cite{mu12} investigated the impact of $\epsilon_c$ on the net torque acting on their planet. Varying $\epsilon_c$ from $0.4H$ to $1.0H$, they find that the torque density in the planet's influence region decreases by a factor $\sim 2$ to $3$ for a Neptune mass planet in a disk with scale height, $H=0.05R$.

\begin{figure}
\begin{center}
\begin{minipage}[b]{0.48\textwidth}
\includegraphics[width=\textwidth]{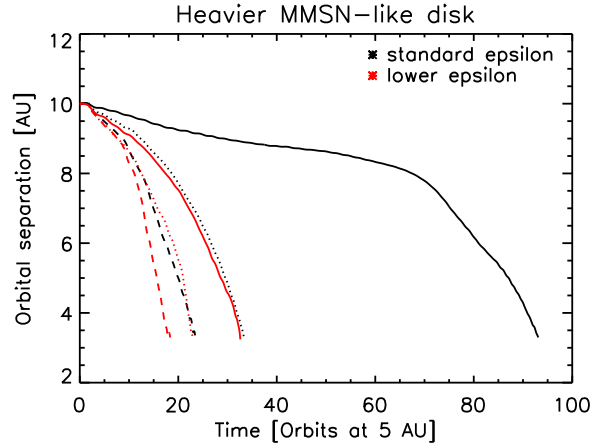}
\end{minipage}
\vspace{-0.6cm}
\caption{Evolution of the orbital separation of a migrating 0.5 $M_{\rm Jup}$ (solid line), 1 $M_{\rm Jup}$ (dotted line) and 2 $M_{\rm Jup}$ companion (dashed line) in the heavier MMSN-like disk with different softening lengths. \textbf{Black:} $\epsilon_{\rm c} = 0.7$, $\epsilon_{\rm SG} = 0.8$. \textbf{Red:} $\epsilon_{\rm c} = 0.3$, $\epsilon_{\rm SG} = 0.4$. A lower softening length causes a more rapid migration.}
\label{tinylamsoft}
\end{center}
\end{figure}

Since the net torque on the companion is the "motor" of migration we perform softening length tests by conducting another suite of simulations using the heavier MMSN disk set-up as in section \ref{mmsndisks} but with $\epsilon_{\rm c} = 0.3$ and $\epsilon_{\rm SG} = 0.4$. We find that a lower softening length increases the angular momentum exchange between the companion and disk by diminishing the smoothening out in the companion's horseshoe region. This causes the companions to migrate even more rapidly through the disk (see Fig. \ref{tinylamsoft}).

Our results are by such consistent with the findings of \cite{mu12}. This shows the importance of using an accurate softening length in migration simulations. Based on \cite{mu12} we choose $\epsilon_{\rm c} = 0.7 H$ and $\epsilon_{\rm SG} = 0.8 H$ since it provides a reasonable description of the gravity force.

\vspace{0.8cm}

\subsubsection{Growth time after companion introduction}

We introduce the companions by linearly increasing their mass over one orbit (see section \ref{intro}). To establish the effect of this numerical choice we repeat the simulations with the 30 $\MJup$ companion in the reference SG disk and the 2 $\MJup$ companion in the heavier MMSN-like disk with introductory growth times of 10 and 600 orbits, respectively. We find that as soon as the companion accumulates enough mass to establish an effective torque on the surrounding disk the subsequent migration is as rapid as in our standard runs. Hence we conclude that the salient features of our simulation results are independent of the numerical growth time.

%%%%%%%%%%%%%%%%%%%%%%%%%%%%%%%%%%%%%%%%%%%%%%%%%%%%%%%%
%%%%%%%%%%%%%%%%%%%% discussion %%%%%%%%%%%%%%%%%%%%%%%%
%%%%%%%%%%%%%%%%%%%%%%%%%%%%%%%%%%%%%%%%%%%%%%%%%%%%%%%%

\section{Discussion}
\label{discussion}

The gap-opening criterion is fundamental to understand planet survival from theoretical and observational perspectives.  We find that not only is it important to consider whether a gap can open in a disk, but more importantly whether it can open \emph{quickly} enough. The frequently used gap-opening criterion \citep{cr06} was derived for a planet on a fixed orbit and not for a migrating planet. Our results show that even when the gap-opening criterion is satisfied there are still instances when a gap does not open. The timescale on which the planet carves out the gap, $t_{gap}$, is the gap-opening timescale. However, at the same time, the planet is also migrating through the region that it is carving out on a timescale, $t_{cross}$. If the planet crosses this region quicker than the time it takes to carve out the gap, i.e. if $t_{cross} < t_{gap}$, a gap will not form. Indeed, when comparing the gap-opening timescales to the migration timescales in our simulations, we find that in many cases even when the gap-opening criterion according to \cite{cr06} is satisfied the gap simply cannot open because the timescale constraint is not satisfied.  Our results show that both a torque balance criterion and a timescale condition are necessary.  The combination of the two will always catch the limiting factor, resulting in an equivalent or stricter criterion for gap-opening.  Thus previous assessments of the gap-opening criterion where the planet is kept on a fixed orbit (e.g. \citealt{cr06}, \citealt{du13}) may not be sufficient enough to determine if a gap will form in a disk.  We also find that gap-opening is dependent on the disk mass when this is not a factor in the semi-analytical torque balance equation (but is implicit in the timescale condition).

We can understand from a phenomenological perspective why this additional timescale condition is necessary to consider. The crossing timescale depends on the torques acting on the planet which depend on the planet-to-primary mass ratio, the disk location, aspect ratio, the surface mass density and hence disk mass, the surface mass density profile and the temperature profile. The gap-opening timescale depends on the aspect ratio, the planet-to-primary mass ratio, the viscosity and the radial location (\citealt{ho84}; \citealt{ra02}; \citealt{ed07}; \citealt{ba13}). Out of these parameters, the current gap criterion by \cite{cr06} does not consider the disk mass or the surface mass density and temperature profiles, though intuitively they would be expected to affect gap-opening.  Yet these are crucial to determining a planet's migration rate through the region that it is trying to open a gap in. In fact the migration and gap-opening timescales are affected by all the variables in the \cite{cr06} criterion as well as the aforementioned additional variables.

Furthermore, our results have a broad impact since we show that this timescale condition is important to consider in both high mass self-gravitating disks and low mass T Tauri disks where planets/companions are expected to form by gravitational instability and core accretion, respectively.

Note that a timescale criterion for gap-opening of the form $t_{cross} > t_{gap}$ would need to naturally take into account very rapid migration (e.g. Type III). This occurs for intermediate mass planets or for planets in very turbulent disks which end up having circumplanetary material around them that causes the corotation torque to be more important.  Different aspects of our analysis suggest that the planet evolves in a regime which exhibits some of the features attributed to Type III-like migration, or even runaway migration \citep{ma03}. In these cases the corotation torques yield the most important contribution and govern planet migration. The relatively large effective viscosity arising in gravitationally unstable disks may explain why the corotation torques never saturate in this case, leading to Type III migration: mass may simply be transferred to the co-orbital region faster than the libration timescale.  This in turn prevents gap-opening, not because the gap is refilled by the viscous flow of material through it but because the planet transits too rapidly through the region where it is trying to open a gap.

We note that disks that have different structures will affect a planet's migration rate and direction, and thus its ability to open a gap. Furthermore, interactions between multiple companions, as well as the interactions between companions and other disk structures such as the spiral structures caused by the presence of further companion(s) or by self-gravity in high mass disks will affect the torques, possibly causing fluctuations on short timescales and adding a significant non-linearity to the outcome. These may well cause gap-opening to be harder. Even Type I migration might not display a smooth continuous radial motion throughout the disk, but may consist of periods with increased or decreased migration rates due to inhomogeneities in the disk structure. A capture of a planet into a zero torque location as described by \cite{bi11} or into mean motion resonance with another planet may well aid the gap-opening process due to the long migration timescales in such scenarios.  In all of these cases, the timescale criterion is still a relevant condition to satisfy since these can be reassessed in the context of differing situations occurring in the same background disk.

%%%%%%%%%%%%%%%%%%%%%%%%%%%%%%%%%%%%%%%%%%%%%
\subsection{Comparisons with previous work}%%
%%%%%%%%%%%%%%%%%%%%%%%%%%%%%%%%%%%%%%%%%%%%%

The idea of a timescale constraint has been considered previously.  In this paper we stress its applicability to a wide range of disks including young massive gravitationally unstable disks as well as older low mass T Tauri disks, i.e. disks in which planets may form by gravitational instability or core accretion.

\cite{ho84} and \cite{wa89} suggested that for a gap of a particular width to open, the planet must drift across that region more slowly than the time it takes to open the gap, i.e. $\tcross > \tgap$.  \cite{li86} on the other hand suggested that since the presence of a planet causes the surface density in the disk to vary the relevant timescales to consider are the crossing timescale and the timescale to modify the surface density, given by $t_\Delta = (H/R)t_{\rm gap}$, and that a gap would form if $t_{\rm cross} > t_\Delta$.  This is an easier criterion for gap-opening than what \cite{ho84} suggested.  Given that the migration simulations performed in this work are in the non-linear regime, the analytical expressions provided by these authors cannot be used to directly compare with the empirically-obtained timescales in this paper, nor do our results indicate exactly which timescale condition should be used. We simply point out the importance of considering a timescale criterion in addition to the torque balance criterion of \cite{cr06}

\cite{ta96} also explore the gap-opening criterion and show that gaps are formed when two conditions are satisfied (see their Figure 8): (i) $\tcross > \tgap$ and (ii) $t_{\rm open} < t_{\rm close}$.  The latter criterion is essentially the same as the criterion that the viscous torques must balance the gravitational torques.  They also show that when the latter criterion is satisfied, gaps do not necessarily open unless the former is satisfied.  Our results are therefore in agreement with theirs.

Similar to \cite{ho84}, \cite{mu10} also discuss the idea that a gap must open only if the migration timescale is longer than the gap-opening timescale.  Using the Type I migration timescale and their numerical results for the gap-opening timescale using local 2D shearing sheet simulations, they find that it is \emph{easier} for planets to open gaps, i.e. lower mass planets are able to open gaps compared to what the \cite{cr06} criterion predicts, in contrast to our results.  However, it must be noted that these were for inviscid disks.  Since the viscosity will affect both the gap-opening timescale (lengthening it) and the migration timescale (higher viscosities decrease the migration timescale; \citealt{ed07}), the inclusion of viscosity will make the timescale criterion harder to satisfy than what \cite{mu10} suggest for inviscid disks (though we note that they do specify that the inclusion of viscosity does require an additional condition - also based on timescale arguments - to be satisfied).  Note that the Type I migration timescale used by \cite{mu10} was an appropriate choice for them since they consider low mass (approx. 1-23 $M_\oplus$) planets but is not an appropriate choice for our simulations since the migration of our high mass companions is in the non-linear regime.

\cite{zh12} and \cite{vo13} perform hydrodynamical simulations of fragments in gravitationally unstable disks and find that in some cases they can open a gap and survive.  Both authors showed that the torque balance criterion is satisfied\footnote{Note that the authors use different torque balance criteria: \cite{zh12} require that $R_H > H$ and $M_c > 40 \Mstar \alpha h^2$ \citep{li93} while \cite{vo13} use the criterion by \cite{cr06} except that they use the mass of the fragment plus the mass within a Hill radius (i.e. that of the circumplanetary disk).}.  It would be particularly interesting to determine whether the timescale condition in their simulations is also satisfied.

\cite{zh14} examined how self-gravity and disk mass influence the ability of a Jovian-mass planet to open gaps. They used disks in the mass range of the MMSN model. They found that disks that are too massive may inhibit gap-opening, consistent with our results. Whereas we attribute this to a timescale problem they argued using 1D hydrodynamical simulations that the gravitoturbulence in a disk with a surface density $\Sigma \gtrsim 3.5 MMSN$ suppresses gap formation. In 2D simulations their planet migration time scales inversely to the disk density. In this sense our results are consistent with this trend.

%%%%%%%%%%%%%%%%%%%%%%%%%%%%%%%%%%%%%%%%
\subsection{Additional considerations}%%
\label{add}
%%%%%%%%%%%%%%%%%%%%%%%%%%%%%%%%%%%%%%%%

We note that in the results presented here the planet has a fixed mass. Migration, in any regime, and gap opening both depend on planet mass. As a planet grows, it may pass through a rapid migration phase before slowing down as it starts to open a gap.  Accretion onto the planet (and hence removal of gas in the surrounding region) can promote gap formation \citep{br99} which will also have an effect on the gap-opening timescale and hence on whether a planet can open a gap quickly enough. On the other hand, although the planet's mass affects the migration timescale, if the growth timescale is too long compared to the migration timescale, a planet's growth may not be affected (analogous to the black hole binary regime; \citealt{iv99}). Regardless, the timescale criterion presented here would still be relevant in the case of planet growth, but it simply means the the criterion would have to be re-assessed during the evolution. Recently, \cite{mo14} confirmed earlier studies by finding a substantial mass accretion mechanism onto a planet on a fixed orbit surrounded by a gap. With a 3D model they found the gas flow in the gap varied significantly with vertical height above the mid-plane. Although their result does not describe accretion onto a migrating planet, this suggests that in future studies the use of a 3D model might be ultimately necessary to realistically simulate the torque in the planet's horseshoe region and hence gap-opening.

\begin{figure}[!h]
\begin{center}
\begin{minipage}[b]{0.48\textwidth}
\includegraphics[width=\textwidth]{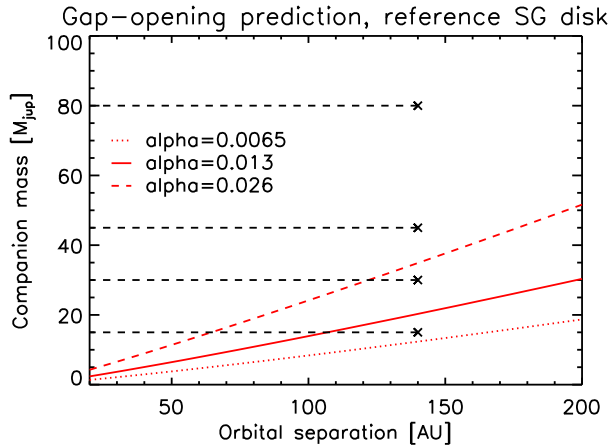}
\end{minipage}
\vspace{-0.6cm}
\caption{Companion's mass versus orbital radius indicating the torque balance criterion of \protect\cite{cr06} in red for the reference SG disk using turbulence $\alpha$ parameters of 0.0065 (dotted), 0.013 (solid) and 0.026 (dashed). Companions are expected to open a gap when situated above the limit. The crosses indicate where in the disk the companions are introduced in our simulations while the dashed lines indicate the inward migration tracks. The criterion is stricter if the turbulence parameter is increased but in all the simulations the companions are expected to open gaps at some stage during their inward migration.}
\label{fig:crit_alpha}
\end{center}
\end{figure}

We also note that in the SG disks the turbulence alpha parameter, while close to $\alpha$ = 0.013 as predicted by \cite{ga01}, may not remain at that value in the whole disk. If $\alpha$ varies then the prediction for gap-opening also changes. Figure~\ref{fig:crit_alpha} shows how the prediction changes in the reference self-gravitating disk if the turbulence varies by a factor of 2 compared to the expectation by \cite{ga01}. While this does change the prediction, the companions should still open gaps during their inwards migration. Thus our conclusion that the gaps have not opened because of the timescale argument is still valid.

%%%%%%%%%%%%%%%%%%%%%%%%%%%%%%%%%%%%%%%%%%%%%%%%%%
\subsection{Implications for planet formation models}%
%%%%%%%%%%%%%%%%%%%%%%%%%%%%%%%%%%%%%%%%%%%%%%%%%%

Our results may have important implications on the use of the criterion in studies using population synthesis models for planets formed by core accretion or gravitational instability, as well as the tidal downsizing model for planets formed by gravitational instability (\citealt{bo10}; \citealt{na10}).  Conventional population synthesis models for standard core accretion planets typically assume that a planet migrates on a Type I migration timescale until it has grown large enough to satisfy the gap-opening criterion. At this stage the models then instantaneously make the planets migrate on a Type II migration timescale (e.g. \citealt{al04}; \citealt{id04}), or more recently involve a smooth transition function between the Type I and Type II migration regimes \citep{di14}.  This is also the case for N-body simulations of planetary system formation in evolving disks \citep{co14}. Population synthesis for disk instability planets has only recently begun to be developed (\citealt{fo13}; \citealt{ga14}).  Both authors instantly change from Type I to Type II migration when a gap is deep enough to have opened.  The former assume this to be the case when the pressure stability condition (eq. \ref{shock}) is satisfied\footnote{Note that their pressure stability condition marginally differs from equation (\ref{shock}) in that the right hand side of their equation is $R_{\rm p} (q/2)^{(1/3)}$.}, while the latter assume that gap-opening occurs when \cite{cr06}'s criterion (eq. \ref{cridacrit}) is satisfied.

Note that population synthesis models do not consider very rapid migration scenarios like Type III, though incorporating a gap-opening timescale condition into such models could naturally account for this, as it would naturally ensure that gap-opening would take longer than the typical migration timescale, thus providing a more accurate estimate of the survivability of planets.  Our results show that gap-opening is the same or harder than that suggested by the torque balance criterion and therefore such population synthesis models are likely to underestimate how quickly a planet transitions from the Type I to Type II migration regimes, and thus have an important effect on a planet's survivability.

From the gravitational instability perspective, \cite{ga14} found that the interplay between the collapse timescale, the migration timescale in the Type I regime, the gap-opening timescale and the tidal mass loss timescale based on the tidal downsizing hypothesis (\citealt{bo10}; \citealt{na10})  is what determines the fraction of surviving gas giants. They used the simulation results by \cite{ga12}, who showed that the collapse to planetary densities occurs on timescales of $\sim 10^4$ yr (comparable to Type I migration timescales), and simply adopted the torque balance criterion of \cite{cr06} to decide when planets open a gap and enter the slow Type II migration regime. By comparing models with and without gap-opening they found very large qualitative differences in the fraction of surviving planets, mass distributions and semi-major axes after less than a million years of disk evolution. In particular, without gap-opening $< 10 \%$ of the initial population of gas giants survived, and most of them ended up on close-in orbits consistent with Hot Jupiters, whereas with gap-opening the fraction of surviving planets increased to $> 50 \%$ , with a sizeable fraction on orbits having $R > 5$ AU. The final mass distribution also differed, being skewed to lower masses when gap-opening is not included as tidal mass loss is more effective when planets migrate inward fast. Other semi-analytical models studying the evolution of a population of disk instability planets, including also dust and planetesimals, were published by \cite{fo13}. They considered an even simpler mass threshold above which gap opening occurs depending on the disk pressure scale height and adopted standard Type I and Type II migration timescales based on the results of Bate et al. (2003). Our results show that gap-opening should occur on timescales shorter than $10^5$ yr, corresponding to the typical clump collapse timescale, in order to affect a planet's orbital evolution, and that this requires relatively low mass disks ($M_{\rm disk} < 0.05 M_\odot$) for planets with masses of a few $\MJup$. Models such as those of \cite{ga14} and \cite{fo13} should thus be revisited by introducing a criterion for the gap-opening timescale, and possibly mechanisms of disk dispersal acting since we show that the disk mass clearly plays a major role.

%%%%%%%%%%%%%%%%%%%%%%%%%%%%%%%%%%%%%%%%%%%%%
\subsection{Implications for observations}%%%%%%%%%%%%%%%%%%%%%%%%
%%%%%%%%%%%%%%%%%%%%%%%%%%%%%%%%%%%%%%%%%%%%%
 
There are many observations that indicate the presence of substantial gaps in gas-rich disks of the order of 10 AU or larger. These gaps have been used to argue for the presence of planetary mass objects embedded within them based on the torque balance gap-opening criterion.  In the context of this work, we can distinguish four cases:

\vspace{0.15cm}

\begin{enumerate}[label=\roman*),nosep]
\item{observed gap with a planet}
\item{observed gap with no planet}
\item{no observed  gap and no planet}
\item{no observed gap but with a planet}
\end{enumerate}

\vspace{0.15cm}

The first case is the major topic addressed in this paper, where we suggest a revision of the conditions needed to open a gap. Since it may be harder for planets to open a gap when considering migration this suggests that higher mass planets may be needed to open a gap in a disk than that based on the pressure stability criterion of \cite{li93} or the torque balance criterion of \cite{cr06} alone. Thus the torque balance criterion provides a \emph{minimum} mass for a single planet to open the gap. Therefore planets in transition disks may be easier to detect due to the higher expected planet masses. In this context it might be worth revisiting published results that searched for planets  embedded within a gap, or investigating those disks whose gaps previously indicated planet masses below the current observable limits. If this is ruled out, one would be tempted then to hypothesize multiple planets.

The second case would require a more sophisticated examination of possible hydrodynamic effects that may cause gaps (that is beyond the scope of this current paper).

The third case, may be common and probably the case for the majority of observed T Tauri systems for most of their evolution: fewer than 20 \% of sun-like stars are expected to form gas giant planets between 0.3 - 20 AU based on extrapolations from \cite{cu08}. It is important to remember that not all disks are easy to resolve in scattered light as observations are limited by the obtainable surface brightness. However, if a disk is resolved in scattered light and the disk structure does not indicate a gap, it is possible that there is no large gap of radial extent in the gas surface density distribution as the scattered light traces small dust particles which are coupled to the gas. But it is an open question whether even small amounts of dust might provide enough scattered light to conceal a partially cleared narrow gap in the gas.

The final case, may be more common than current estimates.  With state-of-the-art instrumentation, it can be difficult to detect a gap that is not completely cleared of material unless it is a significant fraction of the current beam width of mm-wave interferometers (about 10 AU).  A massive planet with a Hill radius of 5 AU that causes a drop in the surface mass density of $> 20 \%$ might just be detectable with high signal-to-noise. Just because we cannot detect a gap with current facilitites, does not mean that one is not there!  In the coming years, observations with the expanded capabilities of ALMA will probe with the best sensitivity and resolution possible the presence or absence of gaps in disks around T Tauri stars (e.g. the extraordinary images of HL Tau from ALMA\footnote{http://www.eso.org/public/news/eso1436/}).   Ultimately the next generation of ELTs will provide additional constraints on the presence/absence of gaps at even higher spatial resolution.

Finally, since gravitationally unstable disks require very high companion masses to open a gap, our results suggest that we should not see many gaps (formed by planets) in young gravitationally unstable disks.  On the other hand if gaps in such disks are observed this suggests that some mechanism must take place that slows down or stalls planet migration in young disks.  Future spatially-resolved observations of these disks, e.g. with ALMA, will inform us about this.

%%%%%%%%%%%%%%%%%%%%%%%%%%%%%%%%%%%%%%%%%%%%%%%%%%%%%%%%
%%%%%%%%%%%%%%%%%%%% Conclusion %%%%%%%%%%%%%%%%%%%%%%%%
%%%%%%%%%%%%%%%%%%%%%%%%%%%%%%%%%%%%%%%%%%%%%%%%%%%%%%%%

\section{Conclusion}
\label{conclusion}

We perform 2D hydrodynamical simulations using the grid-based code, {\sc fargo}, to explore gap-opening in a variety of protoplanetary disks that would be expected to form giant planets either by the core accretion or gravitational instability methods.

The main results of our study are:

\begin{itemize}

\item{In general, we find that gap-opening may be harder than that predicted by the torque balance criterion alone \citep{cr06}, which is the criterion that is commonly used to determine whether gap formation occurs.  

Typically, this has been considered in the context of stationary planets - however, we find that when considering migrating planets, an additional timescale factor is very important to consider. We find that not only is it crucial whether a planet is massive enough to open a gap, but also whether the planet can do so \emph{quickly} enough.  We find that even if the torque balance criterion is satisfied, if the timescale for migration across a region in which a planet is opening a gap, $\tcross$, is faster than the gap-opening timescale, $\tgap$, a planet is not able to open a gap, i.e. for gap-opening we require the additional condition $\tcross > \tgap$. Although the timescale aspect for gap-opening has been investigated before (\citealt{ho84}; \citealt{li93}), we find that the timescale comparison might pose a strict limitation to the occurence of gap-opening. This is particularly the case in migration scenarios where we do not expect conditions for a ''smooth'' Type I migration regime. If a planet's mass is in the \emph{intermediate} stage (when it is opening a partial gap), it may undergo rapid Type III-like migration or even runaway migration.  While the torque balance criterion may suggest that gap-opening should occur, in reality gap-opening may be prevented by the timescale restriction.  We show this to be the case with low mass disks in which planets are likely to form by core accretion, as well as high mass disks in which giant planets or brown dwarfs may form via gravitational instability.}

\item{The disk mass is an important factor that affects gap-opening as this affects the torques on the planet and hence its ability to migrate through a disk. While the disk mass is not considered in the torque balance criterion for gap-opening, it does affect the crossing timescale and is thus an important factor in the timescale condition, $\tcross > \tgap$, which considers all the variables in the torque balance criterion plus more.}

\end{itemize}

Our results are applicable to both observational interpretations and theoretical modelling. Since gaps may be harder to open for migrating planets than when only considering the balance of torques at a particular location, the interpretation on companion masses that open gaps of particular widths in transition disk observations should only be considered as a minimum mass.  Furthermore, population synthesis models may find that planet survival is more difficult with the addition of a timescale constraint.  Finally, since the gravitationally unstable disk simulations show that planets in such disks seem to undergo very rapid migrations, survivability in these disks may only be possible if the SG disks are not very heavy.  Future spatially-resolved observations, such as with ALMA, may constrain if gap-opening in such disks is possible.

%%%%%%%%%%%%%%%%%%%%%%%%%%%%%%%%%%%%%%%%%%%%%%%%%%%%%%%%
%%%%%%%%%%%%%%%%%%%% Acknowledgements %%%%%%%%%%%%%%%%%%
%%%%%%%%%%%%%%%%%%%%%%%%%%%%%%%%%%%%%%%%%%%%%%%%%%%%%%%%

\section*{Acknowledgments}
We thank Richard Nelson, Alessandro Morbidelli, John Papaloizou, Cl\'ement Baruteau, Aur\'elien Crida, Bertram Bitsch \& Lia Sartori for interesting discussions and the referee for a thorough review. The calculations reported here were performed using the {\sc brutus} cluster at ETH Z\"urich.  FM was supported by the ETH Zurich Postdoctoral Fellowship Programme as well as the Marie Curie Actions for People COFUND program. Also, this work has been supported by the DISCSIM project, grant agreement 341137 funded by the European Research Council under ERC-2013-ADG.

\bibliographystyle{apj}

\bibliography{manuscript}

%%%%%%%%
\appendix%%%
%%%%%%%%

\vspace{0.5cm}
\counterwithin{figure}{section}

\section{Further simulation results}
\label{plots}

\begin{figure}
\begin{center}
\begin{minipage}[t]{0.48\textwidth}
\includegraphics[width=\textwidth]{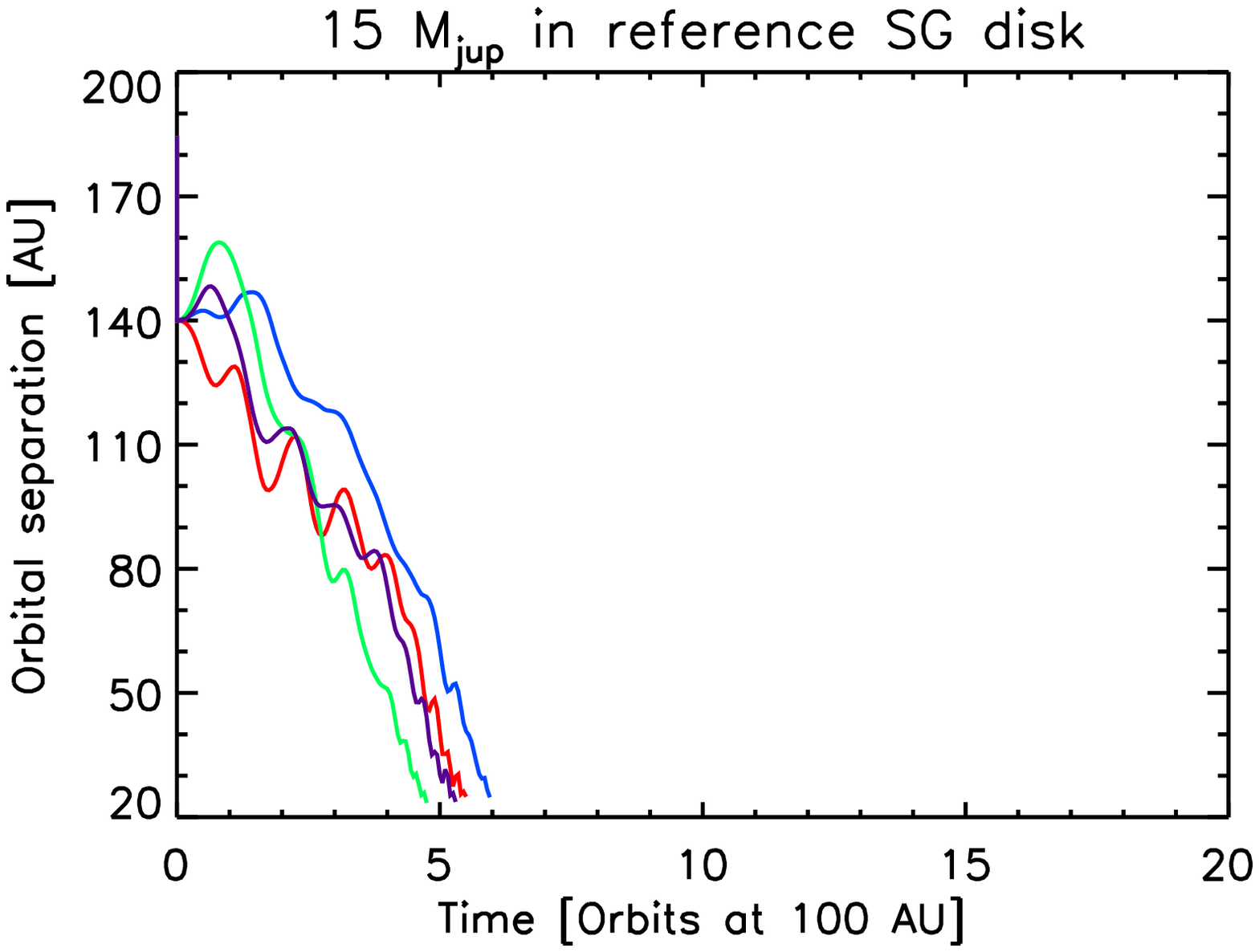}
\end{minipage}
\hfill
\begin{minipage}[t]{0.48\textwidth}
\includegraphics[width=\textwidth]{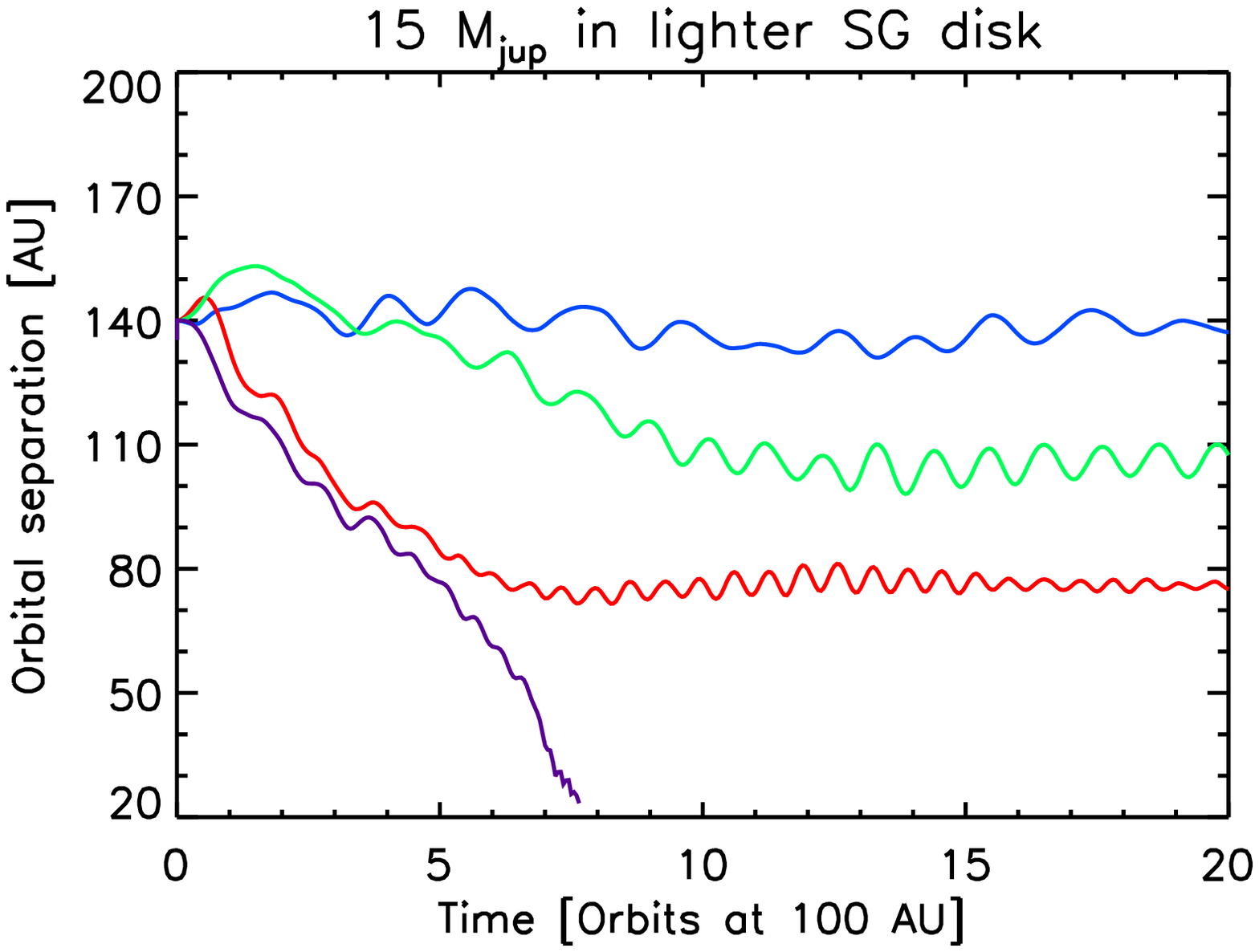}
\end{minipage}
\begin{minipage}[t]{0.48\textwidth}
\includegraphics[width=\textwidth]{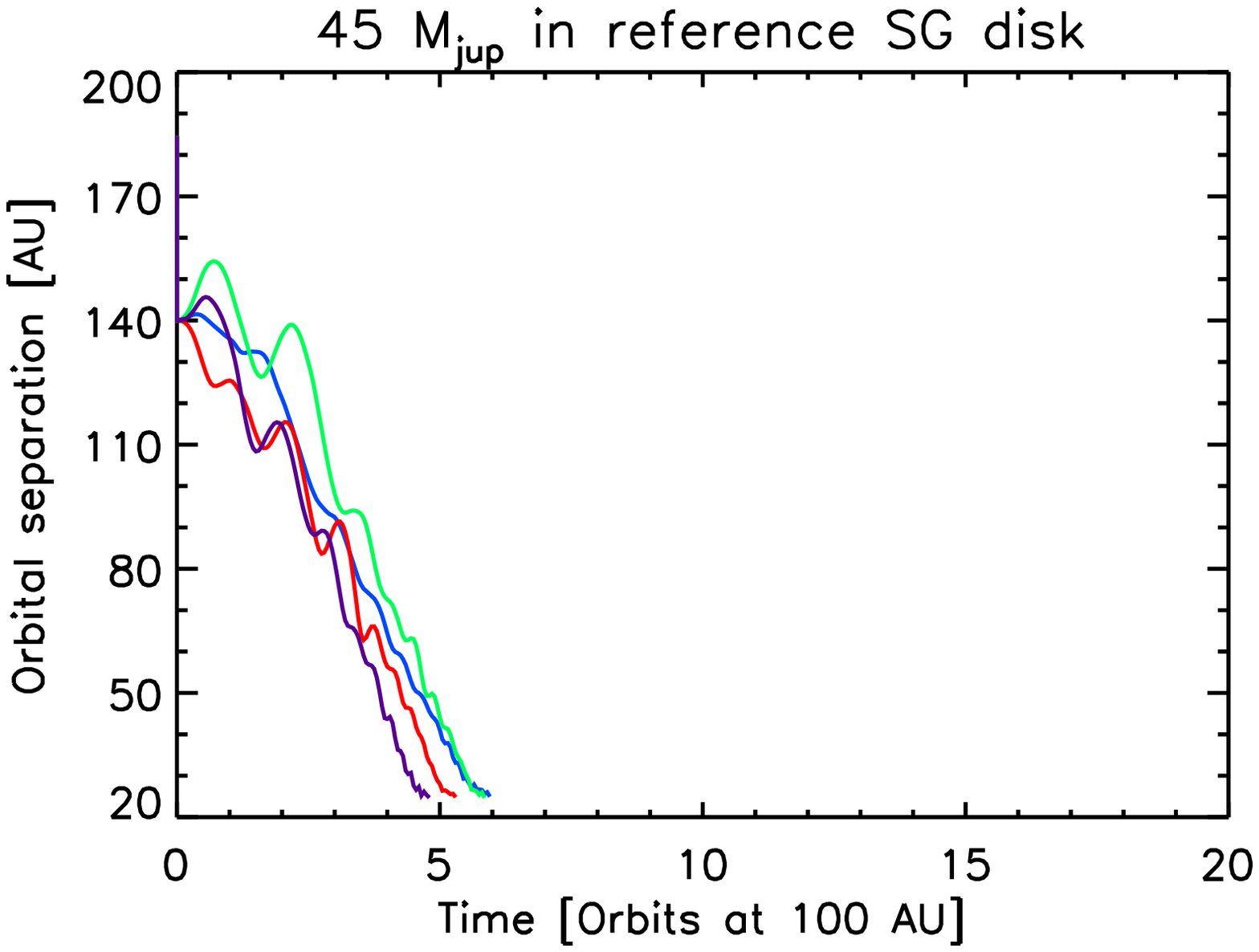}
\end{minipage}
\hfill
\begin{minipage}[t]{0.48\textwidth}
\includegraphics[width=\textwidth]{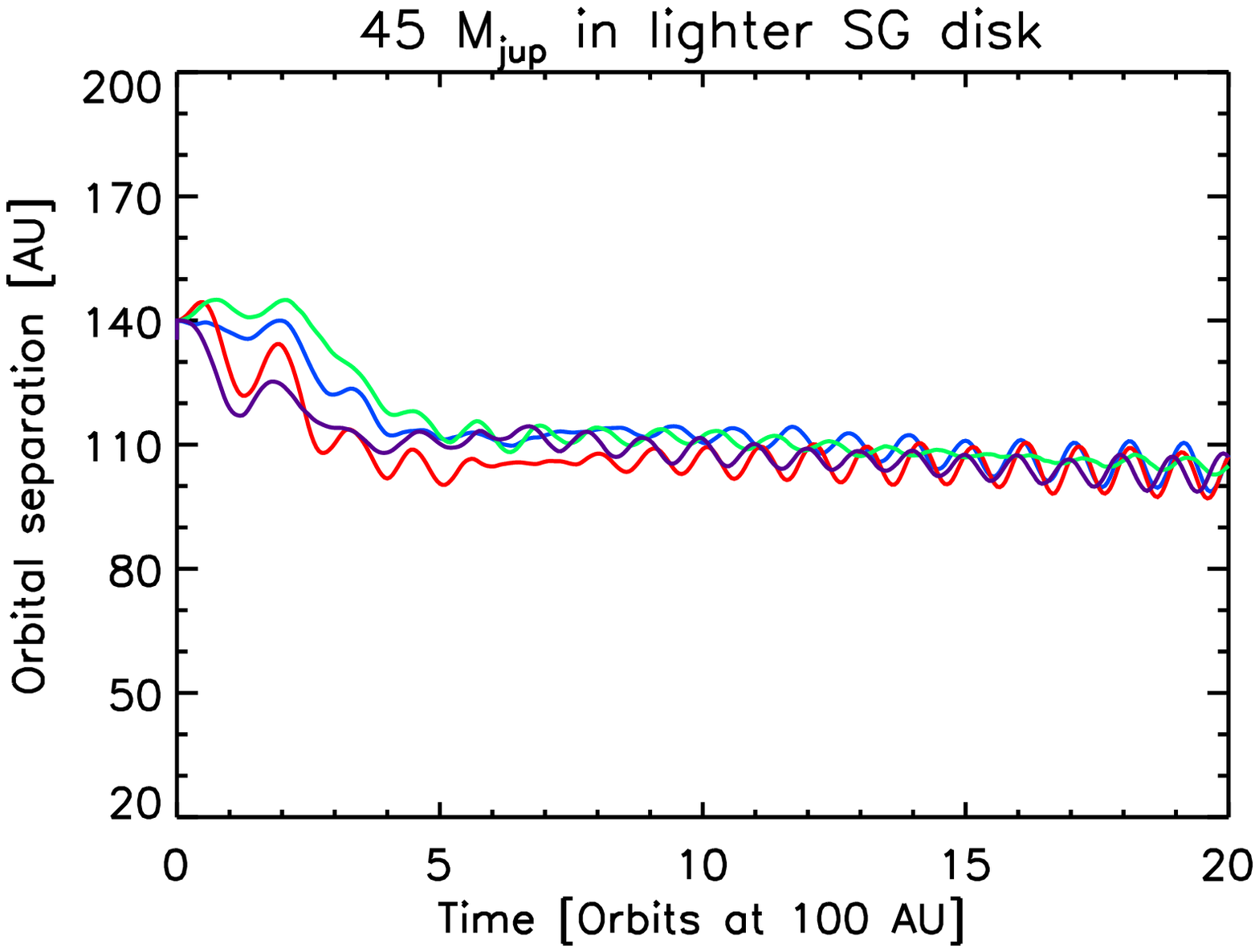}
\end{minipage}
\vspace{-0.3cm}
\caption{Evolution of the orbital separation of a migrating 15 $M_{\rm Jup}$ (top) and 45 $M_{\rm Jup}$ (bottom) companion in the reference (left) and lighter (right) SG disks. The simulation is performed four times each time starting at the same orbital radius but at a different azimuth angle leading to four separate migration trails (displayed with different colors). In the reference SG disk the companions migrate rapidly through the disk preventing any potential gap-opening. In the lighter SG disk, they decelerate enough or even remain at the radial location of introduction so that a gap might evolve. Only the 15 $M_{\rm Jup}$ reaches the inner grid boundary once, likely due to a stochastic effect.}
\label{fig:SG_15_45}
\end{center}
\end{figure}

For the curious reader we present further visualizations of the simulations conducted. Figure \ref{fig:SG_15_45} shows the migrations of a 15 $\MJup$ and 45 $\MJup$ companions in the reference and light SG disks analogous to Figure \ref{SGdisk}. As shown in Figure \ref{fig:SG_15_45}, an increased planet mass or decreased disk mass improves the ability for gap-opening.  At some stage during the inwards migration a gap should open in these simulations (see Figure \ref{criteria}) but the very rapid migration in the reference SG disks prevents this. In the lighter SG disks the migrations are generally much slower and gap formation appears possible.

\begin{figure}
\begin{center}
\begin{minipage}[t]{0.48\textwidth}
\includegraphics[width=\textwidth]{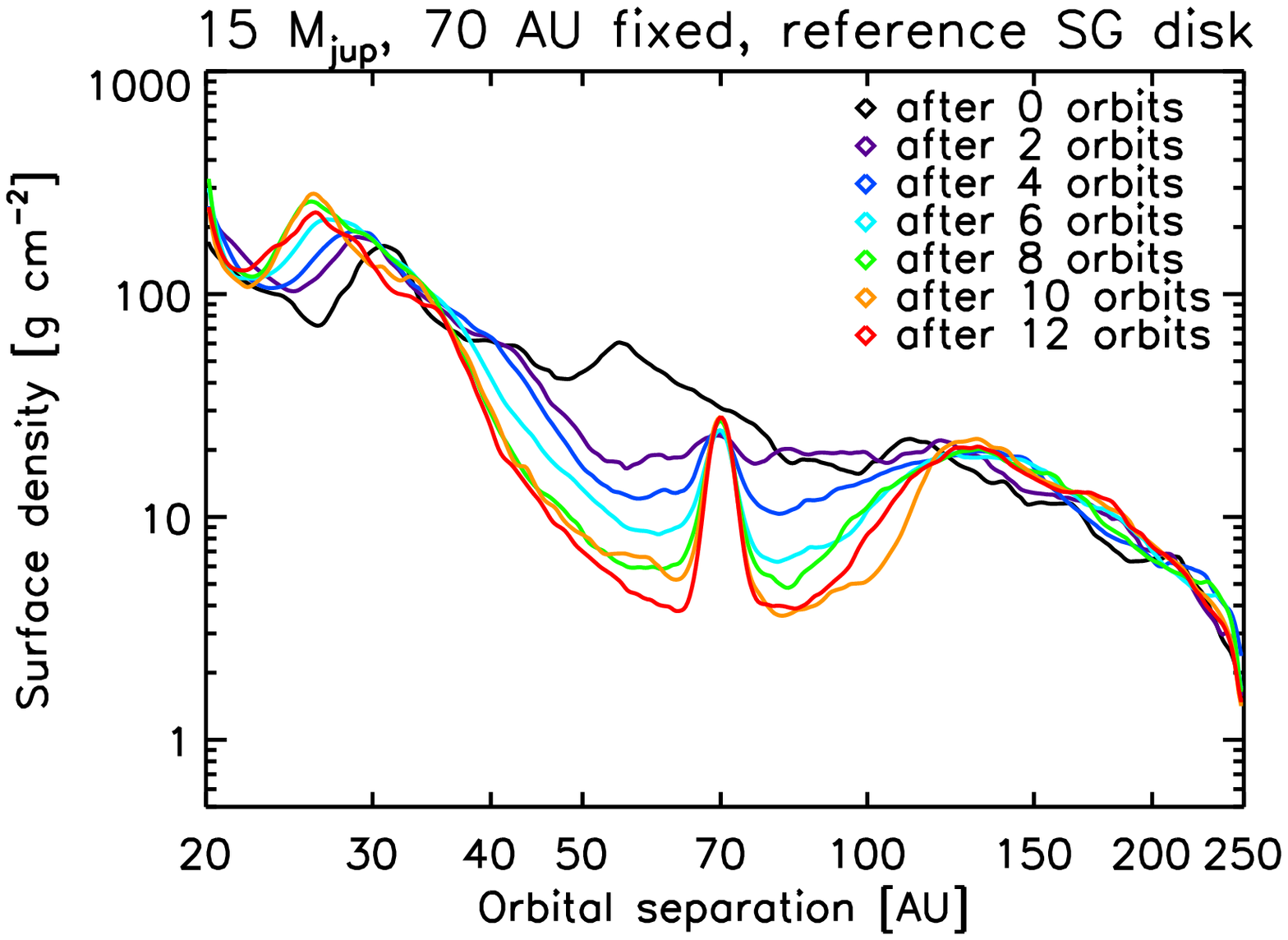}
\end{minipage}
\hfill
\begin{minipage}[t]{0.48\textwidth}
\includegraphics[width=\textwidth]{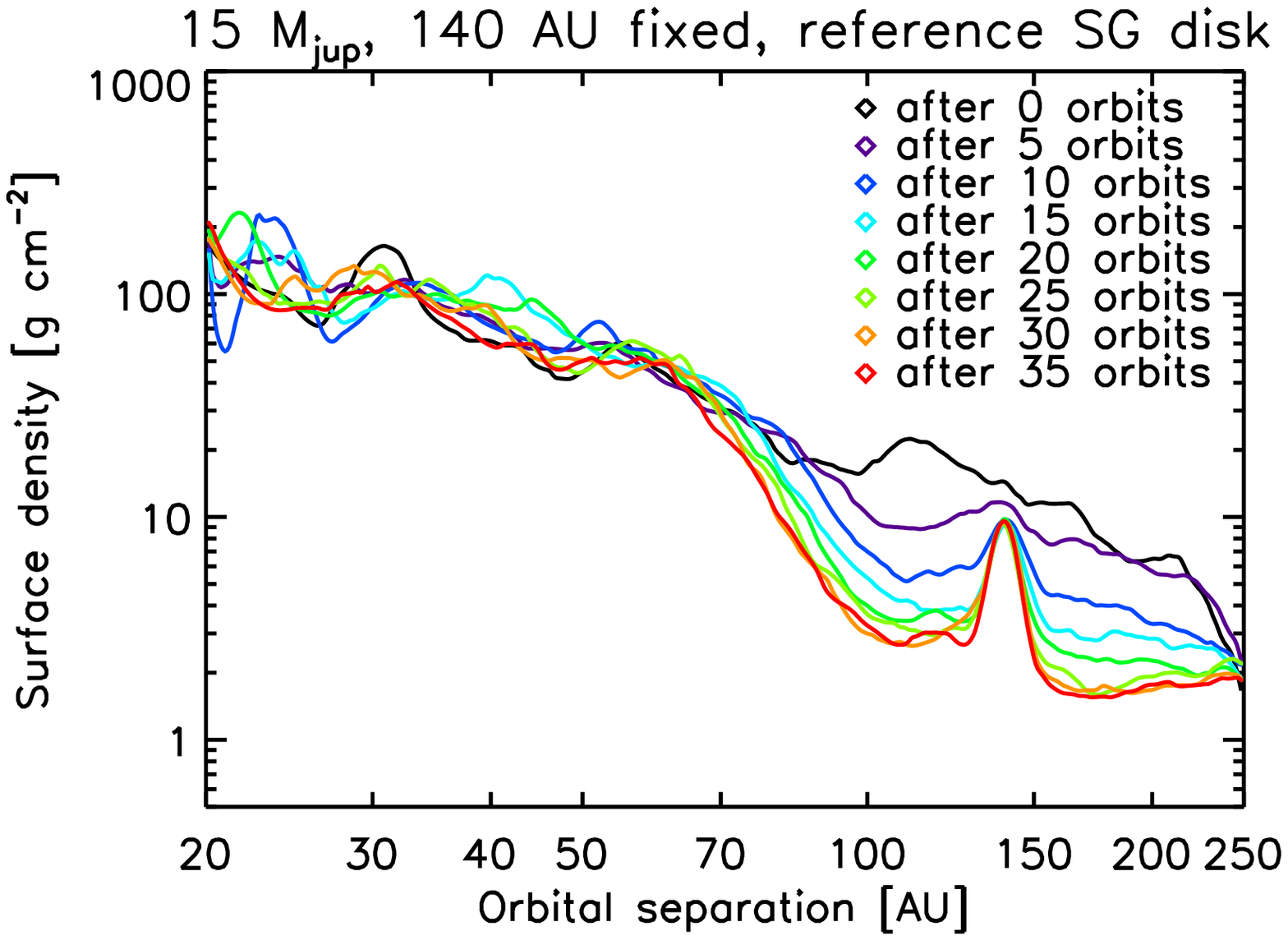}
\end{minipage}
\begin{minipage}[t]{0.48\textwidth}
\includegraphics[width=\textwidth]{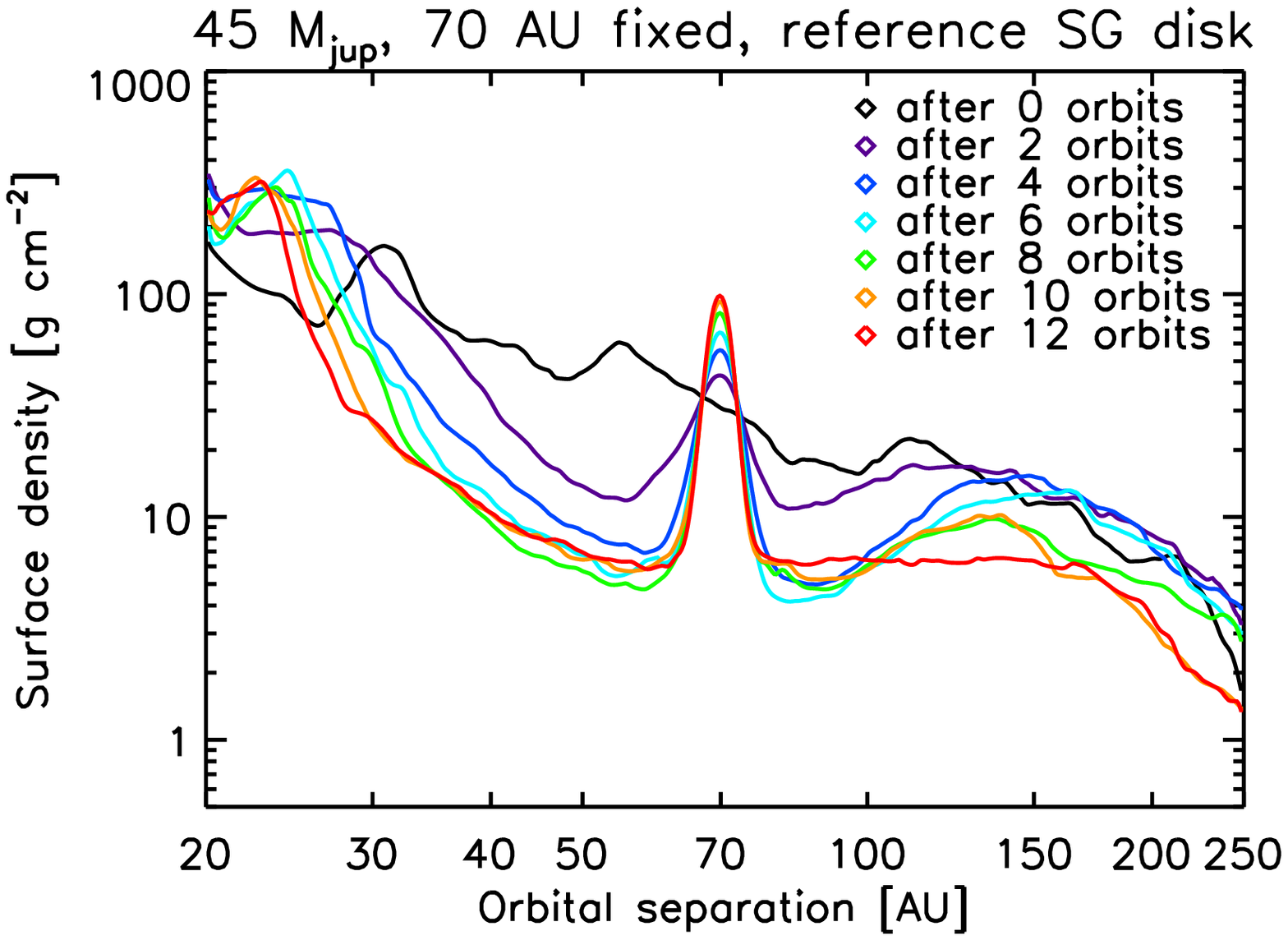}
\end{minipage}
\hfill
\begin{minipage}[t]{0.48\textwidth}
\includegraphics[width=\textwidth]{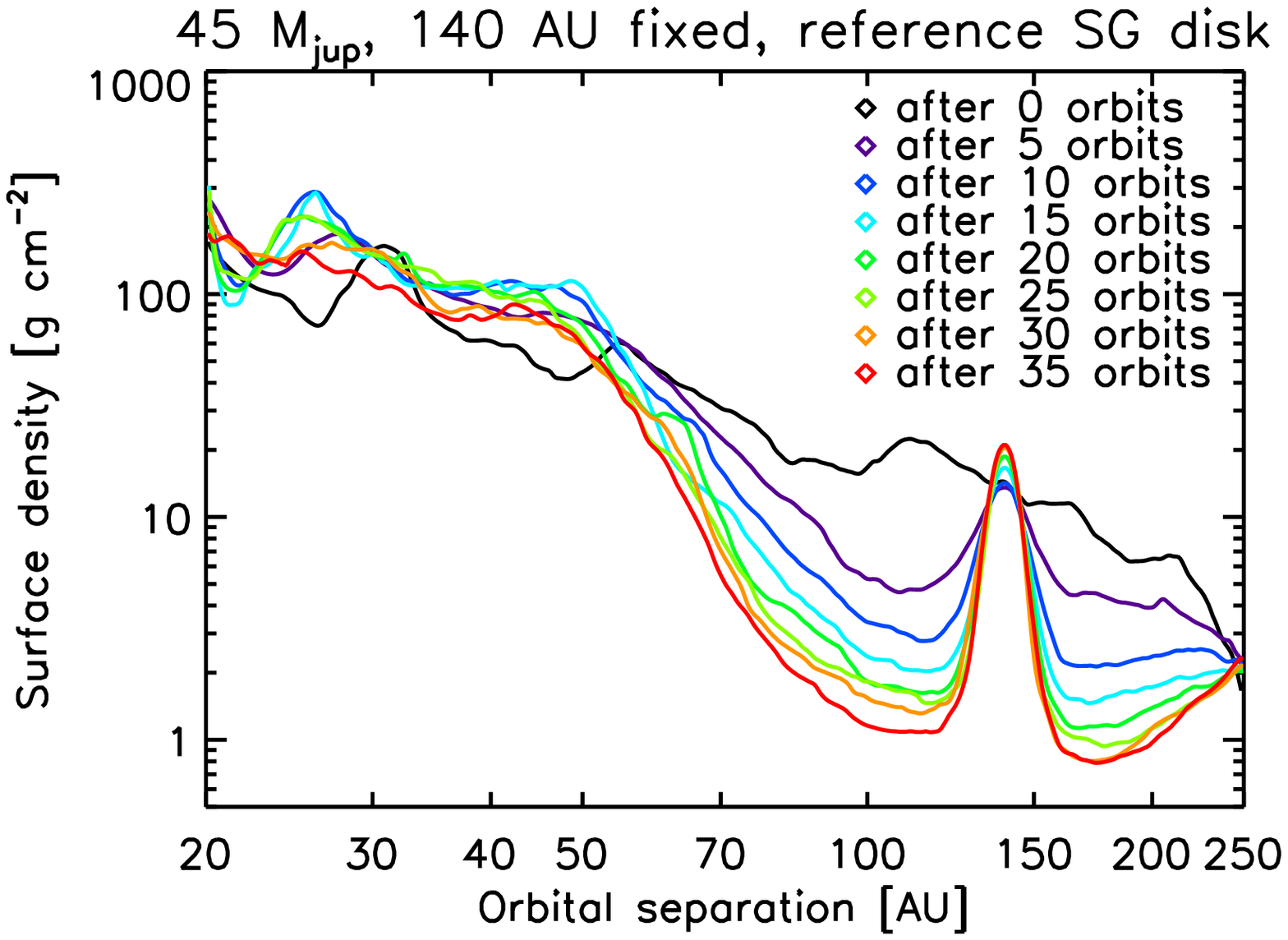}
\end{minipage}
\vspace{-0.3cm}
\caption{Evolution of a gap clearing by a 15 $M_{\rm Jup}$ (top) and 45 $M_{\rm Jup}$ (bottom) companion held on a fixed orbital radius at 70 AU (left) and 180 AU (right) in the reference SG disk. For the embedded 15 $M_{\rm Jup}$ (45 $M_{\rm Jup}$) companion a gap forms after approximately 12 (8) orbits at 70 AU and after 65 (20) orbits at 140 AU. The time it takes for a gap to form is larger than the respective migration timescale shown in Figure~\ref{fig:SG_15_45} for both companions (also see Table \ref{tab:timescales}).}
\label{fig:fixed_15_45}
\end{center}
\end{figure}

Figure \ref{fig:fixed_15_45} illustrates the evolution of the surface density around an embedded 15 $\MJup$ and 45 $\MJup$ companion in the reference SG disk analogous to Figure \ref{fixed}. The timescale to open a gap (8 - 12 orbits at 70 AU and 20 - 65 orbits at 140 AU, where \emph{orbits} is defined as an orbital period at 100 AU) is much longer than the crossing timescale ($\lesssim$ few orbits; Figure~\ref{fig:SG_15_45} left). As with the results presented in Section~\ref{fixie}, this may be the reason why gaps do not open in these disks.

\begin{figure}
\begin{center}
\begin{minipage}[t]{0.48\textwidth}
\includegraphics[width=\textwidth]{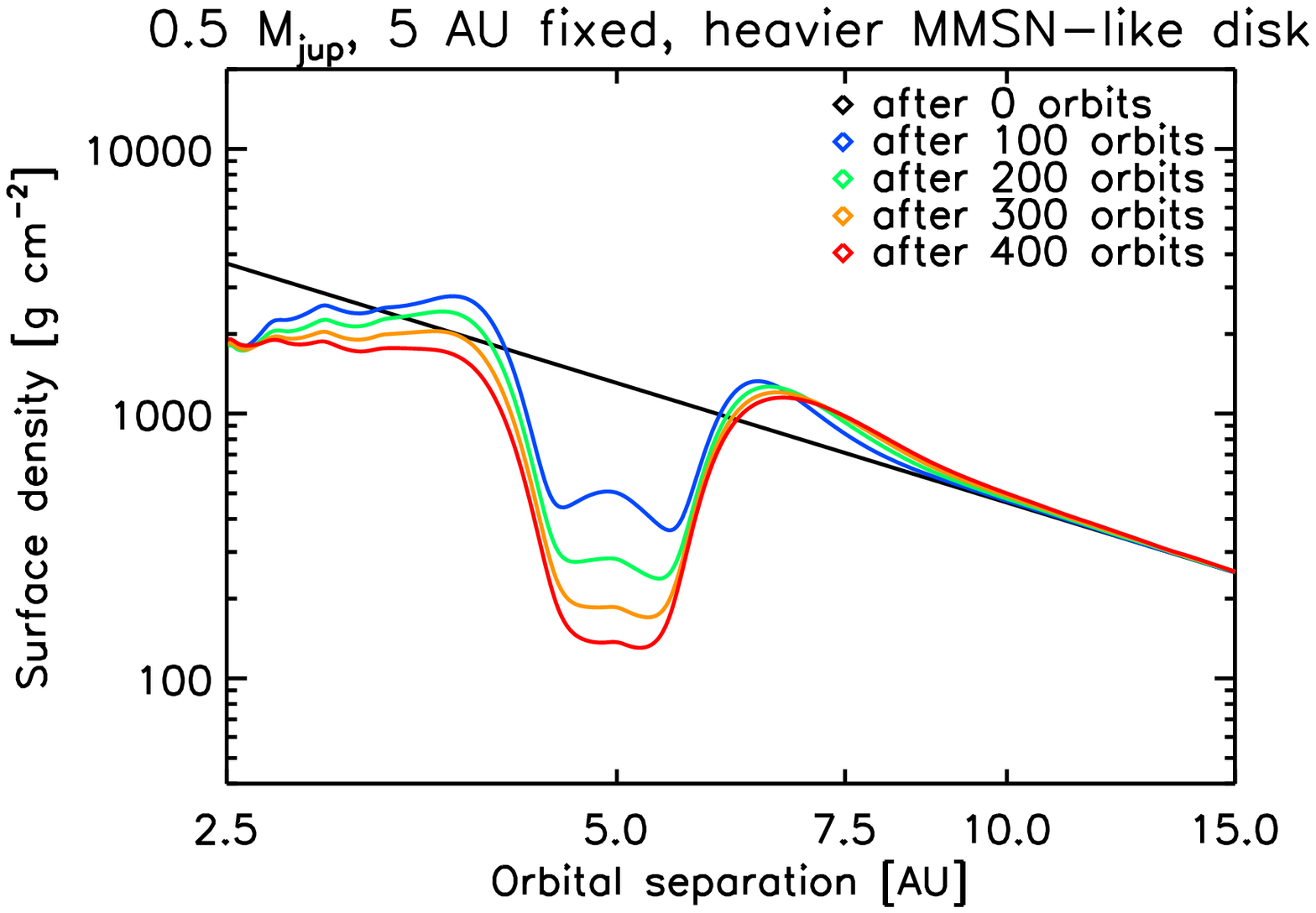}
\end{minipage}
\hfill
\begin{minipage}[t]{0.48\textwidth}
\includegraphics[width=\textwidth]{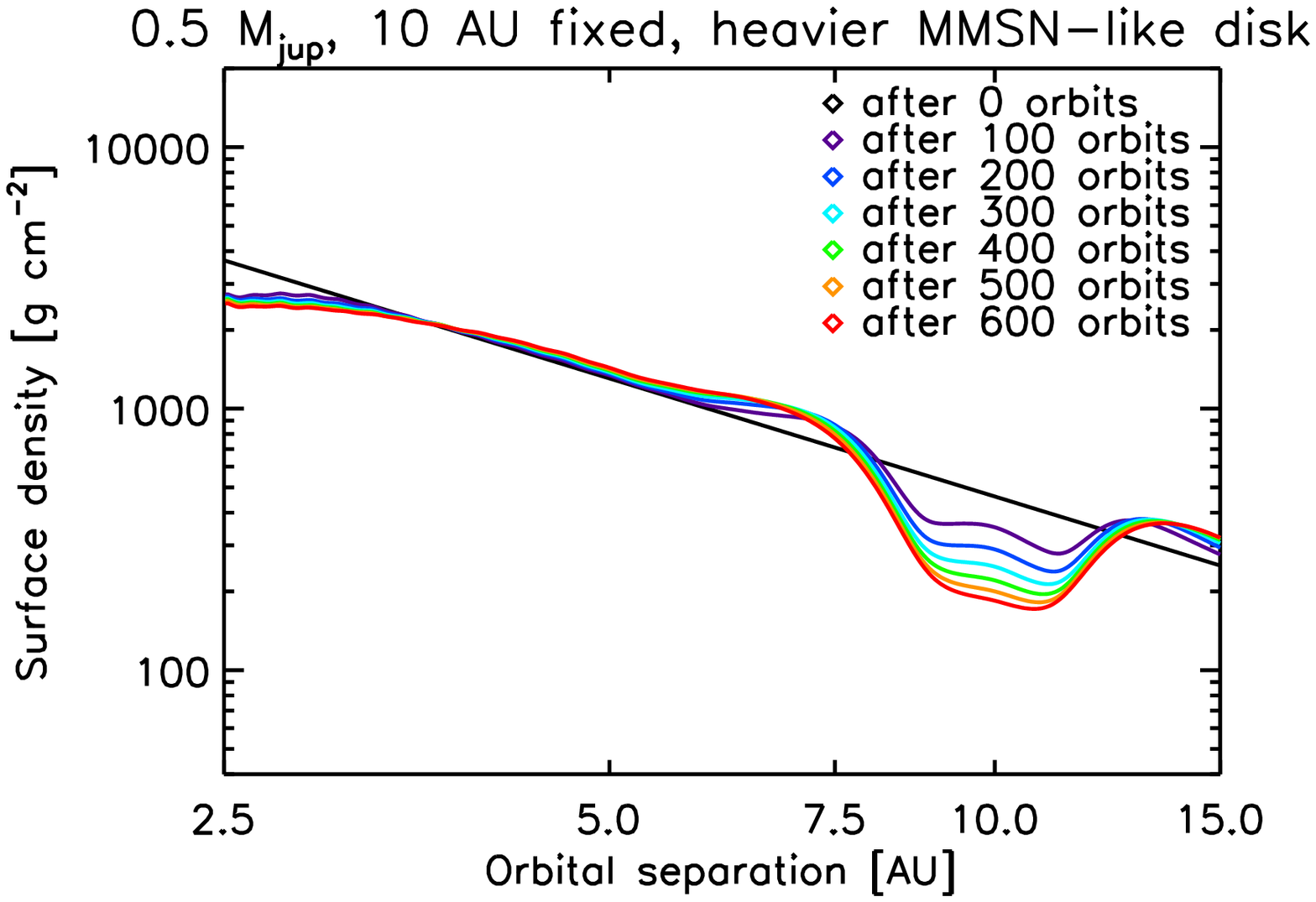}
\end{minipage}
\begin{minipage}[t]{0.48\textwidth}
\includegraphics[width=\textwidth]{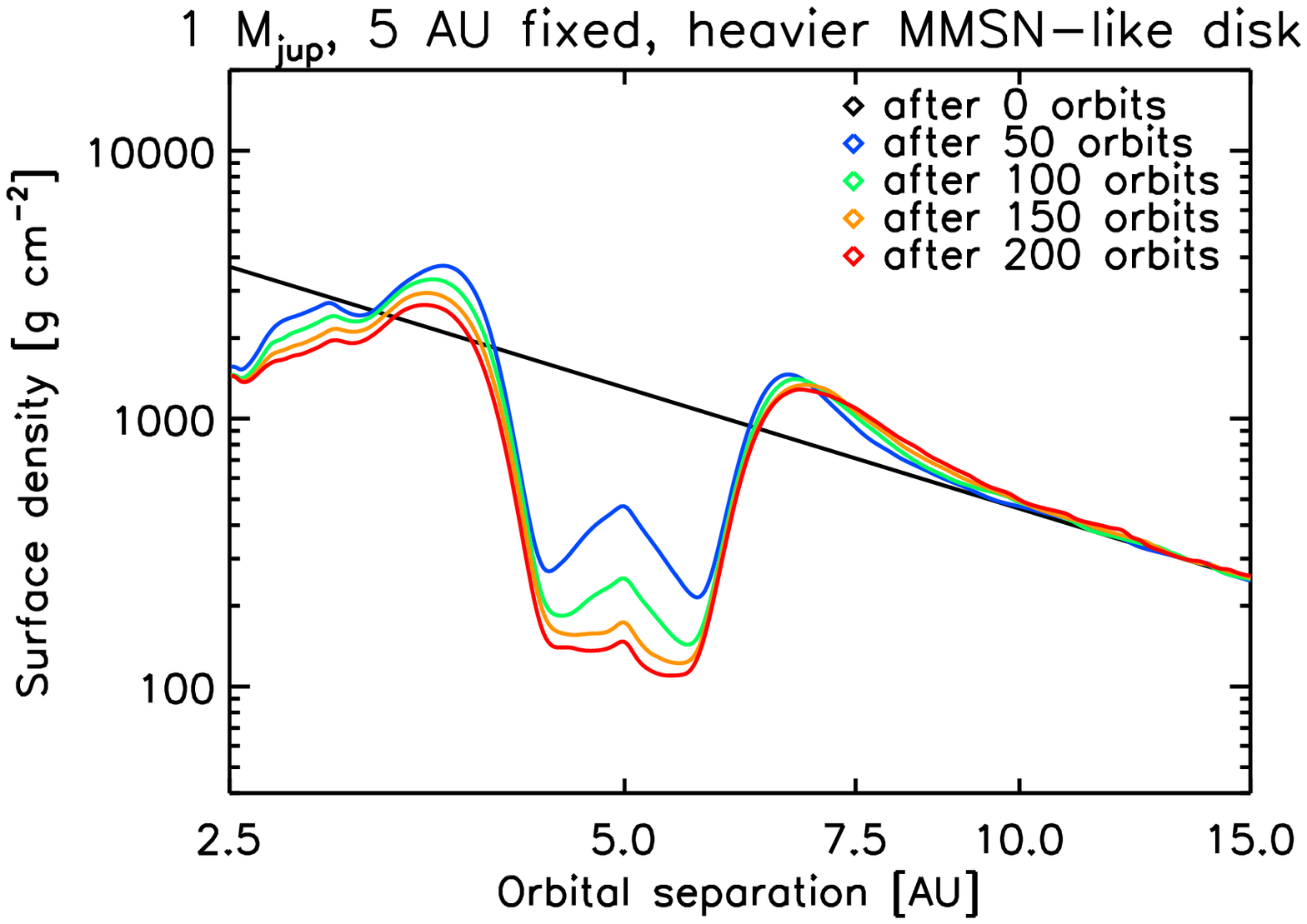}
\end{minipage}
\hfill
\begin{minipage}[t]{0.48\textwidth}
\includegraphics[width=\textwidth]{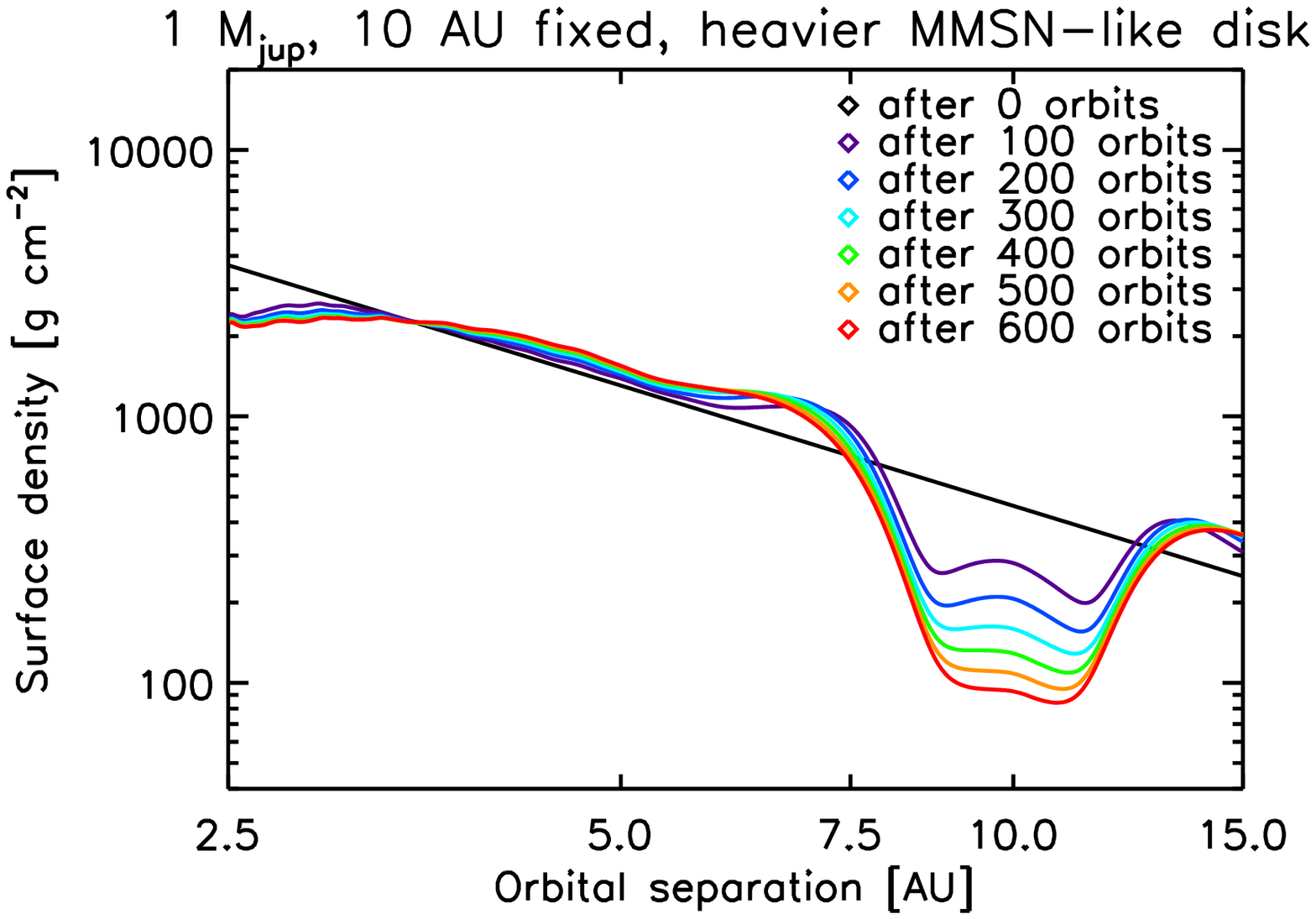}
\end{minipage}
\begin{minipage}[t]{0.48\textwidth}
\includegraphics[width=\textwidth]{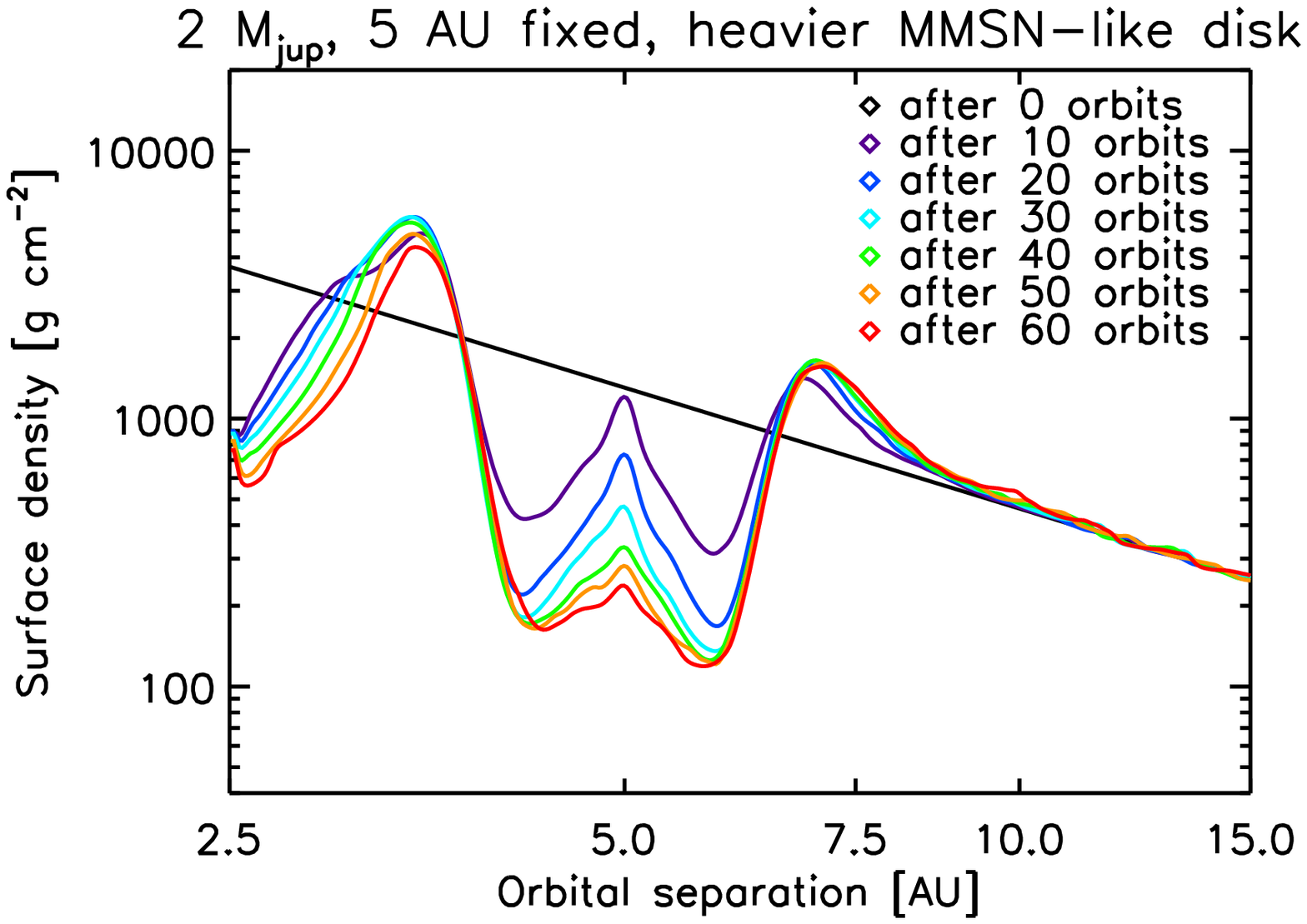}
\end{minipage}
\hfill
\begin{minipage}[t]{0.48\textwidth}
\includegraphics[width=\textwidth]{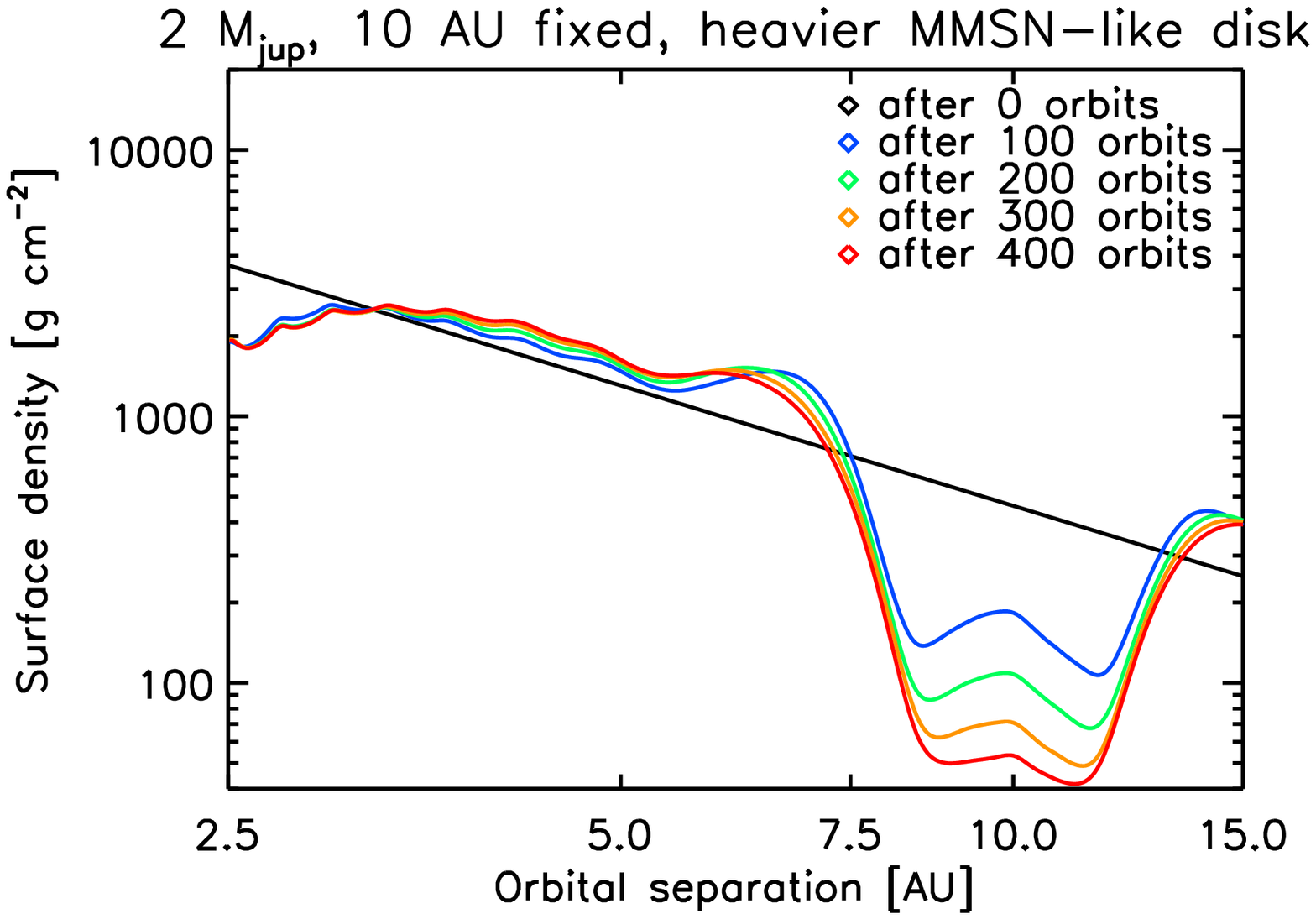}
\end{minipage}
\vspace{-0.3cm}
\caption{Evolution of a gap clearing due to a 0.5 $M_{\rm Jup}$ (top), 1 $M_{\rm Jup}$ (middle) and 2 $M_{\rm Jup}$ companion (bottom) held on a fixed orbital radius at 5 AU (left) and 10 AU (right) in the heavier MMSN-like disk. For the embedded 2 $M_{\rm Jup}$ companion a gap forms after approximately 60 orbits at 5 AU and after 400 orbits at 10 AU. At 5 AU it takes the 0.5 $M_{\rm Jup}$ (1 $M_{\rm Jup}$) companion approximately 400 (200) orbits to form a gap. At 10 AU the 0.5 and 1 $M_{\rm Jup}$ companions do not manage to clear the gap entirely in the 600 orbits simulated. In general, the gap-opening timescale for all companions is much larger than the respective migration timescale shown in Figure~\ref{tinylam}, right panel (also see Table \ref{tab:timescales}).}
\label{densvhfix}
\end{center}
\end{figure}

Figure \ref{densvhfix} illustrates the evolution of the surface density around an embedded 0.5 $\MJup$, 1 $\MJup$ and 2 $\MJup$ companion at 5 AU and 10 AU in the heavier MMSN-like disk. In this disk the gap-opening timescales are much longer than the respective migration timescales (see Fig. \ref{tinylam} on the right), thus preventing gap-opening.

\begin{figure}
\begin{center}
\begin{minipage}[t]{0.48\textwidth}
\includegraphics[width=\textwidth]{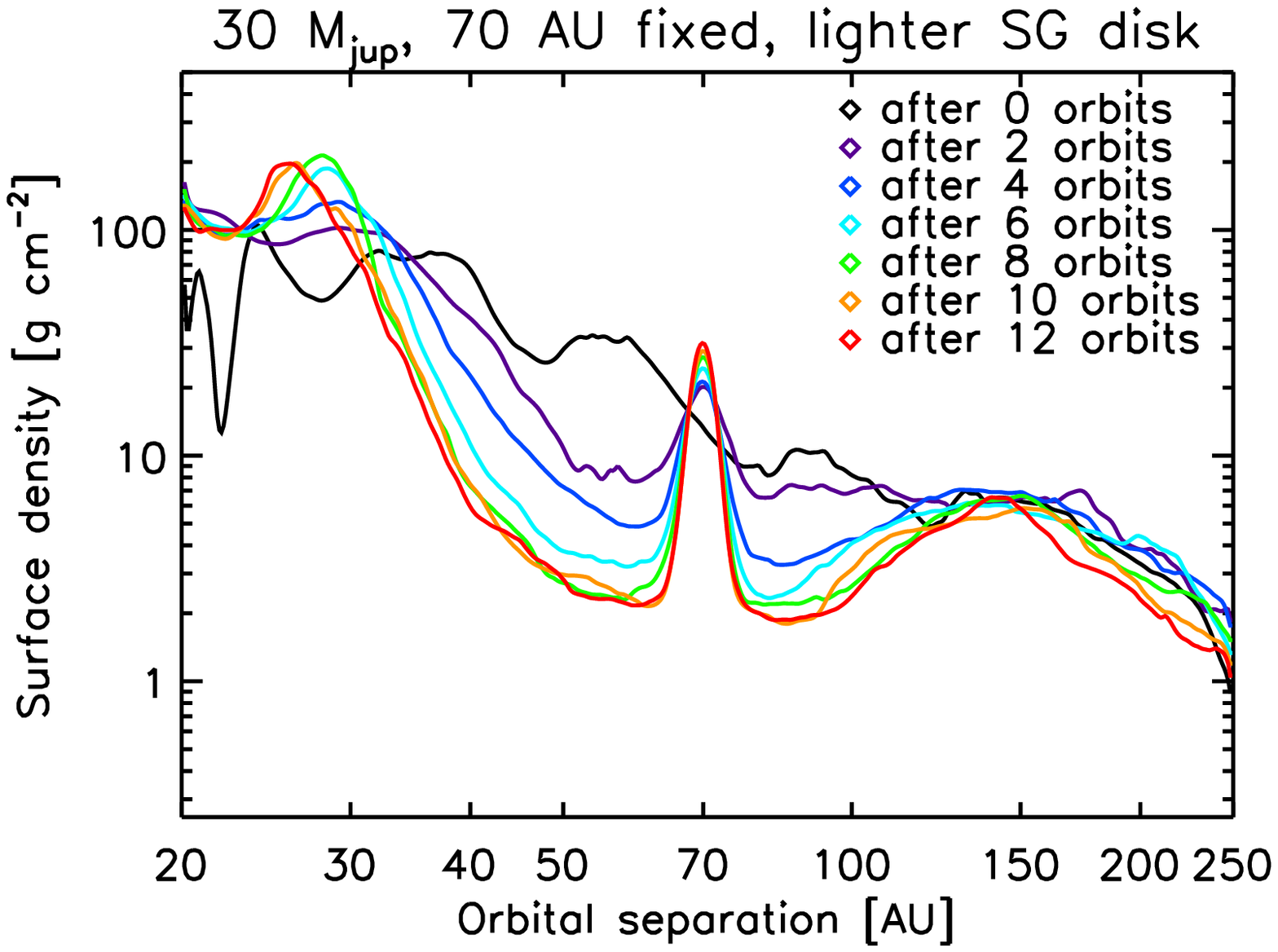}
\end{minipage}
\hfill
\begin{minipage}[t]{0.48\textwidth}
\includegraphics[width=\textwidth]{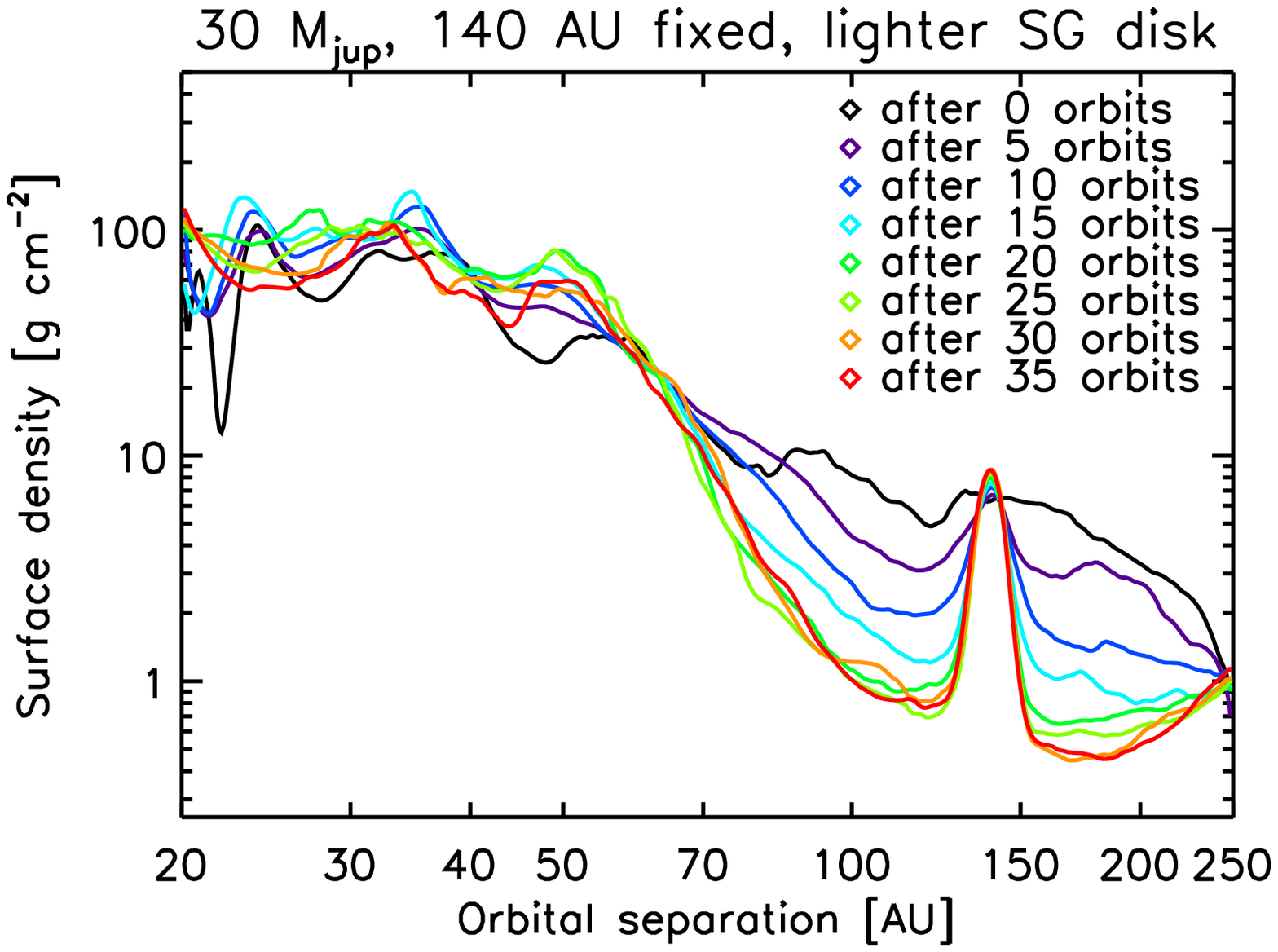}
\end{minipage}
\begin{minipage}[t]{0.48\textwidth}
\includegraphics[width=\textwidth]{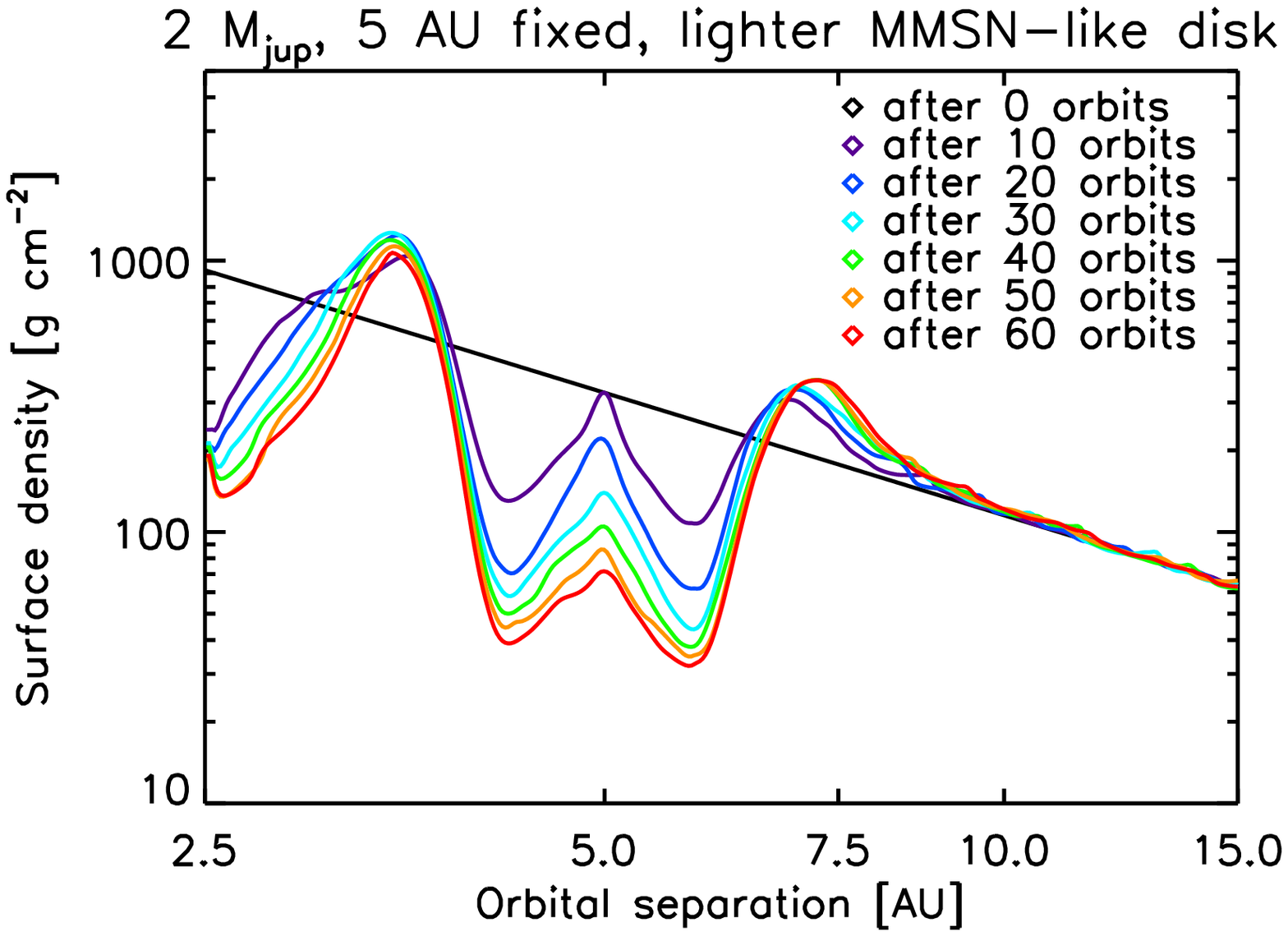}
\end{minipage}
\hfill
\begin{minipage}[t]{0.48\textwidth}
\includegraphics[width=\textwidth]{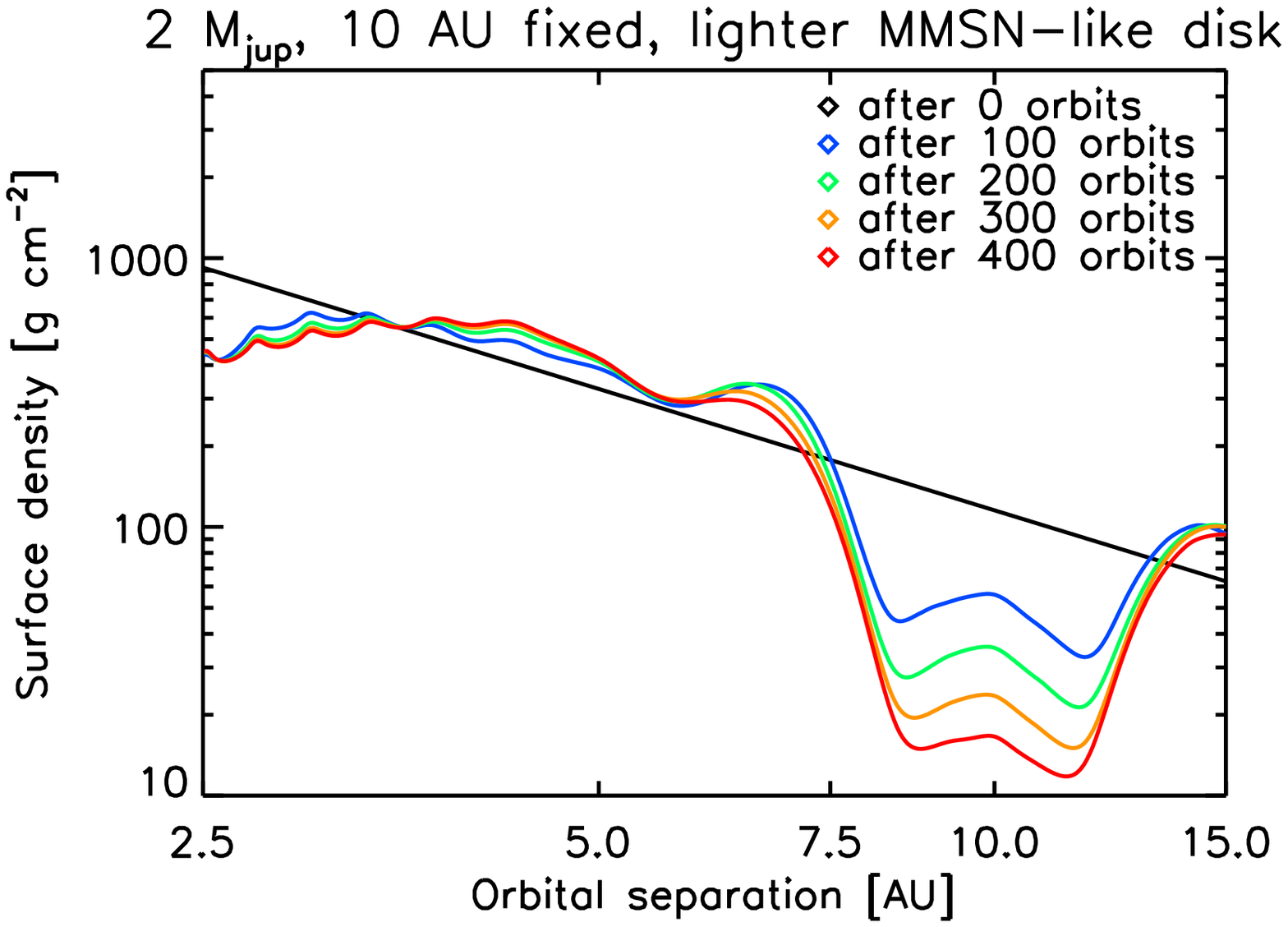}
\end{minipage}
\vspace{-0.3cm}
\caption{Evolution of a gap clearing by a 30 $M_{\rm Jup}$ companion in the lighter SG disk (top) and a 2 $M_{\rm Jup}$ companion in the lighter MMSN-like disk (bottom) held on a fixed orbital radius at 70 AU (top left), 140 AU (top right), 5 AU (bottom left) and 10 AU (bottom right), respectively. The gap-opening timescale appears to be independent of disk mass as can be seen when comparing the density evolutions here with those in the reference SG disk (Fig \ref{fixed}, top panels) and the heavier MMSN-like disk (Fig. \protect\ref{densvhfix}, bottom panels).}
\label{denslighters}
\end{center}
\end{figure}

Figure \ref{denslighters} illustrates the evolution of the surface density around an embedded 30 $\MJup$ companion in the lighter SG disk and a 2 $\MJup$ companion in the lighter MMSN-like disk at 70 AU, 140 AU, 5 AU and 10 AU, respectively. A comparison of the gap evolution with Figures \ref{fixed} (top panel) and \ref{densvhfix} (bottom panels) shows that the gap-opening timescale appears to be independent of mass.

\end{document}